\documentclass[11pt]{article}

\usepackage[utf8]{inputenc}
\usepackage{amsmath,amssymb,latexsym}
\usepackage[T1]{fontenc}
\usepackage{epsfig}
\usepackage{fullpage}
\usepackage{graphics}
\usepackage{graphicx}
\usepackage{makeidx}
\usepackage{slashed}
\usepackage{color}
\usepackage{colortbl}
\usepackage[dvipsnames]{xcolor}
\usepackage{subcaption}
\usepackage{feynmp}
\usepackage{cite}
\usepackage{cancel}
\usepackage{slashed}
\usepackage{mathtools}
\usepackage{bm}

\def\beq{\begin{equation}}
\def\eeq{\end{equation}}
\def\beqa{\begin{eqnarray}}
\def\eeqa{\end{eqnarray}}

\renewcommand{\vec}[1]{\mbox{\boldmath$ #1 $}}

\newcommand{\be}{\begin{equation}}
\newcommand{\ee}{\end{equation}}
\newcommand{\bea}{\begin{eqnarray}}
\newcommand{\eea}{\end{eqnarray}}
\newcommand{\nn}{\nonumber}

\newcommand{\ord}{{\cal O}}
\newcommand{\as}{\alpha_s}
\newcommand{\eps}{\epsilon}
\newcommand{\M}{\mathcal{M}}
\newcommand{\s}{\hat{s}}

\newcommand\Eqns[2]    {Eqs.\,(\ref{#1}) and~(\ref{#2})}

\newcommand\eqn[1]     {eq.\,(\ref{#1})}
\newcommand\eqns[2]    {eqs.\,(\ref{#1}) and~(\ref{#2})}
\newcommand\eqnss[2]   {eqs.\,(\ref{#1})--(\ref{#2})}

\newcommand\refr[1]      {ref.\,\cite{#1}}

\allowdisplaybreaks

\usepackage{jheppub}

\title{Next-to-leading power resummed rapidity distributions near threshold for Drell-Yan and diphoton production}
\author[a]{Robin van Bijleveld,}
\author[a,b,c]{Eric Laenen,} 
\author[d]{Leonardo Vernazza}
\author[a,b,e]{and Guoxing Wang}

\affiliation[a]{Nikhef, Theory Group, Science Park 105, 1098 XG, Amsterdam, The Netherlands}
\affiliation[b]{IoP/ITFA, University of Amsterdam, Science Park 904, 1098 XH Amsterdam, The Netherlands}
\affiliation[c]{ITF, Utrecht University, Leuvenlaan 4, 3584 CE Utrecht, The Netherlands}
\affiliation[d]{INFN, Sezione di Torino, Via P. Giuria 1, I-10125 Torino, Italy}
\affiliation[e]{Zhejiang Institute of Modern Physics, School of Physics, Zhejiang University, \\
No. 866 Yuhangtang Road, Hangzhou 310058, China}

\abstract{We consider Drell-Yan production and QCD-induced diphoton production and compute their rapidity distributions up to next-to-leading power (NLP) in the threshold variable. We give results for rapidity distributions of the  Drell-Yan process up to NNLO accuracy and show that a factorised structure occurs for the leading logarithms (LL) at NLP, generalising the result at leading power. For diphoton production, we generalise methods based on kinematical shifts to find the NLO cross section up to NLP for rapidity distributions. From the results for these two processes, we derive resummed cross sections at NLP LL accuracy that are double differential in the threshold variable and the rapidity variable, which generalise results for single differential resummed cross sections.}

\begin{document}
\maketitle
\flushbottom


\section{Introduction}
\label{intro}

The increasing precision of experimental 
measurements from the Large Hadron Collider 
requires similar improvements in the accuracy 
of Standard Model theoretical predictions, 
especially for perturbative QCD. Progress 
can be made by including ever higher 
order contributions in the strong coupling 
constant, and by resumming certain kinematically 
enhanced contributions to all orders in perturbation 
theory, e.g.\ the production of particles near 
their kinematic threshold. In such processes, 
one can define a (partonic) threshold variable 
$\xi$, such that $\xi \to 0$ when approaching 
threshold. In this regime partonic cross sections 
take the generic form
\be\label{threshold-general}
\frac{d\hat \sigma}{d\xi} = \sigma_0 \sum_{n = 0}^{\infty}
\bigg(\frac{\alpha_s}{\pi} \bigg)^n \sum_{m = 0}^{2n-1}
\bigg[ c_{nm}^{(-1)} \bigg( \frac{\log^m \xi}{\xi}\bigg)_+
+ c_{n}^{(\delta)} \delta(\xi) + c_{nm}^{(0)} \log^m \xi
+\ord(\xi) \bigg],
\ee
where $\sigma_0$ represents the Born-level 
cross section. The summation of logarithms 
related to the first contribution in the 
square brackets on the r.h.s.\ of 
\eqn{threshold-general} has been much 
studied, see e.g.\ refs.~\cite{Altarelli:1979ub,Parisi:1980xd,Curci:1979am,Sterman:1987aj,Catani:1989ne,Catani:1990rp,Korchemsky:1993xv,Korchemsky:1993uz,Forte:2002ni,Contopanagos:1997nh,Becher:2006nr,Schwartz:2007ib,Bauer:2008dt,Chiu:2009mg}, 
and also the summation of the second 
contribution proportional to $\delta(\xi)$
is known for processes with electroweak 
final states (see, for example, ref.~\cite{Eynck:2003fn,Ahrens:2009cxz}). 
These two terms define the \emph{leading power} 
(LP) (in $\xi$) contribution to the partonic 
cross section. Next-to-leading power (NLP) 
contributions are given in the third term, 
and their summation has seen much recent 
study. While they are not as divergent as 
their LP counterparts, they are still 
singular in the limit $\xi\to 0$, and 
can be numerically sizeable in this 
region (see e.g.\ \cite{vanBeekveld:2021hhv}). 
Mapping the structure of large NLP logarithms 
is more challenging than at LP, where large 
logarithms can be related to the emission 
of soft and collinear gluons \cite{DelDuca:1990gz,Bonocore:2014wua,Bonocore:2015esa},
and the factorisation of the cross section 
into universal jet and soft functions, together 
with a process-dependent hard function. At NLP 
radiation becomes sensitive to additional features 
of the underlying hard scattering process, such
as the spin of the emitting particles, and it 
starts to resolve the structure of the hard 
scattering kernel, as well as the structure 
of clusters of virtual particles collinear 
to the directions of the incoming/outgoing 
hard particles.
These features have been investigated
both by means of direct-QCD (or diagrammatic) 
\cite{Laenen:2008ux,Laenen:2008gt,Laenen:2010uz,Bonocore:2015esa,Bonocore:2016awd,DelDuca:2017twk,Gervais:2017yxv,Bahjat-Abbas:2018hpv,vanBeekveld:2019prq,Laenen:2020nrt,Bonocore:2020xuj,Bonocore:2021qxh,Bonocore:2021cbv,Agarwal:2023fdk,Ajjath:2020ulr,Ajjath:2020sjk,Ajjath:2020lwb,Ajjath:2022kyb,Engel:2021ccn,Engel:2023ifn,Czakon:2023tld}
and effective field theory methods, the latter 
based on soft-collinear effective theory (SCET)
\cite{Larkoski:2014bxa,Moult:2016fqy,Moult:2017rpl,Moult:2017jsg,Ebert:2018lzn,Moult:2019mog,Moult:2019uhz,Beneke:2017ztn,Beneke:2018rbh,Beneke:2019kgv,Beneke:2019oqx,Broggio:2021fnr,Broggio:2023pbu}.

Currently, resummation of the large threshold 
logarithms at NLP is achieved at LL accuracy 
for scattering processes with a colour-singlet 
final state. This is done both in the direct-QCD 
approach \cite{Bahjat-Abbas:2019fqa,vanBeekveld:2019cks,vanBeekveld:2021hhv,vanBeekveld:2021mxn,Ajjath:2021pre,Ajjath:2021lvg,Ajjath:2021bbm,Czakon:2023tld}
as in the SCET approach \cite{Moult:2018jjd,Beneke:2018gvs,Beneke:2019mua,Beneke:2020ibj,Beneke:2022obx}. However, these discussions are 
limited to only a few observables, such as an 
inclusive cross section or invariant mass 
distribution. It is also important to consider 
other observables, in particular more 
differential distributions. We note that 
the Drell-Yan rapidity distribution near 
threshold has been recently investigated 
at NLP from a somewhat different perspective 
in \cite{Ajjath:2020lwb,Ajjath:2021pre}.

In this paper we look specifically at rapidity 
distributions and cross sections that are 
double differential in the threshold variable 
(or invariant mass), and in the rapidity of 
(one of) the final state particle(s). It is 
well known that at leading power the partonic 
cross section for rapidity distributions takes 
on a factorised form \cite{Laenen:1992ey,Bolzoni:2006ky,Bonvini:2010tp,Bonvini:2012sh,Bonvini:2023mfj}, 
which enables the result for the invariant 
mass distribution to be conveniently extended 
to cross sections that are differential in the 
rapidity of one of the outgoing particles as well. 
In this paper we investigate whether, or to what 
extent, the results obtained in 
refs.~\cite{DelDuca:2017twk,Bahjat-Abbas:2019fqa}, 
which are both next-to-leading power results, can 
be extended to include rapidity distributions for 
scattering processes with colour-singlet final 
states. One of the main results of the latter 
references is a certain  universality of the 
NLO partonic cross section up to NLP accuracy 
for colour-singlet final states, in that they 
be written as a universal factor multiplied by 
the Born-level partonic cross section with 
shifted Born kinematics. We investigate here 
whether such universality also exists in the 
case of rapidity distributions, which would 
moreover aid the resummation of the leading 
logarithms at next-to-leading power for 
rapidity distributions.

The structure of our paper is as follows. In Section \ref{sectionDY}, we investigate the perturbative structure of the Drell-Yan cross section differential in the invariant mass and the rapidity of the virtual photon, in particular whether a factorised form occurs.  In Section \ref{diphoton}, we analyse the QCD-induced production of two photons, where we stay differential in the rapidity of one of the photons. We compute the NLP corrections at NLO both directly and by exploiting the method of ref.~\cite{DelDuca:2017twk}. In Section \ref{resum} we perform the resummation of the leading threshold logarithms at NLP for both diphoton production and the Drell-Yan process rapidity distributions. Section \ref{conclusions} contains our conclusions, while appendices contain useful expressions for phase space measures.


\section{Rapidity distribution for fixed-order Drell-Yan process}
\label{sectionDY}
\begin{figure}[h]
\begin{center}
\includegraphics[width=0.25\textwidth]{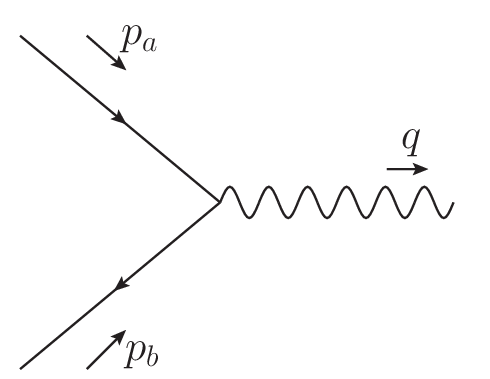} 
\end{center}
\caption{Tree-level contribution to the Drell-Yan process.}
\label{fig:DYLO}
\end{figure} 
\noindent
In this section we consider the 
Drell-Yan process 
\be
A(p_A) + B(p_B) \to \gamma^*(q) + X(p_X),
\ee
where $A(p_A)$ and $B(p_B)$ represent the 
incoming hadrons with respectively momenta $p_A$ and $p_B$,
with $(p_A+p_B)^2 =s$, producing an off-shell 
photon followed by the decay $\gamma^*(q) \to l^- l^+$, 
and where $X(p_X)$ denotes the unobserved QCD final 
state. We focus on the cross section differential 
in the invariant mass and rapidity of the off-shell 
photon in the collider centre of mass frame
\be
Q^2 \equiv q^2, 
\qquad \qquad 
Y \equiv  \frac{1}{2} 
\log \bigg(\frac{q^0 + q^3}{q^0 - q^3}\bigg),
\ee
which reads 
\be\label{eqHadron7b}
\frac{d\sigma}{d Q^2\, dY} = 
\sigma_0  \sum_{ab} 
\int_{\tau}^{1} \frac{dz}{z} \int_0^1 dy \, 
{\cal L}_{ab}(z,y) \, \Delta_{ab}(z,y),
\ee
where 
\be\label{sigma0deltaDY}
\sigma_0 = \frac{4\pi \alpha^2 e_q^2}{3 N_c Q^2 s}\,,
\ee
and $\alpha =e^2/4\pi$ is the electromagnetic coupling.
Following \cite{Anastasiou:2003yy,Becher:2007ty}, 
the luminosity function ${\cal L}_{ab}(z,y)$ 
and the partonic cross section $\Delta_{ab}(z,y)$
are expressed in terms of the variables 
\be\label{zydef}
z \equiv \frac{Q^2}{\hat s} = \frac{Q^2}{x_a x_b s} 
= \frac{\tau}{x_a x_b},
\qquad\qquad 
y = \frac{u-z}{(1-z)\big(1+u\big)},
\ee
where in turn 
\be\label{yhatudef}
u \equiv e^{-2 \hat Y}, 
\qquad {\rm with} \qquad  
\hat Y = Y - \frac{1}{2}\log \bigg(\frac{x_a}{x_b}\bigg).
\ee
Here $\hat Y$ represents the rapidity in the
partonic centre of mass frame; similarly, 
$\hat s = (p_a + p_b)^2$ in \eqn{zydef}
represents the partonic centre of mass energy.
The variables $x_{a}$, $x_{b}$ represent the momentum 
fractions relating the partonic to the hadronic incoming
momenta via $p_a = x_a p_A$, $p_b = x_b p_B$. For 
future reference, we note that the variables 
above are well-defined in the range
\be
\tau \leq z \leq 1,
\qquad
\log \sqrt{z} \leq \hat Y \leq \log \frac{1}{\sqrt{z}}, 
\qquad 
z \leq u \leq \frac{1}{z}, 
\qquad 
0\leq y\leq 1\,.
\ee
For Born kinematics one has
$z = u = 1$, $y = 1/2$. Though \eqn{eqHadron7b} 
contains a sum over the partonic channels 
$a$, $b$, we consider here only 
the leading quark-antiquark annihilation channel
$q(p_a) + \bar q(p_b) \to \gamma^*(q)$ 
(plus the symmetric contribution 
$q\leftrightarrow \bar q$), see figure \ref{fig:DYLO}, and therefore drop 
the indices $a$, $b$. Given these 
definitions, the luminosity function 
in \eqn{eqHadron7b} reads 
\bea\label{luminosityRap} 
{\cal L}(z,y)  \\  \nn
&&\hspace{-1.2cm}=\,  
f_{q/A}\bigg(e^{Y}\sqrt{\frac{\tau}{z} 
\frac{1-(1-y)(1-z)}{1-y (1-z)}}\bigg) \, 
f_{\bar q/B}\bigg(e^{-Y} \sqrt{\frac{\tau}{z}
\frac{1-y(1-z)}{1-(1-y)(1-z)}}\bigg)
+ (q\leftrightarrow \bar q).
\eea
In the threshold region, defined by the 
limit $(1-z) \to 0$, the cross section
is conveniently approximated by a power 
expansion in $(1-z)$,
\be\label{zPowerExpansion}
\Delta(z,y) = \Delta_{\rm LP}(z,y)
+\Delta_{\rm NLP}(z,y) + \ord(1-z),
\ee
and it develops large logarithms 
of $(1-z)$ that need to be resummed.
At leading power in $(1-z)$, the resummation of 
large logarithms in $(1-z)$ for the 
rapidity distribution is relatively easy 
because of the factorisation 
\cite{Laenen:1992ey,Mukherjee:2006uu,Bolzoni:2006ky}
\be\label{zyFactLP}
\Delta_{\rm LP}(z,y) = 
\frac{\delta(y)+\delta(1-y)}{2} 
\Delta_{\rm LP}(z),
\ee
where $\Delta_{\rm LP}(z)$ represents 
the partonic cross section integrated 
over the rapidity, i.e.\ the partonic
invariant mass distribution. Thus, 
the resummation of large logarithms 
in $(1-z)$ trivially follows from 
the resummation obtained for the 
invariant mass distribution. 

Recently, much effort has been 
devoted to the development of 
resummation techniques for the 
NLP term of the invariant mass
distribution, $\Delta_{\rm NLP}(z)$, 
\cite{Beneke:2018gvs,Bahjat-Abbas:2019fqa}.
It is therefore interesting to ask 
to what extent such techniques may 
be applicable to $\Delta_{\rm NLP}(z,y)$
in \eqn{zPowerExpansion}, provided
a factorisation such as the one
in \eqn{zyFactLP} may be established
for $\Delta_{\rm NLP}(z,y)$. In order 
to pursue this investigation, we
start by analysing the perturbative 
structure of $\Delta_{\rm NLP}(z,y)$. 

In general, the partonic cross section 
$\Delta(z,y)$ in $d = 4-2\eps$ spacetime
dimensions takes the form 
\bea\label{PartonicRapidityDef}
\Delta(z,y) =  
\frac{1}{(2\pi)^d}\, 
\frac{(-g^{\mu\nu}) W_{\mu\nu}}{1-\eps}
\, \delta(q^2 - Q^2) \, 
\delta\bigg[y - \frac{p_a \cdot q - z \, p_b \cdot q}{(1-z)
(p_a \cdot q + p_b \cdot q)} \bigg],
\eea
where $W_{\mu\nu}$ represents 
the Drell-Yan hadronic tensor, 
and the normalisation is fixed 
such that the tree-level partonic 
cross section for the invariant 
mass reads $\Delta^{(0)}(z) = 
\delta (1-z)$. Here $\Delta(z)$
is given by the identity
\be
\int_0^1 dy \, 
{\cal L}(z,y) \, \Delta(z,y)
= {\cal L} \bigg(\frac{\tau}{z}\bigg) \, \Delta(z),
\ee
where
\be\label{luminosityInvMass}
{\cal L}(v) = \int_{v}^{1} \frac{dx}{x} 
\, f_{q/A}(x) \, f_{\bar q/B}\bigg(\frac{v}{x}\bigg)
+ (q\leftrightarrow \bar q),
\ee
such that 
\be\label{eqHadron7a}
\frac{d\sigma}{d Q^2} = \sigma_0 \int_{\tau}^{1} \frac{dz}{z} \, 
{\cal L} \bigg(\frac{\tau}{z}\bigg) \, \Delta(z).
\ee
The leading order contribution to 
$\Delta(z,y)$ can be easily calculated
in terms of the squared matrix element 
of the process $q\bar q \to \gamma^*$: 
\be\label{PartonicRapTree0}
\Delta^{(0)}(z,y) =
\frac{1}{4N_c}\frac{1}{2\pi}  
\int d\Phi_{\gamma^*} 
\sum_{\rm s,c,p}
\big|{\cal M}_{q\bar q \to \gamma^*}^{(0)}\big|^2,
\ee
where 
\be\label{M0sqDY}
\sum_{\rm s,c,p} 
\big|{\cal M}_{q\bar q \to \gamma^*}^{(0)}\big|^2 
= 4(1-\eps)\hat s N_c,
\ee
where the sum is over spin, colour and 
polarisation, and the phase space for 
the production of the off-shell photon 
$\int d \Phi_{\gamma^*}$ is defined in 
Appendix \ref{PhaseSpace-DY}. In 
\eqn{M0sqDY} and in what follows we set $e_q =1$, 
$e = 1$, given that these factors are already 
included in $\sigma_0$, see 
\eqns{eqHadron7b}{sigma0deltaDY}. 
Upon integration one easily obtains 
\be\label{PartonicRapTree} 
\Delta^{(0)}(z,y) = \delta(1-z) \, 
\delta\bigg(y - \frac{1}{2} \bigg).
\ee
It can be shown \cite{Bonvini:2012sh} 
that the tree result in \eqn{PartonicRapTree} 
is indeed compatible, up to NLP, with the 
structure in \eqn{zyFactLP}. Inserting 
\eqn{PartonicRapTree} into \eqn{eqHadron7b} 
gives 
\be\label{eqHadron8Tree}
\frac{d\sigma}{d Q^2\, dY} = 
\sigma_0   
\int_{\tau}^{1} \frac{dz}{z} \, 
{\cal L}\bigg(z,\frac{1}{2}\bigg) 
\, \Delta^{(0)}(z),
\ee
while using the factorised form of 
\eqn{zyFactLP} into \eqn{eqHadron7b}
leads to 
\be\label{eqHadron8TreeB}
\frac{d\sigma}{d Q^2\, dY} = 
\sigma_0   
\int_{\tau}^{1} \frac{dz}{z} \, 
\frac{{\cal L}(z,0) + {\cal L}(z,1)}{2} 
\, \Delta^{(0)}(z).
\ee
As discussed in \cite{Bonvini:2012sh}, 
near threshold the luminosity functions 
involved in \eqns{eqHadron8Tree}{eqHadron8TreeB}
are equivalent, up to corrections starting at 
NNLP:
\be\label{yEquivalenceNNLP}
\frac{{\cal L}(z,0) + {\cal L}(z,1)}{2} 
= {\cal L}\bigg(z,\frac{1}{2}\bigg) 
+\ord\big[(1-z)^2\big],
\ee
so that for our purposes,
\eqns{eqHadron8Tree}{eqHadron8TreeB}
can be considered to be equivalent.
For future reference, let us define 
\be\label{PartonicRapTreeBar} 
\Delta^{(0)}(z,y) = 
\frac{\delta(y)+\delta(1-y)}{2} \,
\delta(1-z) \equiv 
\bar \Delta^{(0)}(y) \, \delta(1-z).
\ee
With these results at hand, we have 
now the tools to investigate the 
structure of the perturbative
corrections to \eqn{PartonicRapTree}.
Our goal is to determine the contribution
to the partonic differential distribution 
up to second order in perturbation theory:
indeed, this is necessary as the structure 
of the soft expansion within the method of 
regions \cite{Beneke:1997zp} is fully 
revealed only starting at NNLO. In this 
regard, let us notice that, to the best of 
our knowledge, the analytic results 
presented for the NNLO contribution 
in Section \ref{sec:NNLO1r1v} are given 
here for the first time, and provide a 
useful database for investigations of 
the Drell-Yan rapidity distribution 
at NLP, at all logarithmic accuracy.


\subsection{Next-to-leading order}
\label{sec:NLO}

At NLO one needs to take into account 
the emission of a virtual gluon or a real 
soft gluon. The two contributions 
take the form 
\bea
\label{PartonicRapidityDefB-1virtual}
\Delta^{(1)}(z,y)|_{\rm virtual} &=& 
\frac{1}{4N_c} \frac{1}{2\pi} 
\int d \Phi_{\gamma^*} 
\sum_{\rm s,c,p}
2{\rm Re}\Big[{\cal M}_{q\bar q \to \gamma^*}^{(0)*}
{\cal M}_{q\bar q \to \gamma^*}^{(2)}\Big], \\ 
\label{PartonicRapidityDefB-1real}
\Delta^{(1)}(z,y)|_{\rm real} &=& 
\frac{1}{4N_c}\frac{1}{2\pi}  
\int d\Phi_{\gamma^*g} 
\sum_{\rm s,c,p}
\big|{\cal M}_{q\bar q \to \gamma^* g}^{(1)}\big|^2 ,
\eea
where the phase space for the production of 
the off-shell photon $\int d \Phi_{\gamma^*}$,
as well as the phase space involving an additional 
soft gluon $\int d\Phi_{\gamma^*g}$  
are defined in Appendix \ref{PhaseSpace-DY}.
In \eqn{PartonicRapidityDefB-1virtual}, 
${\cal M}_{q\bar q \to \gamma^*}^{(0)*}$ 
represents the Drell-Yan Born amplitude,
${\cal M}_{q\bar q \to \gamma^*}^{(2)*}$
the one-loop correction involving a virtual
gluon, and ${\cal M}_{q\bar q \to \gamma^* g}^{(1)}$
the tree-level amplitude with emission of a 
real gluon into the final state. An easy 
calculation gives 
\bea\label{PartonicRapidityRes-1virtual} \nn
\Delta^{(1)}(z,y)|_{\rm virtual} &=& 
\delta(1-z) \, \delta\bigg(y - \frac{1}{2} \bigg)
\frac{\alpha_s C_F}{4\pi}  
\bigg(\frac{\bar{\mu}^2}{Q^2}\bigg)^{\eps} 
\bigg[ - \frac{4}{\eps^2} - \frac{6}{\eps} -16 + 14\zeta_2 \\ 
&&\hspace{2.0cm}+\, \eps \bigg(-32 + 21 \zeta_2 
+ \frac{28 \zeta_3}{3} \bigg) +\ldots \bigg]
\eea
for the virtual contribution, where we introduced 
$\bar\mu^2=4\pi e^{-\gamma_E}\mu^2$ as the 
$\overline{\text{MS}}$ renormalisation scale. 
Concerning now the real emission, a simple 
calculation gives
\be\label{M1nlpOriginal}
\sum_{\rm s,c,p} \big|
{\cal M}_{q\bar q \to \gamma^* g}^{(1)}\big|^2 
= \frac{\alpha_s}{4\pi} 64\pi^2 N_c C_F
(1-\eps)\frac{\hat s^2}{k\cdot p_a k\cdot p_b}
\left[1-\frac{2}{\hat s}\big(k\cdot p_a+k\cdot p_b\big)
+\mathcal{O}[(1-z)^2]\right],
\ee
where $k$ is the momentum of the emitted soft gluon. 
Inserting this into \eqn{PartonicRapidityDefB-1real} and 
integrating against the phase space, up to NLP one 
obtains\footnote{Notice that the 
term defined as $\Delta^{(1)}(z,y)|_{\rm real}^{\rm LP}$
in \eqn{PartonicRapidityRes-1real-LP}
refers to the LP squared matrix element. 
It actually contains a NLP correction
of kinematic origin, due to the expansion 
of the phase space.}
\be\label{PartonicRapidityRes-1real-LP} 
\Delta^{(1)}(z,y)|_{\rm real}^{\rm LP}
= \frac{\as C_F}{4\pi} 
\bigg( \frac{\bar{\mu}^2}{Q^2}\bigg)^{\eps} 
\big(1-z\big)^{-1-2\eps} y^{-1-\eps}(1-y)^{-1-\eps}
\Big[ 1-\eps (1-z) \Big] 
\bigg[4-2\zeta_2\eps^2 + \ord(\eps^3)\bigg],
\ee
and
\be\label{PartonicRapidityRes-1real-NLP} 
\Delta^{(1)}(z,y)|_{\rm real}^{\rm NLP} 
= - \frac{\as C_F}{4\pi} \,
\bigg( \frac{\bar{\mu}^2}{Q^2}\bigg)^{\eps} 
\big(1-z\big)^{-2\eps}
y^{-1-\eps}(1-y)^{-1-\eps}
\bigg[4-2\zeta_2\eps^2 + \ord(\eps^3)\bigg].
\ee
The real emission contribution contains poles
that arise upon integration over 
$z$ and $y$, at $z = 1$, $y = 0$ 
and $y = 1$. We can isolate them
by means of the standard expansion 
formula
\be\label{plusDef}
\xi^{-1+a \eps} = \frac{\delta(\xi)}{a\eps} 
+ \frac{1}{\xi}\bigg|_{+} + a \eps \frac{\log\xi}{\xi}\bigg|_{+}
+ \frac{(a \eps)^2}{2!} \frac{\log^2\xi}{\xi}\bigg|_{+} + \ord(\eps^3),
\ee
with $\xi = 1-z$, $y$ or $1-y$. 
Setting $\bar{\mu}^2 = Q^2$ 
for simplicity, this leads to
\bea\label{PartonicRapidityRes-1real-LPb} \nn
\Delta^{(1)}(z,y)|_{\rm real}^{\rm LP} &=& 
\frac{\as C_F}{4\pi} \bigg\{
\bigg[ \frac{2}{\eps^2} -\zeta_2 \bigg] 
\Big[\delta(y)+ \delta(1-y)\Big] \delta(1-z) \\ \nn
&&\hspace{0.0cm} 
+\, \Big[\delta(y)+ \delta(1-y)\Big] 
\bigg( - \frac{4}{\eps} \frac{1}{1-z}\bigg|_+ 
+ 4  + 8 \frac{\log(1-z)}{1-z}\bigg|_+ \bigg) \\ 
&&\hspace{1.0cm}
+\, \frac{4}{y(1-y)}\bigg|_+ \frac{1}{1-z}\bigg|_+ 
+\ldots \bigg\},
\eea
for the LP squared matrix element.
Notice that terms multiplied by $\delta(1-z)$ must have $y$ equal to $0$ or $1$. At NLP we have
\be\label{PartonicRapidityRes-1real-NLPb} 
\Delta^{(1)}(z,y)|_{\rm real}^{\rm NLP} = 
\frac{\as C_F}{4\pi} \bigg\{
\Big[\delta(y)+ \delta(1-y)\Big] 
\bigg( \frac{4}{\eps} - 8 \log(1-z) \bigg)
- \frac{4}{y(1-y)}\bigg|_+ 
+\ldots \bigg\},
\ee
from the NLP contribution to the 
squared matrix element. The virtual 
and real corrections can be combined,
exploiting \eqn{yEquivalenceNNLP}.
After PDF renormalisation, and setting $\bar{\mu}^2=Q^2$,
one arrives at the finite result
\bea\label{PartonicRapidityRes-NLO-LP} \nn
\Delta^{(1)}(z,y)|_{\rm ren}^{\rm LP} &=& 
\frac{\as C_F}{4\pi} \bigg\{
\big(6\zeta_2 -8\big) 
\Big[\delta(y)+ \delta(1-y)\Big] \delta(1-z) \\ 
&&\hspace{-1.0cm} 
+\, \Big[\delta(y)+ \delta(1-y)\Big] 
\bigg(8 \frac{\log(1-z)}{1-z}\bigg|_+ + 4\bigg)
+\frac{4}{y(1-y)}\bigg|_+ \frac{1}{1-z}\bigg|_+ \bigg\},
\eea
and
\be\label{PartonicRapidityRes-NLO-NLP} 
\Delta^{(1)}(z,y)|_{\rm ren}^{\rm NLP} = 
\frac{\as C_F}{4\pi} \bigg\{
\Big[\delta(y)+ \delta(1-y)\Big] 
\bigg(- 8 \log(1-z) \bigg)
- \frac{4}{y(1-y)}\bigg|_+ \bigg\}.
\ee
A few comments are in order. Concerning the LP contribution, 
we see that it takes the factorised form 
of \eqn{zyFactLP} except for the last term.
However, one still needs to take into account 
that the $y$-dependence of the parton distribution
function in the luminosity defined in \eqn{luminosityRap}
is indeed subleading in the $z\to 1$ limit: 
\be\label{PDFrapLP}
f_{q/A}\Bigg(e^{Y}\sqrt{\frac{\tau}{z} 
\frac{1-(1-y)(1-z)}{1-y (1-z)}}\Bigg)\bigg|_{z\to 1}
\to f_{q/A}\big(e^{Y}\sqrt{\tau}\big)
+\ord(1-z).
\ee
At LP the PDFs can be approximated with 
the first term on the r.h.s. of \eqn{PDFrapLP}, 
and the integration against the last term in \eqn{PartonicRapidityRes-NLO-LP} gives zero. 
It is thus possible \cite{Becher:2007ty}
to rearrange \eqn{PartonicRapidityRes-NLO-LP}
such that the LP term always take the factorised 
form of \eqn{zyFactLP}. Another important aspect is that, because of the consideration above, 
it is evident that the NLP contribution 
in \eqn{PartonicRapidityRes-NLO-NLP} in 
general does \emph{not} factorise according to
\eqn{zyFactLP}. However, we see that the 
leading logarithmic contribution at NLP
factorises in rapidity, just as the LP contribution.
This is because the leading logarithms 
(LLs) are associated to the maximally soft and collinear momentum configurations, which give 
rise to the leading poles in the unrenormalised partonic 
cross section. As a consequence of \eqn{plusDef}, 
the LLs in $(1-z)$ can only arise from the first 
term in the expansion of the factor 
$y^{-1-\eps}(1-y)^{-1-\eps}$ in 
\eqn{PartonicRapidityRes-1real-NLP}.

\Eqns{PartonicRapidityRes-1real-LP}{PartonicRapidityRes-1real-NLP}
yield the $z$- and $y$-dependence in
case of a soft gluon emission. In general, 
it can be shown that a similar structure
will appear at higher orders, i.e., 
such corrections will involve factors 
of $\big(1-z\big)^{-1-(n_1+n_2)\eps}$
and $\big(1-z\big)^{-(n_1+n_2)\eps}$
respectively for the LP and NLP contribution, 
multiplied by factors of 
$y^{-1-n_1\eps}(1-y)^{-1-n_2\eps}$.
By the same reasoning as above, 
we can expect the LLs in $(1-z)$ at 
NLP to have the same factorised form
as the LP contribution:
\be\label{zyFactLP-NLP}
\Delta(z,y) = 
\frac{\delta(y)+\delta(1-y)}{2} 
\Big[\Delta_{\rm LP}(z) +
\Delta_{\rm NLP,LLs}(z) \Big]
+\Delta_{\rm NLP,rest}(z,y) + \ord(1-z).
\ee
In the next section we explicitly check 
this expectation at the next order in perturbation 
theory. 


\subsection{Next-to-next-to-leading order}
\label{sec:NNLO1r1v}

At NNLO one needs to consider three contributions, 
namely the double-virtual, the virtual-real and the 
double-real terms. They are given by, respectively, 
\bea
\label{PartonicRapidityDefB-2virtual}
\Delta^{(2)}(z,y)|_{\rm 2v} &=& 
\frac{1}{4N_c} \frac{1}{2\pi} 
\int d \Phi_{\gamma^*} 
\sum_{\rm s,c,p}
\Big\{
2\,{\rm Re}\Big[{\cal M}_{q\bar q \to \gamma^*}^{(0)*}
{\cal M}_{q\bar q \to \gamma^*}^{(4)} \Big]
+\Big|{\cal M}_{q\bar q \to \gamma^*}^{(2)}
\Big|^2\Big\}, \\ 
\label{PartonicRapidityDefB-1v1r}
\Delta^{(2)}(z,y)|_{\rm 1v1r} &=& 
\frac{1}{4N_c} \frac{1}{2\pi} 
\int d\Phi_{\gamma^*g} 
\sum_{\rm s,c,p}
2\,{\rm Re}\Big[{\cal M}_{q\bar q \to \gamma^*g}^{(1)*}
{\cal M}_{q\bar q \to \gamma^*g}^{(3)}\Big], \\ 
\label{PartonicRapidityDefB-2real}
\Delta^{(2)}(z,y)|_{\rm 2r} &=& 
\frac{1}{4N_c}\frac{1}{2\pi}  
\int d\Phi_{\gamma^*(gg + q\bar q)} 
\sum_{\rm s,c,p}
\big|{\cal M}_{q\bar q \to \gamma^* (gg + q\bar q)}^{(2)}\big|^2.
\eea
The phase spaces $\int d\Phi_{\gamma^*}$
and $\int d\Phi_{\gamma^*g}$ have already 
been mentioned in the previous section, while 
$\int d\Phi_{\gamma^*(gg + q\bar q)}$ represents the 
phase space for the emission of two gluons 
(or a quark-antiquark pair) in the final state, 
in addition to the off-shell photon. It is given 
in Appendix \ref{3PhaseSpace-DY}. Of the three 
contributions above, the first involves the 
two-loop (virtual) Drell-Yan amplitude 
${\cal M}_{q\bar q \to \gamma^*}^{(4)}$, 
and does not present new conceptual issues 
compared to the corresponding term at one loop, 
namely $\Delta^{(1)}(z,y)|_{\rm virtual}$ in 
\eqns{PartonicRapidityDefB-1virtual}{PartonicRapidityRes-1virtual}.
Starting from the two-loop quark 
form factor available in literature \cite{Matsuura:1987wt,Gehrmann:2010ue} 
and setting $\bar{\mu}^2 = Q^2$
for simplicity, one has
\bea\label{PartonicRapidityRes-2virtual} \nn
\Delta^{(2)}(z,y)|_{\rm 2v} &=& 
\left(\frac{\alpha_s}{4\pi}\right)^2\delta(1-z) \, \delta\bigg(y - \frac{1}{2} \bigg)
\bigg\{C_F^2 \bigg[\frac{8}{\eps^4}+\frac{24}{\eps^3} 
+\frac{82 - 56 \zeta_2}{\eps^2} +\frac{1}{\eps} 
\bigg(\frac{445}{2} - 156 \zeta_2 \\ \nn
&&\hspace{2.0cm} - \frac{184 \zeta_3}{3}\bigg) 
+\,\frac{2303}{4} - 516 \zeta_2
- 172 \zeta_3 + 274 \zeta_4 \bigg] \\ \nn
&&\hspace{1.0cm}+\, C_F C_A \bigg[ -\frac{11}{3 \eps^3} 
 + \frac{1}{\eps^2}\bigg(-\frac{166}{9} + 2 \zeta_2\bigg) 
+ \frac{1}{\eps}\bigg(-\frac{4129}{54}
+ \frac{121 \zeta_2}{3} + 26 \zeta_3\bigg) \\ 
&&\hspace{2.0cm}-\,\frac{89173}{324}+ \frac{1754 \zeta_2}{9}
+ \frac{934 \zeta_3}{9} - 16 \zeta_4 \bigg] \nn \\ 
&&\hspace{0.01cm}+\, C_F n_f 
\bigg[ \frac{2}{3 \eps^3} + \frac{28}{9 \eps^2} 
+ \frac{1}{\eps}\bigg( \frac{353}{27} 
- \frac{22 \zeta_2}{3} \bigg) 
+ \frac{7541}{162} - \frac{308 \zeta_2}{9} 
- \frac{52 \zeta_3}{9} \bigg]\bigg\}.
\eea
The virtual-real and the double-real 
contribution, respectively in 
\eqns{PartonicRapidityDefB-1v1r}{PartonicRapidityDefB-2real},
require more attention. Let us therefore 
consider them separately in what follows.

\subsubsection*{Virtual-real contribution}

The virtual-real contribution in 
\eqn{PartonicRapidityDefB-1v1r}
involves the one-loop amplitude 
${\cal M}_{q\bar q \to \gamma^*g}^{(3)}$, 
with the emission of a soft gluon. It is therefore the first instance where the loop 
integration involves non-trivial momentum
regions. We can cast the squared matrix 
element for this case in the form
\bea\label{Msq1r1v} 
\sum_{\rm s,c,p}
2\,{\rm Re}\Big[{\cal M}_{q\bar q \to \gamma^*g}^{(1)*}
{\cal M}_{q\bar q \to \gamma^*g}^{(3)}\Big]
&=& \bigg( \frac{\as}{4\pi}\bigg)^2 
256 \, \pi^2\, N_c\, (1-\eps)\nn \\[-0.1cm] 
&&\hspace{-4.8cm}\times\, \bigg\{
C_F^2 \bigg[ \bigg( \frac{\hat s^2}{t u} f_{h_1}(\eps)
+\frac{\hat s(t+u)}{t u} f_{h_2}(\eps) \bigg)
\left(\frac{\mu^2}{-\hat s}\right)^{\eps}  
+ \bigg( \frac{\hat s}{t} \left(\frac{\mu^2}{-t}\right)^{\eps}
+\frac{\hat s}{u} \left(\frac{\mu^2}{-u}\right)^{\eps} \bigg) 
f_{c_1}(\eps)\bigg] \nn \\ 
&&\hspace{-4.2cm}+\,C_A C_F \bigg[ 
\left( \frac{\hat s^2}{t u} +\frac{\hat s(t+u)}{t u} \right) 
\left(-\frac{\hat s\, \mu^2}{t\,u}\right)^{\eps} f_{s}(\eps)
+ \bigg( \frac{\hat s}{t} \left(\frac{\mu^2}{-t}\right)^{\eps}
+\frac{\hat s}{u} \left(\frac{\mu^2}{-u}\right)^{\eps} \bigg) 
f_{c_2}(\eps)\bigg]  \bigg\}\,,\nn\\
\phantom{}
\eea
where we have introduced the variables 
$t = -2 p_a \cdot k$, $u = -2 p_b \cdot k$,
with $k$ being the momentum of the emitted 
soft gluon, and 
\bea\label{loopfunctions} \nn
f_{h_1}(\eps) &=& - \frac{2}{\eps^2} - \frac{3}{\eps} -8 
+ \zeta_2 + \eps \bigg(-16 + \frac{3 \zeta_2}{2} 
+ \frac{14 \zeta_3}{3} \bigg) \\ \nn
&& +\, \eps^2 \bigg(-32 + 4 \zeta_2 + 7 \zeta_3 
+ \frac{47 \zeta_4}{8} \bigg) + {\cal O}(\eps^3), \\ \nn
f_{h_2}(\eps) &=& s^{\eps} f_{h_1}(\eps)\, \frac{\partial }{\partial s} s^{1-\eps} 
= (1-\eps)\, f_{h_1}(\eps), \\ \nn
f_{c_1}(\eps) &=& 
 - \frac{2}{\eps}-\frac{5}{2} 
+ \eps \big(-3 + \zeta_2\big) 
+ \eps^2 \bigg(-4 + \frac{5 \zeta_2}{4} + \frac{14 \zeta_3}{3} \bigg) 
+\ord(\eps^3), \\ \nn
f_{c_2}(\eps) &=& \frac{5}{2} + \eps 
+ \eps^2 \bigg(4 - \frac{5 \zeta_2}{4}\bigg) +\ord(\eps^3), \\ 
f_{s}(\eps) &=& -\frac{1}{\eps^2} - \frac{\zeta_2}{2} 
+ \frac{7\zeta_3}{3} \eps 
+ \frac{39\zeta_4}{16} \eps^2 +\ord(\eps^3).
\eea
One may readily identify the term proportional 
to $(-\hat s)^{-\eps}$ as the hard region 
contribution, the terms proportional to 
$(-t)^{-\eps}$ and $(-u)^{-\eps}$ respectively
as the collinear and anti-collinear region
contribution, and the term proportional to 
$(-\hat s/ t u)^{\eps}$ as the soft region 
contribution. Let us notice that leading 
poles are present only for the hard and 
soft region, at the level of the 
squared matrix element. As such, as already 
discussed for the invariant mass distribution
\cite{Bahjat-Abbas:2019fqa} (see also 
\cite{Bahjat-Abbas:2018hpv}), we expect
that, also for 
the rapidity distribution,  leading logarithms will arise only 
from the hard and soft region. The phase space integration
with measure $d\Phi_{\gamma^*g}$
can be performed with the help of the
equations in Appendix \ref{2PhaseSpace-DY}.
Expanding in powers of $1-z$, the integration
over the LP squared matrix element gives
\be\label{PartonicRapidityRes-1r1v-LP}
\Delta^{(2)}(z,y)|_{\rm 1r1v}^{\rm LP} =
\Delta^{(2)}(z,y)|_{\rm 1r1v}^{\rm LP,h}
+\Delta^{(2)}(z,y)|_{\rm 1r1v}^{\rm LP,s}\,,
\ee
where
\bea\label{PartonicRapidityRes-1r1v-LPh} 
\Delta^{(2)}(z,y)|_{\rm 1r1v}^{\rm LP,h}
&=& C_F^2 \bigg(\frac{\as}{4\pi}\bigg)^2
\bigg(\frac{\bar{\mu}^2}{Q^2}\bigg)^{2\eps} 
y^{-1-\eps} (1-y)^{-1-\eps} \nn \\ 
&&\hspace{-2.0cm}\times \,
\bigg\{\big(1-z\big)^{-1-2\eps}\bigg[
-\frac{16}{\eps^2} - \frac{24}{\eps} 
-64 + 64 \zeta_2 + \ldots \bigg]
+ \big(1-z\big)^{-2\eps}\bigg[\frac{32}{\eps} 
+ 48+\ldots \bigg]\bigg\}\,,\nn\\
\phantom{}
\eea
and
\bea\label{PartonicRapidityRes-1r1v-LPs} \nn
\Delta^{(2)}(z,y)|_{\rm 1r1v}^{\rm LP,s} 
&=& C_A C_F \bigg(\frac{\as}{4\pi}\bigg)^2
\bigg(\frac{\bar{\mu}^2}{Q^2}\bigg)^{2\eps} 
y^{-1-2\eps} (1-y)^{-1-2\eps} \\ 
&&\hspace{-2.0cm}\times \,
\bigg\{\big(1-z\big)^{-1-4\eps}\bigg[
-\frac{8}{\eps^2} + 24 \zeta_2 + \ldots \bigg]
+ \big(1-z\big)^{-4\eps}\bigg[\frac{16}{\eps} 
+\ldots \bigg]\bigg\}\,,
\eea
for the hard and soft region 
contribution, respectively. We recall that NLP corrections 
in \eqns{PartonicRapidityRes-1r1v-LPh}{PartonicRapidityRes-1r1v-LPs} originate from the power expansion of the phase 
space. Next, integration of the NLP squared matrix 
element gives 
\be\label{PartonicRapidityRes-1r1v-NLP}
\Delta^{(2)}(z,y)|_{\rm 1r1v}^{\rm NLP} =
\Delta^{(2)}(z,y)|_{\rm 1r1v}^{\rm NLP,h}
+\Delta^{(2)}(z,y)|_{\rm 1r1v}^{\rm NLP,c+\bar c}
+\Delta^{(2)}(z,y)|_{\rm 1r1v}^{\rm NLP,s}\,,
\ee
where
\bea\label{PartonicRapidityRes-1r1v-NLPh} \nn
\Delta^{(2)}(z,y)|_{\rm 1r1v}^{\rm NLP,h}
&=& C_F^2 \bigg(\frac{\as}{4\pi}\bigg)^2
\bigg( \frac{\bar{\mu}^2}{Q^2}\bigg)^{2\eps} 
y^{-1-\eps} (1-y)^{-1-\eps} 
\big(1-z\big)^{-2\eps} \\ 
&& \hspace{3.0cm}\times\, \bigg[ \frac{16}{\eps^2} 
+ \frac{8}{\eps} + 40 - 64 \zeta_2 + \ldots \bigg]\,,
\eea
\bea\label{PartonicRapidityRes-1r1v-NLPccb} \nn
\Delta^{(2)}(z,y)|_{\rm 1r1v}^{\rm NLP,c+\bar{c}}
&=& \bigg(\frac{\as}{4\pi}\bigg)^2
\bigg(\frac{\bar{\mu}^2}{Q^2}\bigg)^{2\eps} 
\Big[y^{-\eps} (1-y)^{-1-2\eps} 
+ y^{-1-2\eps} (1-y)^{-\eps}\Big]
\big(1-z\big)^{-3\eps} \\ 
&&\hspace{2.0cm}\times\,
\bigg\{ C_F^2 \bigg[\frac{16}{\eps} +20 + \ldots \bigg]
+ C_A C_F \big[-20 + \ldots \big] \bigg\}\,,
\eea
and
\be\label{PartonicRapidityRes-1r1v-NLPs} 
\Delta^{(2)}(z,y)|_{\rm 1r1v}^{\rm NLP,s} 
= C_A C_F \bigg(\frac{\as}{4\pi}\bigg)^2
\bigg( \frac{\bar{\mu}^2}{Q^2}\bigg)^{2\eps} 
y^{-1-2\eps} (1-y)^{-1-2\eps}
\big(1-z\big)^{-4\eps}\bigg[
\frac{8}{\eps^2} - 24\zeta_2 + \ldots \bigg]\,.
\ee
As already discussed for the single real 
emission at NLO, note that the $y$-dependence 
arises from the typical pattern $y^{-a_1- b_1\eps} 
(1-y)^{-a_2-b_2\eps}$, while the dependence 
on the threshold variable $z$ follows the 
pattern $z^{-1-(b_1+b_2)\eps}$ at LP, and 
$z^{-(b_1+b_2)\eps}$ at NLP. The 1-virtual 
1-real correction considered here is the 
first instance where contributions from 
different momentum regions of the 
virtual gluon arise. Inspecting 
\eqnss{PartonicRapidityRes-1r1v-LPh}{PartonicRapidityRes-1r1v-NLPs}, 
we see that the exponents $a_i$ and $b_i$ 
are characteristic of each region: at this 
order, $a_1 = a_2 = 1$, $b_1 = b_2 = 1$ for 
the hard region, $a_1 = 0$, $a_2 = 1$, 
$b_1 = 1$, $b_2 = 2$ for the collinear 
region, $a_1 = 1$, $a_2 = 0$, $b_1 = 2$, 
$b_2 = 1$ for the anti-collinear region, 
and $a_1 = a_2 = 1$, $b_1 = b_2 = 2$ for 
the soft region. Note the quite simple
correspondence between regions and the power dependence of the 
threshold and rapidity variables $z$ and $y$, in particular
involving also the dimensional regularisation parameter.
Expansion in powers of $\eps$ of these 
factors can be obtained by means of 
\eqn{plusDef}. Setting again 
$\bar{\mu}^2 = Q^2$, and 
introducing the notation
\be
{\cal D}_n(x) \equiv \frac{\log^n (x)}{x}\bigg|_+, 
\qquad \qquad 
{\cal L}_n(x) \equiv \log^n (x), 
\ee
with $x = \bar z \equiv (1-z)$, or $x = y$, 
or $x = \bar y \equiv (1-y)$, we obtain
\bea\label{PartonicRapidityRes-1r1v-LPh-Expanded} \nn
\Delta^{(2)}(z,y)|_{\rm 1r1v}^{\rm LP,h}
&=& C_F^2 \bigg(\frac{\as}{4\pi}\bigg)^2 
\bigg\{ \Big[\delta(y) + \delta(1 - y)\Big] 
\bigg[ \delta(1 - z) \bigg(-\frac{8}{\eps^4} 
- \frac{12}{\eps^3} - \frac{32 - 32 \zeta_2}{\eps^2} \\ \nn
&&\hspace{0.0cm}-\, \frac{1}{\eps}\bigg(64 - 48 \zeta_2 
- \frac{64 \zeta_3}{3}\bigg) - 128 + 128 \zeta_2 
+ 32 \zeta_3 - 72 \zeta_4 \bigg)
+\frac{16 {\cal D}_0(\bar z)}{\eps^3} \\ \nn
&&\hspace{0.0cm}+\, \frac{24 {\cal D}_0(\bar z) 
- 32 {\cal D}_1(\bar z) - 32}{\eps^2} 
+ \frac{1}{\eps}\bigg(64 {\cal D}_0(\bar z) (1 - \zeta_2) 
- 48 {\cal D}_1(\bar z) + 32 {\cal D}_2(\bar z) \\ \nn
&&\hspace{0.0cm}-\, 48 + 64 {\cal L}_1(\bar z)\bigg)
+ 128 {\cal D}_0(\bar z) 
\bigg(1 - \frac{3 \zeta_2}{4} - \frac{\zeta_3}{3}\bigg) 
- 128 {\cal D}_1(\bar z) (1 - \zeta_2) \\ \nn
&&\hspace{0.0cm} +\, 48 {\cal D}_2(\bar z) 
- \frac{64}{3} {\cal D}_3(\bar z) 
+ 96 {\cal L}_1(\bar z) - 64 {\cal L}_2 (\bar z) 
- 128 (1 - \zeta_2) \bigg] \\ \nn
&&\hspace{0.0cm}+ \, {\cal D}_0(y) {\cal D}_0(\bar y) 
\bigg[- \frac{16 {\cal D}_0(\bar z)}{\eps^2}
+\frac{32 - 24 {\cal D}_0(\bar z) 
+ 32 {\cal D}_1(\bar z)}{\eps}+ 48 \\ \nn
&&\hspace{0.0cm}-\,64 {\cal D}_0(\bar z) 
(1 - \zeta_2) + 48 {\cal D}_1(\bar z) 
- 32 {\cal D}_2(\bar z) 
- 64 {\cal L}_1(\bar z) \bigg] \\ \nn
&&\hspace{0.0cm}+\, 
\Big[{\cal D}_0(y) {\cal D}_1(\bar y) 
+ {\cal D}_1(y) {\cal D}_0(\bar y) \Big]
\bigg[\frac{16 {\cal D}_0(\bar z)}{\eps}
- 32 +24 {\cal D}_0(\bar z) 
-32 {\cal D}_1(\bar z)\bigg] \\
&&\hspace{0.0cm}-\, 
16 {\cal D}_0(\bar z) {\cal D}_1(y) {\cal D}_1(\bar y) 
- 8 {\cal D}_0(\bar z) \Big({\cal D}_0(y) {\cal D}_2(\bar y) 
+ {\cal D}_2(y) {\cal D}_0(\bar y) \Big) \bigg\}\,,
\eea
and
\bea\label{PartonicRapidityRes-1r1v-LPs-Expanded} \nn
\Delta^{(2)}(z,y)|_{\rm 1r1v}^{\rm LP,s} 
&=& C_A C_F \bigg(\frac{\as}{4\pi}\bigg)^2
\bigg\{ \Big[\delta(y) + \delta(1 - y)\Big] 
\bigg[\delta (1 - z) \bigg(
-\frac{1}{\eps^4} 
+ \frac{3 \zeta_2}{\eps^2} 
+ \frac{8 \zeta_3}{3 \eps} 
- \frac{3 \zeta_4}{4} \bigg)  \\ \nn 
&&\hspace{-1.0cm}+\,\frac{4 {\cal D}_0(\bar z)}{\eps^3} 
- \frac{8 + 16 {\cal D}_1(\bar z)}{\eps^2} 
- \frac{1}{\eps}\bigg(12 {\cal D}_0(\bar z) \zeta_2 
- 32 {\cal D}_2(\bar z) - 32 {\cal L}_1(\bar z) \bigg) \\ \nn
&&\hspace{-1.0cm}-\, \frac{32}{3} {\cal D}_0(\bar z) \zeta_3 
+ 48 {\cal D}_1(\bar z) \zeta_2 
- \frac{128}{3} {\cal D}_3(\bar z) 
+ 24 \zeta_2 - 64 {\cal L}_2(\bar z)\bigg] \\ \nn
&&\hspace{-1.0cm}+\, {\cal D}_0(y){\cal D}_0(\bar y)
\bigg[-\frac{8 {\cal D}_0(\bar z)}{\eps^2}
+\frac{16+32{\cal D}_1(\bar z)}{\eps} 
+ 24 \zeta_2 {\cal D}_0(\bar z) 
- 64{\cal D}_2(\bar z) -64 {\cal L}_1(\bar z)\bigg] \\ \nn
&&\hspace{-1.0cm}+\, 
\Big[{\cal D}_0(y) {\cal D}_1(\bar y) 
+ {\cal D}_0(\bar y) {\cal D}_1(y)\Big]
\bigg[\frac{16 {\cal D}_0(\bar z)}{\eps}  
 - 32 - 64 {\cal D}_1(\bar z)\bigg]  \\
&&\hspace{-1.0cm} -\, 32 {\cal D}_0(\bar z)
{\cal D}_1(y){\cal D}_1(\bar y) 
- 16 {\cal D}_0(\bar z) 
\Big({\cal D}_0(y) {\cal D}_2(\bar y) 
+ {\cal D}_0(\bar y) {\cal D}_2(y)\Big)\bigg\}\,, 
\eea
for the hard- and soft-region 
contribution to the LP squared matrix element. Next, 
the NLP squared matrix element gives
\bea\label{PartonicRapidityRes-1r1v-NLPh-Expanded} \nn
\Delta^{(2)}(z,y)|_{\rm 1r1v}^{\rm NLP,h}
&=& C_F^2 \bigg(\frac{\as}{4\pi}\bigg)^2
\bigg\{ \Big[\delta(y) + \delta(1 - y)\Big] 
\bigg[-\frac{16}{\eps^3} 
 - \frac{8 - 32 {\cal L}_1(\bar z)}{\eps^2} 
 - \frac{1}{\eps}\Big(40 - 64 \zeta_2 \\ \nn
&&\hspace{-1.0cm}-\, 16 {\cal L}_1(\bar z) 
 + 32 {\cal L}_2(\bar z)\Big) 
 - 64 + 32 \zeta_2  + \frac{128 \zeta_3}{3} 
 + {\cal L}_1(\bar z) (80 - 128 \zeta_2)  \\ \nn
&&\hspace{-1.0cm}-\, 16 {\cal L}_2(\bar z)
+ \frac{64}{3} {\cal L}_3(\bar z) \bigg] 
 + {\cal D}_0(\bar y) {\cal D}_0(y)
 \bigg[\frac{16}{\eps^2}+\frac{8 
 - 32 {\cal L}_1(\bar z)}{\eps}
+ 40 - 64 \zeta_2 \\ \nn  
&& \hspace{-1.0cm}-\, 16 {\cal L}_1(\bar z)  
+32 {\cal L}_2(\bar z) \bigg]
+ \Big[{\cal D}_0(y) {\cal D}_1(\bar y)
+{\cal D}_0(\bar y) {\cal D}_1(y)\Big]
\bigg[- \frac{16}{\eps} - 8 
+ 32 {\cal L}_1(\bar z)\bigg] \\ 
&& \hspace{-1.0cm}+\, 16 {\cal D}_1(\bar y) {\cal D}_1(y) 
+ 8 \Big({\cal D}_0(y) {\cal D}_2(\bar y) 
 + {\cal D}_0(\bar y) {\cal D}_2(y)\Big) 
\bigg\},
\eea
\bea\label{PartonicRapidityRes-1r1v-NLPccb-Expanded} \nn
\Delta^{(2)}(z,y)|_{\rm 1r1v}^{\rm NLP,c+\bar{c}}
&=& \bigg(\frac{\as}{4\pi}\bigg)^2
\bigg\{ C_F^2 \bigg[ 
\Big(\delta(y) + \delta(1 - y)\Big) 
\bigg(-\frac{8}{\eps^2} 
- \frac{10 - 24 {\cal L}_1(\bar z)}{\eps} 
- 12 + 30 {\cal L}_1(\bar z) \\ \nn  
&& \hspace{-2.0cm}-\, 36 {\cal L}_2(\bar z)
+ 8 \zeta_2 \bigg) + {\cal D}_0(y)
\bigg(\frac{16}{\eps} + 20 
- 16{\cal L}_1(\bar y) 
- 48{\cal L}_1(\bar z) \bigg) \\ 
&& \hspace{-2.0cm}+\, {\cal D}_0(\bar y) 
\bigg(\frac{16}{\eps} + 20 - 16{\cal L}_1(y) 
- 48{\cal L}_1(\bar z) \bigg)
- 32 \Big({\cal D}_1(y) + {\cal D}_1(\bar y)\Big) 
\bigg]\nn \\  
&&\hspace{-2.0cm}+\, C_A C_F \bigg[ 
\Big(\delta(y) + \delta(1 - y)\Big) 
\bigg(\frac{10}{\eps} + 4 - 30 {\cal L}_1(\bar z)\bigg) 
- 20 \Big({\cal D}_0(y) + {\cal D}_0(\bar y)\Big) 
\bigg]\bigg\}\,,
\eea
and
\bea\label{PartonicRapidityRes-1r1v-NLPs-Expanded} \nn
\Delta^{(2)}(z,y)|_{\rm 1r1v}^{\rm NLP,s} 
&=& C_A C_F \bigg(\frac{\as}{4\pi}\bigg)^2
\bigg\{
 \Big[\delta(y) + \delta(1 - y)\Big] 
 \bigg[-\frac{4}{\eps^3} 
 + \frac{16 {\cal L}_1(\bar z)}{\eps^2} 
 - \frac{32 {\cal L}_2(\bar z) - 12 \zeta_2}{\eps} \\ \nn
&&\hspace{0.0cm}+\, \frac{32 \zeta_3}{3} 
 - 48 {\cal L}_1(\bar z) \zeta_2 
 + \frac{128}{3} {\cal L}_3(\bar z) \bigg]
 +{\cal D}_0(\bar y) {\cal D}_0(y) 
 \bigg[\frac{8}{\eps^2}-\frac{32{\cal L}_1(\bar z)}{\eps} \\ \nn
&&\hspace{0.0cm}-\, 
24\zeta_2 + 64 {\cal L}_2(\bar z)\bigg] 
+\Big({\cal D}_0(y) {\cal D}_1(\bar y) 
+{\cal D}_0(\bar y) {\cal D}_1(y)\Big)
\bigg[- \frac{16}{\eps} + 64 {\cal L}_1(\bar z) \bigg]  \\
&&\hspace{0.0cm}  
+\, 32 {\cal D}_1(\bar y) {\cal D}_1(y)
+16 \Big({\cal D}_0(y) {\cal D}_2(\bar y) 
+ {\cal D}_0(\bar y) {\cal D}_2(y)\Big) 
\bigg\}\,,
\eea
for the hard, collinear plus 
anti-collinear and soft region, 
respectively. 

Let us comment on these results. 
First of all, the LP squared matrix element in
\eqns{PartonicRapidityRes-1r1v-LPh-Expanded}{PartonicRapidityRes-1r1v-LPs-Expanded} 
receives contributions from the hard and 
the soft region, but no contribution from 
the collinear region. This is a well-known 
result for the invariant mass distribution 
\cite{Bonocore:2014wua}, which clearly 
remains true in case of more differential 
distributions, like the double-differential
distribution in invariant mass and rapidity 
considered here. This is because, near 
threshold, contributions from the collinear 
and anti-collinear regions at LP decouple 
at the level of the matrix element, 
regardless of the exact form of the 
phase space \cite{Collins:1981uw,Collins:1985ue,Collins:1988ig,Bauer:2001yt}.
Moreover, we can single out three different 
types of contributions in the LP squared matrix element, 
\eqns{PartonicRapidityRes-1r1v-LPh-Expanded}{PartonicRapidityRes-1r1v-LPs-Expanded}: 
the first is given by terms proportional 
to $\delta(y) + \delta(1 - y)$, which 
corresponds to the LP factorised 
terms of \eqn{zyFactLP-NLP}. Then we 
have terms which are proportional to
$y$- and $\bar y$-plus distributions.
These do not factorise. However, as 
discussed around \eqn{PDFrapLP}, the 
behaviour of the PDFs near $z \to 1$
is such that $y$- and $\bar y$-plus 
distributions in the LP squared matrix element 
actually contribute at NLP in $1-z$. In 
this respect, the important thing to 
notice concerning our assumption in 
\eqn{zyFactLP-NLP} is that such $y$- and 
$\bar y$-plus distributions in 
\eqns{PartonicRapidityRes-1r1v-LPh-Expanded}{PartonicRapidityRes-1r1v-LPs-Expanded}
do not contain leading ${\cal D}_3(\bar z)$
distributions, nor leading ${\cal L}_3(\bar z)$
logarithms. As such, these terms 
contribute to the last term $\Delta_{\rm NLP,rest}(z,y)$
in \eqn{zyFactLP-NLP}. A third type 
of contribution in 
\eqns{PartonicRapidityRes-1r1v-LPh-Expanded}{PartonicRapidityRes-1r1v-LPs-Expanded}
is given by logarithms ${\cal L}_n(\bar z)$
arising from the expansion of the phase space.
More precisely, the power expansion of the 
partonic cross section is given by the sum 
of three terms:
\be\label{powerExpansion+PS}
\Delta \sim \underbrace{\int d\Phi_{\rm LP} \,|\mathcal{M}|_{\rm LP}^2
+ \int d\Phi_{\rm NLP} \,|\mathcal{M}|_{\rm LP}^2}_{\Delta_{\rm LP}} 
+ \underbrace{\int d\Phi_{\rm LP} 
\,|\mathcal{M}|_{\rm NLP}^2}_{\Delta_{\rm NLP}}.
\ee
In \eqns{PartonicRapidityRes-1r1v-LPh-Expanded}{PartonicRapidityRes-1r1v-LPs-Expanded} 
we include, with a slight abuse of language, 
both the first and second term of 
\eqn{powerExpansion+PS} into 
$\Delta_{\rm LP}$. Thus, $\Delta_{\rm LP}$
contains logarithms ${\cal L}_n(\bar z)$,
originating from the second term of 
\eqn{powerExpansion+PS}. However, only 
logarithms with $n = 1,2$ appear. Thus, 
power corrections from the phase space 
measure do not give rise to leading 
logarithms at NLP. This result has been 
already exploited to construct the 
resummation of LLs in the Drell-Yan 
invariant mass distribution, see section 
3 of \cite{Bahjat-Abbas:2019fqa}, and 
it remains true for the distribution 
differential in both invariant
mass and rapidity. Lastly, in regard to the NLP squared matrix element in 
\eqnss{PartonicRapidityRes-1r1v-NLPh-Expanded}{PartonicRapidityRes-1r1v-NLPs-Expanded},
we see that leading logarithms 
${\cal L}_3(\bar z)$ arise only in the 
hard and soft region, 
\eqns{PartonicRapidityRes-1r1v-NLPh-Expanded}{PartonicRapidityRes-1r1v-NLPs-Expanded}, 
while the collinear and anti-collinear 
region contribution in 
\eqn{PartonicRapidityRes-1r1v-NLPccb-Expanded}
contain at most next-to-leading logarithms.
Also in this case the 
result is quite general, i.e.\ 
is independent of the particular 
differential distribution considered, 
because the collinear momenta 
configurations contain only subleading 
poles already at the level of the 
squared matrix element (cf.\ 
\eqns{Msq1r1v}{loopfunctions} with 
\eqn{PartonicRapidityRes-1r1v-NLPccb}),
and thus upon integration no LLs 
can be generated. We conclude that we are left with the NLP 
LLs ${\cal L}_3(\bar z)$ from 
the hard and soft region, 
\eqns{PartonicRapidityRes-1r1v-NLPh-Expanded}{PartonicRapidityRes-1r1v-NLPs-Expanded}, 
which indeed have the factorised 
form of \eqn{zyFactLP-NLP}.

\subsubsection*{Double-real contribution}

We are now left with the double-real 
correction listed in \eqn{PartonicRapidityDefB-2real}:
\bea \label{PartonicRapidityDefB-2realB}
\Delta^{(2)}(z,y)|_{\rm 2r} &=& 
\frac{1}{4N_c}\frac{1}{2\pi}  
\int d\Phi_{\gamma^*(gg + q\bar q)} 
\sum_{\rm s,c,p}
\big|{\cal M}_{q\bar q \to \gamma^* (gg + q\bar q)}^{(2)}\big|^2,
\eea
which is given in terms of the tree-level amplitude 
${\cal M}_{q\bar q \to \gamma^* (gg + q\bar q)}^{(2)}$,
describing the emission of two soft gluons (or a 
quark-antiquark pair) in the final state, and the 
corresponding phase space $\int d\Phi_{\gamma^*(gg + q\bar q)}$.
The expression of the squared matrix element 
$\big|{\cal M}_{q\bar q \to \gamma^* (gg + q\bar q)}^{(2)}\big|^2$
is rather lengthy, and we do not report it here. 
Instead, it is interesting to spend some words
on the phase space integration, which is obviously
more involved compared to the case of a single 
real emission discussed in the previous section.
The phase space integral for double-real emission reads
\bea
\int d\Phi_{\gamma^*(gg + q\bar q)} &=& \big(\mu^2\big)^{4-d}\int \frac{d^dq}{(2\pi)^{d-1}} 
\frac{d^d k_1}{(2\pi)^{d-1}} \frac{d^d k_2}{(2\pi)^{d-1}}
\,(2\pi)^d\, \delta^{(d)}(p_a + p_b - q - k_1 - k_2) 
\nonumber \\
&&\hspace{1.0cm}\times  \, \delta_+(k_1^2) \, \delta_+(k_2^2) \,\delta(q^2 - Q^2)\, 
\delta\bigg[y - \frac{p_a \cdot q - z \, p_b \cdot q}{(1-z)
(p_a \cdot q + p_b \cdot q)} \bigg]\,.
\eea
In this case it proves useful to follow 
the parametrisation used in \cite{Hamberg:1990np,Laenen:2010uz}, 
in which the three-particle phase space is 
factorised into two two-body phase spaces, 
one involving the off-shell photon and the other
the vector sum of the emitted gluon 
momenta $K = k_1 + k_2$:
\be
\int d\Phi(p_a + p_b \to q + k_1 + k_2) 
= \int_0^{\infty} \frac{dK^2}{2\pi}  
\int d\Phi(p_a + p_b \to q + K)
\times \, \int d\Phi(K \to k_1 + k_2).
\ee
Subsequently, the phase space 
$\int d\Phi(K \to k_1 + k_2)$ is 
evaluated in the centre of mass frame
of the two-gluon system, where the 
momenta are parameterised as follows:
\bea \nn
p_a &=& \frac{\hat s-\tilde t}{2\sqrt{s_{12}}}
(1,0,\ldots,0,1), \\ \nn
p_b &=& \left(\frac{ \tilde t +s_{12} - Q^2}{2\sqrt{s_{12}}},0,
\ldots ,0, |{\vec q}|\sin \psi,  |{\vec q}| \cos \psi 
-  \frac{\hat s-\tilde t}{2\sqrt{s_{12}}} \right), \\ \nn
q &=& \left(\frac{\hat s -s_{12} - Q^2}{2\sqrt{s_{12}}},0,
\ldots ,0, |{\vec q}|\sin \psi,  |{\vec q}| \cos \psi \right), \\ \nn
k_1 &=& \frac{\sqrt{s_{12}}}{2}(1,0,\ldots, 
\sin\theta_2 \sin\theta_1, 
\cos\theta_2 \sin\theta_1,\cos\theta_1), \\ 
k_2 &=& \frac{\sqrt{s_{12}}}{2}(1,0,\ldots, 
-\sin\theta_2 \sin\theta_1, 
-\cos\theta_2 \sin\theta_1,-\cos\theta_1), 
\eea
where  
\be
\tilde t = 2 p_a \cdot q,  
\qquad \qquad 
\tilde u = 2 p_b \cdot q,
\qquad \qquad 
s_{12} = 2 k_1 \cdot k_2 = \hat s - \tilde t - \tilde u + Q^2,
\ee
and
\be
\cos \psi = \frac{(\hat s - Q^2)(\tilde u - Q^2)
- s_{12}(\tilde t +Q^2)}{(\hat s - \tilde t)
\sqrt{\Lambda(\hat s,Q^2,s_{12})}}\,,  
\qquad \qquad 
|{\vec q}| =
\frac{\sqrt{\Lambda(\hat s,Q^2,s_{12})}}{2\sqrt{s_{12}}}\,,
\ee
where $\Lambda$ is the standard K\"{a}llen function
$\Lambda(x,y,z) = a^2 + b^2 + c^2 - 2ab -2ac -2bc$. 
The Mandelstam variables $\tilde t$ and $\tilde u$ 
can in turn be expressed as functions of the photon 
energy fraction $z$ and of two further variables 
$0 < v_1 < 1$ and $0 < v_2 < 1$, such that
\bea 
\tilde t = \hat s \bigg[ z + v_2 (1-z) - 
\frac{v_2(1-v_2) v_1 (1-z)^2}{1-v_2(1-z)}\bigg], \qquad
\tilde u = \hat s \big[1 - v_2(1-z) \big].
\eea
With this parametrisation, $\int d\Phi_{\gamma^*(gg + q\bar q)}$
reads 
\bea\label{PS3} \nn
\int d\Phi_{\gamma^*(gg + q\bar q)} &=& \frac{1}{(4\pi)^d} 
\frac{\hat s^{d-3} (\mu^2)^{4-d}}{\Gamma(d-3)} 
(1-z)^{2d-5} \int_0^{\pi} d\theta_1 
\int_0^{\pi} d\theta_2\, 
(\sin\theta_1)^{d-3} (\sin\theta_2)^{d-4} \\ \nn
&&\hspace{-2.0cm} \times \, \int_0^1 dv_1 
\int_0^1 dv_2 \, \left[ v_1 (1-v_1) \right]^{d/2-2}  
\left[ v_2 (1-v_2) \right]^{d-3} 
\left[ 1- v_2 (1-z) \right]^{1-d/2} \\
&&\hspace{-2.0cm} \times \,
\delta\Bigg\{y - \frac{\hat s \Big(z+v_2(1-z)
-\frac{v_2(1-v_2)v_1 (1-z)^2}{1-v_2(1-z)}\Big) 
- z \hat s \big(1-v_2(1-z)\big)}{(1-z)
\Big[\hat s \Big(z+v_2(1-z)
-\frac{v_2(1-v_2)v_1 (1-z)^2}{1-v_2(1-z)}\Big)
+ \hat s \big(1-v_2(1-z)\big)\Big]} \Bigg\}.
\eea
The Dirac delta function in the last 
line is rather involved, but it greatly 
simplifies in the $z \to 1$ limit. In this 
case, following \cite{Laenen:1992ey}, we expand
\bea\label{yrap2pexpanded} \nn
\delta\Bigg\{y - \frac{\hat s \Big(z+v_2(1-z)
-\frac{v_2(1-v_2)v_1 (1-z)^2}{1-v_2(1-z)}\Big) 
- z \hat s \big(1-v_2(1-z)\big)}{(1-z)
\Big[\hat s \Big(z+v_2(1-z)
-\frac{v_2(1-v_2)v_1 (1-z)^2}{1-v_2(1-z)}\Big)
+ \hat s \big(1-v_2(1-z)\big)\Big]} \Bigg\} && \\[0.1cm]
&&\hspace{-9.0cm} = \delta(y-v_2) 
+ \frac{v_2 (1-v_2) v_1}{2} (1-z) \delta'(y-v_2) + 
\ord\big[(1-z)^2 \big].
\eea
Inserting this power expansion into the 
three-body phase space of \eqn{PS3}, it is 
possible to perform the integral of the two-real 
squared matrix element in \eqn{PartonicRapidityDefB-2realB}
with the standard methods discussed in 
\cite{Hamberg:1990np,Laenen:2010uz}. After some 
elaboration, we obtain the LP and NLP contribution 
before expanding in $\eps$ and the scale factors. The 
contribution of the LP squared matrix element reads 
\bea\label{PartonicRapidityRes-2r-LP} \nn 
\Delta^{(2)}(z,y)|_{\rm 2r}^{\rm LP}
&=& \bigg(\frac{\as}{4\pi}\bigg)^2
\bigg(\frac{\bar{\mu}^2}{Q^2}\bigg)^{2\eps} 
y^{-1-2\eps} (1-y)^{-1-2\eps} \\ \nn
&&\hspace{-2.0cm}\times \,
\bigg\{\big(1-z\big)^{-1-4\eps}\bigg[
C_F^2 \bigg(\frac{32}{\eps^2} - 96 \zeta_2 + \ldots\bigg) 
+ C_A C_F \bigg(\frac{8}{\eps^2} + \frac{44}{3\eps} 
+\frac{268}{9} - 32 \zeta_2 + \ldots\bigg) \\ 
&& \hspace{2.0cm}+\, n_f C_F \bigg( - \frac{8}{3\eps} 
-\frac{40}{9} + \ldots\bigg) \bigg]\nn \\ 
&& \hspace{-2.0cm}+\, \big(1-z\big)^{-4\eps}\bigg[
C_F^2 \bigg( -\frac{80}{\eps} + \ldots\bigg) 
+ C_A C_F \bigg( - \frac{16}{\eps} 
-\frac{88}{3} + \ldots\bigg) 
+ n_f C_F \bigg( \frac{16}{3} + \ldots\bigg) \bigg]\bigg\}\,,\nn\\
\phantom{}
\eea
and the NLP contribution reads 
\bea\label{PartonicRapidityRes-2r-NLP} \nn
\Delta^{(2)}(z,y)|_{\rm 2r}^{\rm NLP} 
&=& \bigg(\frac{\as}{4\pi}\bigg)^2
\bigg(\frac{\bar{\mu}^2}{Q^2}\bigg)^{2\eps} 
y^{-1-2\eps} (1-y)^{-1-2\eps} \, \big(1-z\big)^{-4\eps} \\ \nn
&&\hspace{-2.0cm}\times \,
\bigg[
C_F^2 \bigg(-\frac{32}{\eps^2} -8 + 96 \zeta_2 + \ldots\bigg) 
+ C_A C_F \bigg(- \frac{8}{\eps^2} - \frac{44}{3\eps} 
-\frac{220}{9} + 32 \zeta_2 + \ldots\bigg) \\ 
&& \hspace{2.0cm}+\, n_f C_F \bigg( \frac{8}{3\eps} 
+\frac{64}{9} + \ldots\bigg) \bigg].
\eea
We see that the scale factors are consistent
with our expectation, discussed below 
\eqn{PartonicRapidityRes-1r1v-NLPs}. Namely, 
the $y$-dependence arises from the typical 
pattern $y^{-a_1- b_1\eps} (1-y)^{-a_2-b_2\eps}$,
with $a_1 = a_2 = 1$, $b_1 = b_2 = 2$, which is 
characteristic of the soft region: indeed, 
in this case we are considering the emission of 
two soft gluons, and the same pattern arises
in case of the virtual-real contribution, 
where both the virtual and the real gluon
are taken to be soft, cf.\
\eqn{PartonicRapidityRes-1r1v-NLPs}. 
This immediately allows us to conclude 
that, upon expanding in $\eps$, the double-real 
contribution, too, has the structure of 
\eqn{zyFactLP-NLP}, as we have already 
verified in case of \eqn{PartonicRapidityRes-1r1v-NLPs}, 
compare with \eqns{PartonicRapidityRes-1r1v-LPs-Expanded}{PartonicRapidityRes-1r1v-NLPs-Expanded}.
In this case, expanding the scale factors 
in \eqns{PartonicRapidityRes-2r-LP}{PartonicRapidityRes-2r-NLP}
in $\eps$ by means of \eqn{plusDef} and setting $\bar{\mu}^2 = Q^2$
we explicitly obtain 
\bea\label{PartonicRapidityRes-2r-LP-Expanded} \nn
\Delta^{(2)}(z,y)|_{\rm 2r}^{\rm LP}
&=& \bigg(\frac{\as}{4\pi}\bigg)^2 
\Bigg\{ C_F^2  \Bigg[ 
\Big[\delta(y) + \delta(1 - y)\Big] 
\bigg[ \delta(1 - z) 
\bigg(\frac{4}{\eps^4} 
- \frac{12\zeta_2}{\eps^2} 
- \frac{56\zeta_3}{3\eps} +15\zeta_4 \bigg) \\ \nn
&&\hspace{-2.0cm}-\,
\frac{16 {\cal D}_0(\bar z)}{\eps^3} 
+ \frac{64 {\cal D}_1(\bar z) +40}{\eps^2} 
+ \frac{1}{\eps}\bigg(
-128 {\cal D}_2(\bar z)
+ 48 \zeta_2 {\cal D}_0(\bar z) 
- 160 {\cal L}_1(\bar z) \bigg) \\ \nn
&&\hspace{-2.0cm}+\,
\frac{512}{3} {\cal D}_3(\bar z) 
- 192 \zeta_2 {\cal D}_1(\bar z) 
+ \frac{224}{3} \zeta_3 {\cal D}_0(\bar z) 
+ 120 \zeta_2
+ 320 {\cal L}_2(\bar z) \bigg] \\ \nn
&&\hspace{-2.0cm}+ \, {\cal D}_0(y) {\cal D}_0(\bar y) 
\bigg[\frac{32 {\cal D}_0(\bar z)}{\eps^2}
-\frac{128 {\cal D}_0(\bar z) +80}{\eps} 
+ 256 {\cal D}_2(\bar z) 
- 96 \zeta_2 {\cal D}_0(\bar z) 
+ 230 {\cal L}_1(\bar z) \bigg] \\ \nn
&&\hspace{-2.0cm}+\, 
\Big[{\cal D}_0(y) {\cal D}_1(\bar y) 
+ {\cal D}_1(y) {\cal D}_0(\bar y) \Big]
\bigg[-\frac{64 {\cal D}_0(\bar z)}{\eps}
+256 {\cal D}_1(\bar z) + 160 \bigg] \\ \nn
&&\hspace{-2.0cm}-\, 
128 {\cal D}_0(\bar z) {\cal D}_1(y) {\cal D}_1(\bar y) 
+ 64 {\cal D}_0(\bar z) \Big[{\cal D}_0(y) {\cal D}_2(\bar y) 
+ {\cal D}_2(y) {\cal D}_0(\bar y) \Big] \Bigg] \\ \nn
&&\hspace{-2.0cm}+\, C_A C_F \Bigg[ 
\Big[\delta(y) + \delta(1 - y)\Big] 
\bigg[ \delta(1 - z) 
\bigg(\frac{1}{\eps^4} 
+ \frac{11}{6\eps^3} 
+ \frac{1}{\eps^2}\bigg(
\frac{67}{18} - 4\zeta_2\bigg)  \\ \nn
&&\hspace{-2.0cm}+\,
\frac{1}{\eps}\bigg(\frac{202}{27} 
- \frac{11\zeta_2}{2}
- \frac{29\zeta_3}{3}\bigg)
+\frac{1214}{81}-\frac{67\zeta_2}{6}
-\frac{77\zeta_3}{9}-\frac{17\zeta_4}{4} \bigg]
-\frac{4 {\cal D}_0(\bar z)}{\eps^3}  \\ \nn
&&\hspace{-2.0cm}+\,
\frac{1}{\eps^2} \bigg( 16{\cal D}_1(\bar z)
-\frac{22}{3}{\cal D}_0(\bar z) +8 \bigg) 
+ \frac{1}{\eps}\bigg(
-32 {\cal D}_2(\bar z)
+\frac{88}{3} {\cal D}_1(\bar z)
+\bigg(16\zeta_2 - \frac{134}{9} 
\bigg){\cal D}_0(\bar z) \\ \nn
&&\hspace{-2.0cm}+\,
\frac{44}{3} - 32{\cal L}_1(\bar z) \bigg) 
+\frac{128}{3} {\cal D}_3(\bar z) 
- \frac{176}{3} {\cal D}_2(\bar z) 
+ \bigg(\frac{536}{9} -64\zeta_2
\bigg){\cal D}_1(\bar z) \\ \nn
&&\hspace{-2.0cm}+\,
\bigg(\frac{116}{3} \zeta_3 + 22\zeta_2
-\frac{808}{27}\bigg){\cal D}_0(\bar z) 
+\frac{238}{9} - 32 \zeta_2
+ 64 {\cal L}_2(\bar z)
- \frac{176}{3} {\cal L}_1(\bar z)\bigg] \\ \nn
&&\hspace{-2.0cm}+ \, {\cal D}_0(y) {\cal D}_0(\bar y) 
\bigg[\frac{8 {\cal D}_0(\bar z)}{\eps^2}
+\frac{1}{\eps} \bigg(-32 {\cal D}_1(\bar z) 
+\frac{44}{3} {\cal D}_0(\bar z) - 16 \bigg)
+ 64 {\cal D}_2(\bar z)  \\ \nn
&&\hspace{-2.0cm}-\,
\frac{176}{3} {\cal D}_1(\bar z)
+\bigg(\frac{268}{9} - 32 \zeta_2
\bigg){\cal D}_0(\bar z) - \frac{88}{3} 
+ 64{\cal L}_1(\bar z) \bigg] \\ \nn
&&\hspace{-2.0cm}+\, 
\Big[{\cal D}_0(y) {\cal D}_1(\bar y) 
+ {\cal D}_1(y) {\cal D}_0(\bar y) \Big]
\bigg[-\frac{16 {\cal D}_0(\bar z)}{\eps}
+64 {\cal D}_1(\bar z) 
-\frac{88}{3} {\cal D}_0(\bar z) + 32\bigg] \\ \nn
&&\hspace{-2.0cm}+\, 
32 {\cal D}_0(\bar z) {\cal D}_1(y) {\cal D}_1(\bar y) 
+ 16 {\cal D}_0(\bar z) \Big[{\cal D}_0(y) {\cal D}_2(\bar y) 
+ {\cal D}_2(y) {\cal D}_0(\bar y) \Big] \Bigg] \\ \nn
&&\hspace{-2.0cm}+\, n_f C_F  \Bigg[ 
\Big[\delta(y) + \delta(1 - y)\Big] 
\bigg[ \delta(1 - z) 
\bigg(-\frac{1}{3\eps^3} 
- \frac{5}{9\eps^2} 
+ \frac{1}{\eps}\bigg( 
\zeta_2 - \frac{28}{27}\bigg)
-\frac{164}{81} +\frac{5\zeta_2}{3} \\ \nn
&&\hspace{-2.0cm}+\, 
\frac{14\zeta_3}{9}\bigg)
+\frac{4{\cal D}_0(\bar z)}{3\eps^2} 
+ \frac{1}{\eps}\bigg(
-\frac{16}{3} {\cal D}_1(\bar z)
+ \frac{20}{9} {\cal D}_0(\bar z) 
- \frac{8}{3} \bigg) +
\frac{32}{3} {\cal D}_2(\bar z) 
- \frac{80}{9} {\cal D}_1(\bar z) \\ \nn
&&\hspace{-2.0cm}+\,\bigg(\frac{112}{27} 
- 4 \zeta_2\bigg){\cal D}_0(\bar z)
- \frac{28}{9} +\frac{32}{3} 
{\cal L}_1(\bar z) \bigg]
+{\cal D}_0(y) {\cal D}_0(\bar y) 
\bigg[-\frac{8 {\cal D}_0(\bar z)}{3\eps} 
+ \frac{32}{3} {\cal D}_1(\bar z) \\ 
&&\hspace{-2.0cm}- 
\frac{40}{9} {\cal D}_0(\bar z)
+ \frac{16}{3} \bigg]
+ \frac{16 {\cal D}_0(\bar z)}{3}
\Big[{\cal D}_0(y) {\cal D}_1(\bar y) 
+ {\cal D}_1(y) {\cal D}_0(\bar y) 
\Big]\Bigg]\Bigg\}\,,
\eea
at leading power, and 
\bea\label{PartonicRapidityRes-2r-NLP-Expanded} \nn
\Delta^{(2)}(z,y)|_{\rm 2r}^{\rm NLP}
&=& \bigg(\frac{\as}{4\pi}\bigg)^2 
\Bigg\{ C_F^2  \Bigg[ 
\Big[\delta(y) + \delta(1 - y)\Big] 
\bigg[ \frac{16}{\eps^3} 
- \frac{64 {\cal L}_1(\bar z)}{\eps^2} 
+ \frac{128 {\cal L}_2(\bar z)
+4 - 48 \zeta_2}{\eps} \\ \nn
&&\hspace{-2.2cm}-\,
\frac{512}{3} {\cal L}_3(\bar z) 
+ \Big(192 \zeta_2 - 16\Big) {\cal L}_1(\bar z) 
+ 12 - \frac{224\zeta_3}{3}\bigg] 
+ {\cal D}_0(y) {\cal D}_0(\bar y) 
\bigg[-\frac{32}{\eps^2}
+\frac{128 {\cal L}_1(\bar z)}{\eps}\\ \nn
&&\hspace{-2.2cm}-\, 8 + 96\zeta_2 
- 256 {\cal L}_2(\bar z) \bigg] +
\Big[{\cal D}_0(y) {\cal D}_1(\bar y) 
+ {\cal D}_1(y) {\cal D}_0(\bar y) \Big]
\bigg[\frac{64}{\eps} -256 {\cal L}_1(\bar z)\bigg] \\ \nn
&&\hspace{-2.2cm}-\, 
128 {\cal D}_1(y) {\cal D}_1(\bar y) 
- 64 \Big[{\cal D}_0(y) {\cal D}_2(\bar y) 
+ {\cal D}_2(y) {\cal D}_0(\bar y) \Big] \Bigg] 
+ C_A C_F \Bigg[ \Big[\delta(y) 
+ \delta(1 - y)\Big] \bigg[\frac{4}{\eps^3}  \\ \nn
&&\hspace{-2.2cm}+\, 
\frac{1}{\eps^2}\bigg(
\frac{22}{3} - 16 {\cal L}_1(\bar z) \bigg) 
+\frac{1}{\eps}\bigg(32{\cal L}_2(\bar z)
- \frac{88}{3} {\cal L}_1(\bar z)
+ \frac{110}{9} - 16 \zeta_2\bigg) 
- \frac{128}{3}{\cal L}_3(\bar z) \\ \nn
&&\hspace{-2.2cm}
+\, \frac{176}{3}{\cal L}_2(\bar z)
+ \bigg(64\zeta_2 - \frac{440}{9} \bigg)
{\cal L}_1(\bar z) + \frac{580}{27} 
-22\zeta_2 - \frac{116\zeta_3}{3}\bigg] 
+ {\cal D}_0(y) {\cal D}_0(\bar y) 
\bigg[-\frac{8}{\eps^2} \\ \nn
&&\hspace{-2.2cm}+\, \frac{1}{\eps} 
\bigg(32 {\cal L}_1(\bar z) 
-\frac{44}{3}\bigg) - 64 {\cal L}_2(\bar z) 
+ \frac{176}{3} {\cal L}_1(\bar z)
- \frac{220}{9} +32\zeta_2 \bigg] 
+\Big[{\cal D}_0(y) {\cal D}_1(\bar y)  \\ \nn
&&\hspace{-2.2cm}+ {\cal D}_1(y) 
{\cal D}_0(\bar y) \Big]
\bigg[\frac{16}{\eps} 
- 64 {\cal L}_1(\bar z) 
+\frac{88}{3} \bigg]
- 32 {\cal D}_1(y) 
{\cal D}_1(\bar y) - 16 \Big[{\cal D}_0(y) 
{\cal D}_2(\bar y) + {\cal D}_2(y) 
{\cal D}_0(\bar y) \Big] \Bigg] \\ 
&&\hspace{-2.2cm}+\, n_f C_F  \Bigg[ 
\Big[\delta(y) + \delta(1 - y)\Big] 
\bigg[ -\frac{4}{3\eps^2} 
+ \frac{1}{\eps}\bigg(
\frac{16}{3} {\cal L}_1(\bar z)
- \frac{32}{9} \bigg) 
-\frac{32}{3} {\cal L}_2(\bar z) 
+\frac{128}{9} {\cal L}_1(\bar z)\nn \\ 
&&\hspace{-2.2cm}-\, 
\frac{244}{27} + 4\zeta_2 \bigg]
+{\cal D}_0(y) {\cal D}_0(\bar y) 
\bigg[\frac{8}{3\eps} 
- \frac{32}{3} {\cal L}_1(\bar z)
+ \frac{64}{9} \bigg]
- \frac{16}{3}
\Big[{\cal D}_0(y) {\cal D}_1(\bar y) 
+ {\cal D}_1(y) {\cal D}_0(\bar y) 
\Big]\Bigg]\Bigg\}\,,\nn\\
\phantom{}
\eea
at NLP.

\subsubsection*{Sum}

The double-real correction completes our 
calculation of the Drell-Yan rapidity 
distribution at NNLO. Summing eqs. 
(\ref{PartonicRapidityRes-2virtual}),
(\ref{PartonicRapidityRes-1r1v-LPh-Expanded}),
(\ref{PartonicRapidityRes-1r1v-LPs-Expanded}),
(\ref{PartonicRapidityRes-1r1v-NLPh-Expanded}),
(\ref{PartonicRapidityRes-1r1v-NLPccb-Expanded}),
(\ref{PartonicRapidityRes-1r1v-NLPs-Expanded}),
(\ref{PartonicRapidityRes-2r-LP-Expanded}) 
and~(\ref{PartonicRapidityRes-2r-NLP-Expanded})
we obtain
\bea\label{PartonicRapidityNNLO-LP-Expanded} \nn
\Delta^{(2)}(z,y)|^{\rm LP}
&=& \bigg(\frac{\as}{4\pi}\bigg)^2 
\Bigg\{ C_F^2  \Bigg[ 
\Big[\delta(y) + \delta(1 - y)\Big] 
\bigg[ \delta(1 - z) 
\bigg( \frac{9-8\zeta_2}{\eps^2} 
+\frac{1}{\eps}\bigg(\frac{189}{4} 
- 30 \zeta_2 - 28 \zeta_3 \bigg) \\ \nn
&&\hspace{-2.0cm}+\,\frac{1279}{8}-130\zeta_2
-54\zeta_3 + 80\zeta_4 \bigg)
+ {\cal D}_0(\bar z)
\bigg( \frac{24}{\eps^2} 
+\frac{64 - 16\zeta_2}{\eps}
+128 - 96\zeta_2 + 32\zeta_3 \bigg) \\ \nn
&&\hspace{-2.0cm}+\, {\cal D}_1(\bar z)
\bigg( \frac{32}{\eps^2} 
-\frac{48}{\eps} -128 - 64\zeta_2 \bigg) 
+ {\cal D}_2(\bar z)
\bigg( 48 -\frac{96}{\eps} \bigg) 
+\frac{448}{3}{\cal D}_3(\bar z)
+\frac{8}{\eps^2}-\frac{48}{\eps} - 128+8\zeta_2 \\ \nn
&&\hspace{-2.0cm}+\, {\cal L}_1(\bar z)
\bigg( 96 -\frac{96}{\eps} \bigg)
+ 256{\cal L}_2(\bar z) \bigg] 
+{\cal D}_0(y) {\cal D}_0(\bar y) 
\bigg[ {\cal D}_0(\bar z)
\bigg( \frac{16}{\eps^2} 
-\frac{24}{\eps} -64 - 32\zeta_2 \bigg) \\ \nn
&&\hspace{-2.0cm}+\, {\cal D}_1(\bar z)
\bigg( 48 -\frac{96}{\eps} \bigg) 
+224{\cal D}_2(\bar z) -\frac{48}{\eps} 
+ 48 + 256{\cal L}_1(\bar z) \bigg] +
\Big[{\cal D}_0(y) {\cal D}_1(\bar y) \\ \nn
&&\hspace{-2.0cm}+\, {\cal D}_1(y) {\cal D}_0(\bar y) \Big]
\bigg[{\cal D}_0(\bar z)
\bigg( 24 -\frac{48}{\eps} \bigg) 
+224{\cal D}_1(\bar z) + 128 \bigg] +
112 {\cal D}_0(\bar z) {\cal D}_1(y) {\cal D}_1(\bar y) \\ \nn
&&\hspace{-2.0cm}+\, 56 {\cal D}_0(\bar z) 
\Big[{\cal D}_0(y) {\cal D}_2(\bar y) 
+ {\cal D}_2(y) {\cal D}_0(\bar y) \Big] \Bigg] 
+ C_A C_F \Bigg[ 
\Big[\delta(y) + \delta(1 - y)\Big] 
\bigg[ \delta(1 - z) 
\bigg( -\frac{11}{2\eps^2}  \\ \nn
&&\hspace{-2.0cm}
+\,\frac{1}{\eps}\bigg(-\frac{123}{4} 
+\frac{44\zeta_2}{3} +6 \zeta_3 \bigg) 
-\frac{981}{8} + \frac{1553\zeta_2}{18}  
+ \frac{130\zeta_3}{3} -13\zeta_4 \bigg) 
+ {\cal D}_0(\bar z) \bigg( -\frac{22}{3\eps^2}  \\ \nn
&&\hspace{-2.0cm}+\, 
\frac{1}{\eps}\bigg( -\frac{134}{9} + 4\zeta_2 \bigg)
- \frac{808}{27} +22\zeta_2 + 28\zeta_3 \bigg) 
+ {\cal D}_1(\bar z)
\bigg( \frac{88}{3\eps} +\frac{536}{9} 
- 16\zeta_2 \bigg) -\frac{176}{3} {\cal D}_2(\bar z) \\ \nn
&&\hspace{-2.0cm}+\, \frac{44}{3\eps} + \frac{238}{9}-8\zeta_2
-\frac{176}{3} {\cal L}_1(\bar z) \bigg]  
+ {\cal D}_0(y) {\cal D}_0(\bar y) 
\bigg[ {\cal D}_0(\bar z)
\bigg( \frac{44}{3\eps} +\frac{268}{9} - 8\zeta_2 \bigg) 
- \frac{176}{3}{\cal D}_1(\bar z)
- \frac{88}{3} \bigg] \\ \nn
&&\hspace{-2.0cm}
-\,\frac{88}{3}{\cal D}_0(\bar z) \Big[{\cal D}_0(y) {\cal D}_1(\bar y) 
+{\cal D}_1(y) {\cal D}_0(\bar y) \Big] \Bigg] 
+ n_f C_F \Bigg[ 
\Big[\delta(y) + \delta(1 - y)\Big] 
\bigg[ \delta(1 - z) \bigg( \frac{1}{\eps^2}  \\ \nn
&&\hspace{-2.0cm}+\, 
\frac{1}{\eps}\bigg(\frac{11}{2} 
-\frac{8\zeta_2}{3} \bigg) +\frac{85}{4} 
- \frac{139\zeta_2}{9} - \frac{4\zeta_3}{3} \bigg) 
+ {\cal D}_0(\bar z) 
\bigg(\frac{4}{3\eps^2} + \frac{20}{9\eps}
+ \frac{112}{27} -4 \zeta_2 \bigg) \\ \nn
&&\hspace{-2.0cm}+\, {\cal D}_1(\bar z)
\bigg( - \frac{16}{3\eps}  -\frac{80}{9}\bigg) 
+\frac{32}{3} {\cal D}_2(\bar z) 
- \frac{8}{3\eps} - \frac{28}{9} + 
\frac{32}{3} {\cal L}_1(\bar z) \bigg]
+{\cal D}_0(y) {\cal D}_0(\bar y) 
\bigg[ {\cal D}_0(\bar z) \bigg(-\frac{8}{3\eps} \\ 
&&\hspace{-2.0cm}-\, \frac{40}{9} \bigg)
+ \frac{32}{3}{\cal D}_1(\bar z) + \frac{16}{3} \bigg] 
+\frac{16}{3}{\cal D}_0(\bar z) \Big[{\cal D}_0(y) {\cal D}_1(\bar y) 
+{\cal D}_1(y) {\cal D}_0(\bar y) \Big] \Bigg]\Bigg\}\,,
\eea
at leading power, and 
\bea\label{PartonicRapidityNNLO-NLP-Expanded} \nn
\Delta^{(2)}(z,y)|^{\rm NLP}
&=& \bigg(\frac{\as}{4\pi}\bigg)^2 
\Bigg\{ C_F^2  \Bigg[ 
\Big[\delta(y) + \delta(1 - y)\Big] 
\bigg[ - \frac{32 {\cal L}_1(\bar z)+16}{\eps^2}  \\ \nn
&&\hspace{-2.2cm}+\, \frac{96 {\cal L}_2(\bar z) 
+ 40 {\cal L}_1(\bar z)
-46 +16 \zeta_2}{\eps}
-\frac{448}{3} {\cal L}_3(\bar z)
-52 {\cal L}_2(\bar z)
+ \Big(94+ 64 \zeta_2\Big) {\cal L}_1(\bar z) \\ \nn
&&\hspace{-2.2cm}-\, 64 +40\zeta_2 - 32 \zeta_3 \bigg] 
+ {\cal D}_0(y) \bigg(\frac{16}{\eps} + 20 
- 16{\cal L}_1(\bar y) - 48{\cal L}_1(\bar z) \bigg)
+ {\cal D}_0(\bar y) \bigg(\frac{16}{\eps} + 20 \\ \nn
&& \hspace{-2.2cm} - 16{\cal L}_1(y) 
- 48{\cal L}_1(\bar z) \bigg)
- 32 \Big({\cal D}_1(y) 
+ {\cal D}_1(\bar y)\Big)
+ {\cal D}_0(y) {\cal D}_0(\bar y) 
\bigg[-\frac{16}{\eps^2} 
+\frac{96{\cal L}_1(\bar z) + 8}{\eps} \\ \nn
&&\hspace{-2.2cm}-\, 224 {\cal L}_2(\bar z)
- 16 {\cal L}_1(\bar z) + 32 + 32\zeta_2 \bigg] 
+ \Big[{\cal D}_0(y) {\cal D}_1(\bar y) 
+ {\cal D}_1(y) {\cal D}_0(\bar y) \Big]
\bigg[\frac{48}{\eps} -224 {\cal L}_1(\bar z) - 8 \bigg] \\ \nn
&&\hspace{-2.2cm}-\, 112 {\cal D}_1(y) {\cal D}_1(\bar y) 
- 56 \Big[{\cal D}_0(y) {\cal D}_2(\bar y) 
+ {\cal D}_2(y) {\cal D}_0(\bar y) \Big] \Bigg] 
+ C_A C_F \Bigg[ \Big[\delta(y) 
+ \delta(1 - y)\Big] \bigg[\frac{22}{3\eps^2}  \\ \nn
&&\hspace{-2.2cm}+\, 
\frac{1}{\eps}\bigg(
- \frac{88}{3} {\cal L}_1(\bar z)
+ \frac{200}{9} - 4 \zeta_2\bigg) 
+ \frac{176}{3}{\cal L}_2(\bar z) 
+ \bigg(16\zeta_2 - \frac{710}{9} \bigg)
{\cal L}_1(\bar z) + \frac{688}{27} 
-22\zeta_2 - 28\zeta_3\bigg]  \\ \nn
&&\hspace{-2.2cm}-\,  20 \Big({\cal D}_0(y) 
+ {\cal D}_0(\bar y)\Big)
+{\cal D}_0(y) {\cal D}_0(\bar y) 
\bigg[-\frac{44}{3\eps}  
+ \frac{176}{3} {\cal L}_1(\bar z)
- \frac{220}{9} +8\zeta_2 \bigg] \\ \nn
&&\hspace{-2.2cm}+\,\frac{88}{3}
\Big[{\cal D}_0(y) {\cal D}_1(\bar y) 
+ {\cal D}_1(y) {\cal D}_0(\bar y) \Big] \Bigg] 
+ n_f C_F  \Bigg[ 
\Big[\delta(y) + \delta(1 - y)\Big] 
\bigg[ -\frac{4}{3\eps^2}+\frac{1}{\eps}\bigg(
\frac{16}{3} {\cal L}_1(\bar z) \\ \nn
&&\hspace{-2.2cm} 
-\, \frac{32}{9} \bigg)
- \frac{32}{3} {\cal L}_2(\bar z) 
+\frac{128}{9} {\cal L}_1(\bar z) 
- \frac{244}{27} + 4\zeta_2 \bigg]
+{\cal D}_0(y) {\cal D}_0(\bar y) 
\bigg[\frac{8}{3\eps} 
- \frac{32}{3} {\cal L}_1(\bar z)
+ \frac{64}{9} \bigg] \\
&&\hspace{-2.2cm} -\, \frac{16}{3}
\Big[{\cal D}_0(y) {\cal D}_1(\bar y) 
+ {\cal D}_1(y) {\cal D}_0(\bar y) 
\Big]\Bigg]\Bigg\}\,,
\eea
at NLP. \Eqns{PartonicRapidityNNLO-LP-Expanded}{PartonicRapidityNNLO-NLP-Expanded} 
still contain $\eps$ poles, which are eventually 
removed by PDF renormalisation. However, 
this result already allows us to conclude 
that the partonic cross section $\Delta(z,y)$
has the structure of \eqn{zyFactLP-NLP}
also at NNLO, i.e.\
\be\label{zyFactLP-NLP2}
\Delta^{(2)}(z,y) = 
\frac{\delta(y)+\delta(1-y)}{2} 
\Big[\Delta_{\rm LP}^{(2)}(z) +
\Delta^{(2)}_{\rm NLP,LLs}(z) \Big]
+\Delta^{(2)}_{\rm NLP,rest}(z,y) + \ord(1-z).
\ee
Indeed, $y$- and $(1-y)$-plus 
distributions appearing in the LP term of 
\eqn{PartonicRapidityNNLO-LP-Expanded} are 
effectively contributing at NLP, after the 
simplification in \eqn{PDFrapLP} is taken 
into account. Furthermore, such terms 
do not contain leading logarithms or 
leading plus distributions in $(1-z)$. These terms can therefore be rearranged
to be part of $\Delta_{\rm NLP,rest}(z,y)$
in \eqn{zyFactLP-NLP2}. Hence the only 
term remaining at LP is the term 
proportional to $(\delta(y)+\delta(1-y))/2$
in \eqn{PartonicRapidityNNLO-LP-Expanded}.
Concerning now the NLP result in
\eqn{PartonicRapidityNNLO-NLP-Expanded}, 
we see that leading logarithms in $(1-z)$
do arise only in the $(\delta(y)+\delta(1-y))/2$
term, and do not contribute to the $y$- and 
$(1-y)$-plus distribution terms. Thus 
also \eqn{PartonicRapidityNNLO-NLP-Expanded}
is consistent with \eqn{zyFactLP-NLP2}.

\subsection{Rapidity distribution 
and kinematic shifts}\label{sec:DYshifted}

The result in \eqn{zyFactLP-NLP2} allows us 
to conclude that near threshold the $y$-dependence
in the partonic cross section factorises into a 
universal factor $[\delta(y)+\delta(1-y)]/2$, 
provided one restricts to the LP term and the 
LLs at NLP. Although proven explicitly up to 
two loops, the discussion in the previous 
sections suggests that the result in 
\eqn{zyFactLP-NLP2} should extend to all 
orders, as in the ansatz of \eqn{zyFactLP-NLP}. 
The $z$-dependence is entirely contained into
the factors $\Delta_{\rm LP}^{(2)}(z)$ and 
$\Delta^{(2)}_{\rm NLP,LLs}(z)$, which are 
thus proportional to the corresponding terms
in the partonic invariant mass distribution. 
This result suggest that it may be possible 
to exploit methods developed in \cite{DelDuca:2017twk},
and obtain the NLP NLO partonic cross section 
not by direct calculation, but rather by means 
of kinematic shifts. To be more specific, 
given a colourless final state produced by 
the annihilation of an initial $q\bar q$
pair with momenta $p_a$, $p_b$, the squared matrix 
element should be given by 
\be\label{M2shiftedDY}
\sum_{\rm s,c,p}\left|\M\right|^2_{\mathrm{NLO, NLP}} 
= g_s^2\, C_F\, \frac{\hat s}{(p_a\cdot k)(p_b\cdot k)}
\sum_{\rm s,c,p} \left|\M^{(0)}_{q\bar q \to \gamma^*}
(p_a+\delta p_a,p_b+\delta p_b)\right|^2,
\ee
where 
\be\label{shifted_momenta}
\delta p_a^{\mu} =
-\frac{1}{2}\bigg(\frac{k \cdot p_b}{p_a \cdot p_b} p_a^{\mu}
- \frac{k \cdot p_a}{p_a \cdot p_b} p_b^{\mu} + k^{\mu} \bigg), 
\quad 
\delta p_b^{\mu} =
-\frac{1}{2}\bigg(\frac{k \cdot p_a}{p_a \cdot p_b} p_b^{\mu}
- \frac{k \cdot p_b}{p_a \cdot p_b} p_a^{\mu} + k^{\mu} \bigg), 
\ee
which implies $\hat s \to z \hat s $. Given the simple 
structure of $\big|{\cal M}_{q\bar q \to \gamma^*}^{(0)}\big|^2$,
see \eqn{M0sqDY}, which depends on the single 
scale $\hat s$, we immediately have
\be\label{M0sqDYshifted}
\sum_{\rm s,c,p}\left|\M^{(0)}_{q\bar q \to \gamma^*}
(p_a+\delta p_a,p_b+\delta p_b)\right|^2 
= 4(1-\eps) z \hat s N_c.
\ee
Inserting this result into
\eqn{M2shiftedDY} we get
\bea\label{M2shiftedDYres} \nn
\sum_{\rm s,c,p}\left|\M\right|^2_{\mathrm{NLO, NLP}} 
&=& \frac{\as}{4\pi} 64 \pi^2 N_c C_F (1-\eps)
\frac{z \hat s^2}{k \cdot p_a\, k \cdot p_b}  \\
&=& \frac{\as}{4\pi} 64 \pi^2 N_c C_F (1-\eps)
\frac{\hat s^2}{k \cdot p_a\, k \cdot p_b} 
\Big(1 - (1-z)\Big).
\eea

This result has to be compared with the 
exact result in \eqn{M1nlpOriginal}. We see 
that the two forms of the NLO correction are 
indeed equal up to the factor $1 - 2/\hat{s}
\big(k \cdot p_a + k \cdot p_b\big)$,
appearing in the exact result of \eqn{M1nlpOriginal}, 
vs the factor $1 - (1-z)$, obtained in the shifted 
result in \eqn{M2shiftedDYres}. Indeed, the two factors
coincide, given that upon phase space integration 
one has $\big(k \cdot p_a + k \cdot p_b\big) 
= \big(\hat{s}-Q^2\big)/2=\hat{s}(1-z)/2$. 
Furthermore, the rapidity distribution 
is symmetric w.r.t.\ the exchange 
$y \leftrightarrow (1-y)$, a feature 
which is not altered between LP and 
NLP. Indeed, near threshold additional 
gluon radiation is constrained to be soft, 
but remains isotropic. Thus phase space 
integration gives rise to the same factor 
$y^{-1-\eps}(1-y)^{-1-\eps}$ both at LP 
and NLP, as can be seen explicitly in 
\eqns{PartonicRapidityRes-1real-LP}{PartonicRapidityRes-1real-NLP},
and the shift procedure gives rise to the correct 
relation between the LP and NLP contribution, 
which within the exact result of 
\eqns{PartonicRapidityRes-1real-LP}{PartonicRapidityRes-1real-NLP} 
arises after phase space integration. 
To be more specific, integrating \eqn{M1nlpOriginal}
\emph{or} \eqn{M2shiftedDYres} against the phase space measure 
$d\Phi_{\gamma^*g} $ as in \eqn{PartonicRapidityDefB-1real}
gives rise to the same NLP LL result
\be\label{PartonicRapidityRes-1real-LP+NLP} 
\Delta^{(1)}(z,y)|_{\rm real}^{\rm LP+NLP,LL}
= \frac{\as C_F}{4\pi} 
\bigg( \frac{\bar{\mu}^2}{Q^2}\bigg)^{\eps} 
y^{-1-\eps}(1-y)^{-1-\eps} 
\big(1-z\big)^{-1-2\eps} 
\Big[1 - (1-z) \Big]
\bigg[4 + \ord(\eps)\bigg].
\ee
In light of using this result for 
resummation, let us keep the exact phase 
space dependence, and expand the $y$-dependent part in powers of 
$\eps$. One has
\bea\label{PartonicRapidityRes-1real-LP+NLPb} \nn
\Delta^{(1)}(z,y)|_{\rm real}^{\rm LP+NLP,LL}
&=& \frac{\as C_F}{\pi} 
\bigg( \frac{\bar{\mu}^2}{Q^2}\bigg)^{\eps} 
\frac{e^{\eps\gamma_E}}{\Gamma(1-\eps)}
\left[-\frac{\delta(y)+\delta(1-y)}{\eps}+\ord(\eps^0)\right]
z \big(1-z\big)^{-1-2\eps} \\ \nn
&=& \frac{\Gamma(-2\eps)}{\Gamma^2(-\eps)}
\left[-\frac{\delta(y)+\delta(1-y)}{\eps}+\ord(\eps^0)\right]
K_{\rm{NLP}}(z,\eps) \\ 
&=& \Big[\bar \Delta^{(0)}(y) +\ord(\eps) \Big]
K_{\rm{NLP}}(z,\eps),
\eea
where the factor $\bar \Delta^{(0)}(y)$ 
has been defined in \eqn{PartonicRapTreeBar},
and the factor 
\be\label{KNLP}
K_{\rm{NLP}}(z,\eps) = \frac{\as C_F}{\pi} 
\bigg( \frac{\bar{\mu}^2}{\s}\bigg)^{\eps}
z \big(1-z\big)^{-1-2\eps}
\frac{e^{\eps \gamma_E}\Gamma^2(-\eps)}{
\Gamma(-2\eps)\Gamma(1-\eps)},
\ee
has been introduced in eq.\ (3.62) of 
\cite{Bahjat-Abbas:2019fqa}. In this respect, the last line of \eqn{PartonicRapidityRes-1real-LP+NLPb} 
constitutes the generalisation of eq.\ (3.61) 
of \cite{Bahjat-Abbas:2019fqa} for 
the DY rapidity distribution. In Section 
\ref{subsec:DYgamma} we will use this 
result to sum the large logarithms in $1-z$
to all order in $\as$ at NLP, at LL accuracy.


\section{Rapidity distribution for fixed order diphoton production}
\label{diphoton}

In this section we move away from the discussion of the Drell-Yan
process and look instead at QCD-induced diphoton production. Also 
for this process we only consider the leading production channel,
given by an incoming quark-antiquark pair. Whereas in the previous 
section we considered the rapidity of the whole final state, i.e.\ 
the virtual photon, here we select the rapidity of one of the final 
state particles. The production of two photons is important in
particle physics, since one of the most important decay channels for
the Higgs boson is $H\to\gamma\gamma$. The QCD production of two
photons forms a large irreducible background in the $H\to\gamma\gamma$
analysis \cite{Chatrchyan:2012ufa,Aad:2012tfa}. In this section 
we look at diphoton production at fixed order, similar to the 
Drell-Yan process in the previous section. This will also serve 
as preparation for the calculation of the corresponding
resummed cross section, which we perform in 
Section \ref{resum}.

\subsection{Leading order}
\label{fixedorder}

\begin{figure}[h!]
\centering
\begin{subfigure}[b]{.35\linewidth}
\includegraphics[width=0.9\textwidth,height=5cm,keepaspectratio]{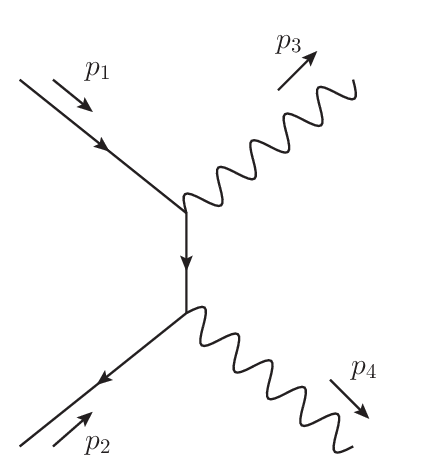}
\end{subfigure}
\hspace{2cm}
\begin{subfigure}[b]{.35\linewidth}
\includegraphics[width=0.9\textwidth,height=5cm,keepaspectratio]{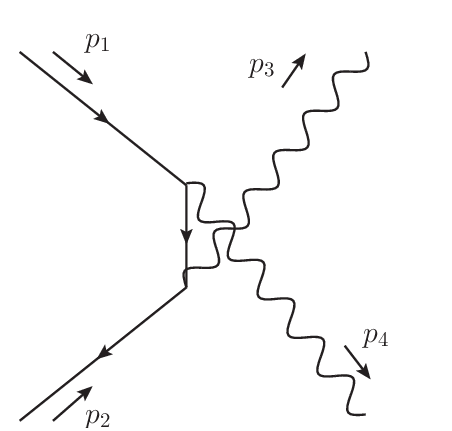}
\end{subfigure}
\caption{Tree-level contributions to diphoton production.}
\label{fig:diph-tree}
\end{figure}
\noindent We consider diphoton production by an incoming quark-antiquark pair
\begin{equation}
    q(p_1)+\bar{q}(p_2) \to \gamma(p_3)+\gamma(p_4)\,.
\end{equation}
At LO the process is represented by the two diagrams in figure \ref{fig:diph-tree}. Taking into account a symmetry factor 
of $\tfrac{1}{2}$ since the final-state photons are indistinguishable,
the corresponding squared matrix element, respectively 
averaged and summed over the spin, colour and polarisation 
degrees of freedom of the initial and final state, indicated by the bar, is given 
by 
\begin{align}
    \overline{\left|\M\right|_\text{LO}^2} 
    &= \frac{2(ee_q)^4}{N_c} \Bigg[\frac{1+\cos^2\theta}{1-\cos^2\theta}(1-\epsilon)^2-\epsilon(1-\epsilon)\bigg]\label{LOsquaredmatrixelement},
\end{align}
where $\theta$ is the angle between the incoming quark 
and the photon with momentum $p_3$ in the centre of mass 
frame of the incoming quarks, and where $e_q$ is the quark fractional
electric charge. For simplicity we set $e_q = 1$ and 
$e^2 = 4\pi \alpha$. At lowest order it does 
not matter which of the two photons is chosen, but at higher orders
we choose the photon with momentum $p_3$ as the one whose rapidity we record.
To calculate the cross section, 
one must integrate \eqn{LOsquaredmatrixelement} over 
the phase space of the two outgoing photons, which involves integrating $\theta$ between $\theta=0$ 
and $\theta=\pi$. This integrand diverges at both the
endpoints. However, since detectors have no coverage at these extremes
one may restrict the integration to a range from $\theta = \delta$ 
to $\theta =\pi - \delta$, with $\delta$ 
fixed by the experimental set-up. More convenient for this purpose
is the  pseudorapidity $\eta$
\begin{align}
\eta \equiv -\log\left(\tan\frac{\theta}{2}\right) = \frac12 \log\left(\frac{p_3^{(0)}+p_3^{(3)}}{p_3^{(0)}-p_3^{(3)}}\right)\label{rapidity},
\end{align}
with $p_3^2=0$ and $p_3^{(3)}=p_3^{(0)}\cos\theta$. 
Integrating the pseudorapidity in the range 
$(-\infty,\infty)$ still yields a divergent integral, 
but the cut-off $\delta$ on the angle $\theta$, which
we have introduced above, implies a corresponding cut-off 
on the pseudorapidity, which then also serves as a regulator of the 
integral. 
The squared amplitude in \eqn{LOsquaredmatrixelement} reads then
\begin{align}
\overline{\left|\M\right|_\text{LO}^2}
&= \frac{2(4\pi\alpha)^2}{N_c}
\left[\cosh(2\eta)(1-\epsilon)^2-\epsilon(1-\epsilon)\right]
\label{diphotonLO} \,.
\end{align}
The double differential cross section is therefore given by
\begin{align}\label{eq:LOsrosssection1}
\frac{d\hat{\sigma}^{(0)}_{q\bar{q}}}{dzd\eta}
(\hat{s},z,\eta,\epsilon,\bar\mu^2) = \frac{1}{2\hat{s}} 
\,\frac{d}{dzd\eta}\int dR_2 \, 
\overline{\left|\M\right|_\text{LO}^2}=\frac{d\bar{\sigma}^{(0)}_{q\bar{q}}}{dzd\eta}
(\hat{s},\eta,\epsilon,\bar\mu^2)\, \delta(1-z)\,,
\end{align}
where the phase space integration $\int dR_2$ 
is given in Appendix~\ref{PSintegral2} and
\begin{align}\label{eq:LOsrosssection2}
\frac{d\bar{\sigma}^{(0)}_{q\bar{q}}}{dzd\eta}
(\hat{s},\eta,\epsilon,\bar\mu^2) &= 
\frac{\pi \alpha^2}{N_c\,\hat{s}}
\frac{e^{\epsilon\gamma_E}}{\Gamma(1-\epsilon)}
\left(\frac{\bar\mu^2}{\hat{s}}\right)^{\epsilon}
(4\cosh^2\eta)^{\epsilon} \nonumber \\
&\quad\times \bigg[(1+\tanh^2\eta)(1-\epsilon)^2-\frac{\epsilon(1-\epsilon)}{\cosh^2\eta}\bigg]\,.
\end{align}
The partonic 
centre of mass energy is $\hat s = (p_1 + p_2)^2$,
we denote the invariant mass of the final state by
$Q^2 = (p_3 + p_4)^2$ and define the
corresponding ratio $z = Q^2/\hat s$. 
The double differential cross section 
in $d=4$ dimensions is therefore given by 
\begin{equation}\label{eq:dsigmagammalo}
\frac{d\hat{\sigma}_{q\bar{q}}^{(0)}}{dzd\eta}(\hat{s},z,\eta) 
= \frac{\pi \alpha^2}{N_c\,\hat{s}}\,(1+\tanh^2\eta)
\,\delta(1-z)\,. 
\end{equation}
This partonic cross section can be translated to a hadronic one. We define the hadronic rapidity
\begin{equation}
    Y=\eta+\frac12\log\left(\frac{x_a}{x_b}\right)\,.
\end{equation}
The hadronic cross section then reads
\begin{align}
\label{HadronicDiphoton}
\frac{d\sigma}{dQ^2dY}=\frac{1}{s}\int_\tau^1\frac{dz}{z}\int_{\log\left(\sqrt{\frac{\tau}{z}}e^Y\right)}^{\log\left(\sqrt{\frac{z}{\tau}}e^Y\right)} d\eta\, \mathcal{L}(z,\eta)\frac{d\hat{\sigma}}{dzd\eta}(\hat{s},z,\eta)\,,
\end{align}
where the luminosity function is given by
\begin{equation}
    \mathcal{L}(z,\eta) =f_{q/A}\left(e^{Y}e^{-\eta}\sqrt{\frac{\tau}{z}}\right)f_{\bar{q}/B}\left(e^{-Y}e^{\eta}\sqrt{\frac{\tau}{z}}\right)+(q\leftrightarrow \bar{q}),
\end{equation}
in analogy with \eqn{luminosityRap}. Note that there 
is a bound on the integration over the rapidity $\eta$ 
for finite $Y$. The divergence that we observed for 
the partonic cross section when one would integrate 
over the full range of $\eta$ is now transformed to a 
divergence if we would integrate over the full range 
of the hadronic rapidity $Y$. In reality there 
is a finite rapidity range for the hadronic rapidity 
$Y$. The partonic differential cross section 
$d\hat{\sigma}/dzd\eta$ can be calculated 
perturbatively and its LO contribution is 
given by \eqn{eq:dsigmagammalo}.

Comparing \eqn{PartonicRapTree}
with \eqn{eq:dsigmagammalo}, we see that 
the main difference between a distribution 
differential in the total rapidity of the final 
state (\eqn{PartonicRapTree}) and a distribution 
differential in the rapidity of a given particle 
in the final state (\eqn{eq:dsigmagammalo}) is 
that the latter exhibits a non-trivial dependence 
on the rapidity considered, given e.g.\ by 
the factor $1+\tanh^2\eta$ in case of diphoton 
production. Therefore, a priori it is not clear 
whether for $z\to 1$ one may express 
the NLO NLP correction to \eqn{eq:dsigmagammalo} 
as obtained in~\cite{DelDuca:2017twk}, namely, in terms of the 
LO partonic distribution with shifted kinematics 
times a universal factor. The purpose of this section 
is indeed to investigate this issue. To this end
we will now proceed in two different ways, both 
aimed at obtaining the NLP partonic distribution 
at NLO. At first we will 
derive this result from a direct calculation in the 
soft limit at NLO. We will then try to obtain the 
same NLP cross section at NLO by expressing it 
in terms of a universal factor times the LO partonic 
differential distribution with shifted kinematics. 
The calculation that we perform serves as an 
extension of the method developed 
in~\cite{DelDuca:2017twk}, which so far 
has been developed only for total cross 
sections, invariant mass distributions 
and (in Section \ref{sec:DYshifted} of the
present paper) for distributions differential 
in the total rapidity of the final state.

\subsection{Next-to-leading order}\label{sec:nlophoton}

At NLO, the partonic process with one 
gluon emission is given by
\begin{align}
q(p_1) + \bar{q}(p_2) \rightarrow \gamma(p_3) + \gamma(p_4) + g(k)\,,
\end{align}
with $k^2=0$. We define the following 
invariant variables
\begin{align}
s_{ij}=(\sigma_{i}p_i+\sigma_{j}p_j)^2\,, &\qquad i,j = 1,\dots ,4, \nonumber \\
s_{i}=(\sigma_{i}p_i-k)^2\,, &\qquad i= 1,\dots ,4 \,,
\end{align}
where $\sigma_{i}=+1$ if the momenta $p_{i}$ are incoming, and
$\sigma_{i}=-1$ otherwise. First we use the direct calculation method. We
calculate the squared matrix element straightforwardly and perform the
power expansion in the soft limit $s_i\ll s_{ij}$. Note that there
are five independent variables in the massless $2 \to 3$
process. Using momentum conservation we
can express $s_{i4}$ and $s_{4}$ as linear combinations of
$s_{ij}\, (i,j=1,2,3)$ and $s_i\, (i=1,2,3)$. We can expand the NLO
squared matrix element in the limit
$s_i\ll s_{ij}$. After the expansion, $s_3$ can be further removed by using the on-shell
condition $p_4^2=(p_1+p_2-p_3-k)^2=0$. As a result, we have the
squared matrix element in the following form
\begin{align}\label{eq:ampnlogeneral}
\overline{\big|\mathcal{M}\big|_{\rm real}^2}\left(s_{12},s_{13},s_{23},s_1,s_2\right)=\sum_{l}c_l(\epsilon)s_{12}^{\alpha_l}s_{13}^{\beta_l}s_{23}^{\gamma_l}s_{1}^{\kappa_l}s_{2}^{-1-\alpha_l-\beta_l-\gamma_l-\kappa_l}\,.
\end{align}
Note that the total power of $s_{ij}$ and $s_i$ is limited by the mass dimension of the squared amplitude. In terms of the light-cone coordinates defined in  Appendix \ref{PhaseSpace-DY} we have
\begin{align}
s_{13}&=-2\sqrt{\frac{\hat{s}}{2}}p_3^-\,,
\qquad \qquad 
s_{23}=-2\sqrt{\frac{\hat{s}}{2}}p_3^+\,, \nonumber \\
s_{1}&=-2\sqrt{\frac{\hat{s}}{2}}k^-\,,
\qquad \qquad 
s_{2}=-2\sqrt{\frac{\hat{s}}{2}}k^+\,,
\end{align}
where $\hat{s}=s_{12}$. To obtain the differential 
cross section at NLO, we need to calculate the 
three-body phase space integral, which is given in Appendix~\ref{PSintegral3}. Up to NLP, 
the contribution from the NLO real emission to 
the differential cross section is given by
\begin{align}\label{eq:NLObare}
    \frac{d\hat{\sigma}_{q\bar{q}}^{(1)}(Q^2,z,\eta,\epsilon,\bar\mu^2)}{dzd\eta}\bigg|_{\text{real}}^{\text{NLP}}&= \frac{\alpha_sC_F}{\pi}\left(\frac{\bar\mu^2}{Q^2}\right)^{\epsilon}\frac{2z^{1+\epsilon}(1-z)^{-1-2\epsilon}e^{\eps\gamma_E}\Gamma(-\epsilon)}{\Gamma(1-2\epsilon)}\frac{d\bar{\sigma}^{(0)}_{q\bar{q}}}{dzd\eta}(Q^2,\eta,\epsilon,\bar\mu^2)\nn\\
    &\hspace{-2.0cm}=\frac{\alpha_sC_F}{\pi}\left(\frac{\bar\mu^2}{Q^2}\right)^\eps\bigg[\delta(1-z)\bigg(\frac{1}{\eps^2}- \frac{\pi^2}{4}\bigg)+\frac{2}{\eps}\left(1-\frac{1}{1-z}\bigg|_+\right) \\  \nn
    &\hspace{0.0cm}+4 \,\frac{\log(1-z)}{1-z}\bigg|_+-4\log(1-z) + 2 +\ord(\eps) \bigg] \frac{d\bar{\sigma}^{(0)}_{q\bar{q}}}{dzd\eta}(Q^2,\eta,\epsilon,\bar\mu^2)
\end{align}
where the factor $z^{1+\epsilon}$ follows from
\begin{align}
    \frac{d\bar{\sigma}^{(0)}_{q\bar{q}}}{dzd\eta}(\hat{s},\eta,\epsilon,\bar\mu^2) \rightarrow \frac{d\bar{\sigma}^{(0)}_{q\bar{q}}}{dzd\eta}(Q^2,\eta,\epsilon,\bar\mu^2) \,.
\end{align}
In the second line of \eqn{eq:NLObare} we expanded 
the result to NLP in $(1-z)$ and to finite order in $\epsilon$. In fact, the NLO differential cross 
section can be calculated up to arbitrary power 
in $(1-z)$ by using \eqn{eq:phaseintnlo}. We will
demonstrate this in Section \ref{beyondNLP}.
Upon adding the virtual contribution at one loop, 
the $1/\eps^2$ pole from \eqn{eq:NLObare} cancels 
with the virtual contribution, leaving a $1/\eps$ 
pole that is removed by so-called mass factorisation, i.e.\ PDF renormalisation. This 
subtraction term arises from \eqn{HadronicDiphoton} 
when we replace the hadrons by the quarks. The 
left-hand side of that equation becomes the 
partonic cross section with $1/\eps$ poles, and 
the PDFs become parton-in-parton distributions, 
which are directly related to the Altarelli-Parisi 
splitting kernels. From this one obtains the finite partonic cross section at $\mathcal{O}(\alpha_s)$ accuracy
\begin{align}\label{eq:massfactorisation}
     \frac{d\hat{\sigma}_{q\bar{q}}^{(1)}(Q^2,z,\eta,\epsilon,\bar\mu^2)}{dzd\eta}\bigg|_{\text{ren}} & = \frac{d\hat{\sigma}_{q\bar{q}}^{(1)}(Q^2,z,\eta,\epsilon,\bar\mu^2)}{dzd\eta}\bigg|_{\text{real}+\text{virtual}} \nn \\
     &\hspace{1.0cm}
     -\int_0^1d\xi\,\Gamma_{qq}^{(1)}(\xi)\frac{d\hat{\sigma}_{q\bar{q}}^{(0)}(\xi\hat{s},\xi,z,\eta,\epsilon,\bar\mu^2)}{dzd\eta}\,,
\end{align}
where $\Gamma_{qq}^{(1)}(\xi)$ is the 
well-known splitting function kernel, 
given by
\begin{align}\label{eq:splitting}
    \Gamma_{qq}^{(1)}(\xi)&= -\frac{\alpha_sC_F}{2\pi}\left(\frac{\bar\mu^2}{Q^2}\right)^{\epsilon}\frac{1}{\epsilon}\left[2\frac{1}{1-\xi}\bigg|_+-2+(1-\xi)+\frac{3}{2}\delta(1-\xi)\right]\,.
\end{align}
The LO cross section entering the subtraction 
term in \eqn{eq:massfactorisation} is defined as
\begin{align} \label{LOpartonic}
   \frac{d\hat{\sigma}_{q\bar{q}}^{(0)}(\xi\hat{s},\xi,z,\eta,\epsilon,\bar\mu^2)}{dzd\eta} &= \frac{1}{8N_c^2\xi \hat{s}}\bigg\{\int d\Phi_2(\xi p_1+p_2;p_3,p_4)
   \big|\mathcal{M}_\text{LO}(\xi p_1+p_2;p_3,p_4)\big|^2 \nonumber \\ 
   &\hspace{-2.5cm}+ \int d\Phi_2(p_1+\xi p_2;p_3,p_4)\big|\mathcal{M}_\text{LO}(p_1+\xi p_2;p_3,p_4)\big|^2\bigg\}  \delta\left[\eta-\frac12\log\left(\frac{p_3^+}{p_3^-}\right)\right]\delta\left[\xi-\frac{Q^2}{\hat{s}}\right] \,.
\end{align}
The calculation of squared matrix elements and the 
two-body phase space is straightforward and is 
similar to \eqn{eq:LOsrosssection1}. Note that 
only LP contributes to \eqn{eq:LOsrosssection1}. 
The LP result of \eqn{LOpartonic} is proportional to \eqn{eq:LOsrosssection2}, while higher power 
corrections appear due to $p_1\to \xi p_1$ or 
$p_2\to \xi p_2$. When combining the two terms in 
\eqn{LOpartonic} NLP corrections cancel each 
other, while even higher power corrections 
remain. As a result, up to NLP, \eqn{LOpartonic} 
is given by
\begin{align}\label{eq:NLOsubtractionnlp}
   \frac{d\hat{\sigma}_{q\bar{q}}^{(0)}(\xi\hat{s},\xi,z,\eta,\epsilon,\bar\mu^2)}{dzd\eta}\bigg|^{\text{NLP}} =2\frac{d\bar{\sigma}^{(0)}_{q\bar{q}}}{dzd\eta}(Q^2,\eta,\epsilon,\bar\mu^2)\delta(\xi-z) \,.
\end{align}
We therefore get the subtraction term in 
\eqn{eq:massfactorisation} by combining 
\eqns{eq:splitting}{eq:NLOsubtractionnlp} 
to be of the following form
\begin{align}\label{eq:subtractionterm}
\int_0^1d\xi\,\Gamma_{qq}^{(1)}(\xi)\frac{d\hat{\sigma}_{q\bar{q}}^{(0)}(\xi\hat{s},\xi,z,\eta,\epsilon,\bar\mu^2)}{dzd\eta}\bigg|^{\text{NLP}} & 
\nn \\ 
&\hspace{-5.0cm}=\,\frac{\alpha_sC_F}{\pi}\left(\frac{\bar\mu^2}{Q^2}\right)^{\epsilon}\frac{1}{\epsilon}\bigg[ 2 \bigg( 1-\frac{1}{1-z}\bigg|_+ \bigg) -\frac{3}{2}\delta(1-z)\bigg]\frac{d\bar{\sigma}^{{(0)}}_{q\bar{q}}}{dzd\eta}(Q^2,\eta,\epsilon,\bar\mu^2) \,.
\end{align}
Focusing on the real contribution of \eqn{eq:NLObare}, 
one immediately sees that the part of the subtraction 
term in \eqn{eq:subtractionterm} not proportional to 
$\delta(1-z)$ does remove the single pole in 
\eqn{eq:NLObare}, while the $\delta(1-z)$ term  
removes the single pole in the virtual 
contribution. The latter is not discussed explicitly here
since we focus on the LL contributions up to NLP 
at NLO. In the end, we obtain the finite part of 
the NLO differential cross section with one 
gluon emission up to NLP LL accuracy, and it reads
\begin{align}
    \frac{d\hat{\sigma}_{q\bar{q}}^{(1)}(Q^2,z,\eta)}{dzd\eta}\bigg|_{\text{ren}}^{\text{NLP, LL}}
     &= \frac{\alpha_s}{4\pi}C_F\frac{(4\pi\alpha)^2}{\pi N_c Q^2} \left(1+\tanh^2\eta\right)\left\{\frac{\log(1-z)}{1-z}\bigg|_+ - \log(1-z)\right\}\,.\label{diphotonfinitecrosssection}
\end{align}
If we compare \eqn{diphotonfinitecrosssection} to \eqn{eq:dsigmagammalo}, we note that the dependence on the rapidity $\eta$ is the same, i.e.\ it completely factorises from the NLO contribution from emissions. This is in complete analogy with Drell-Yan, cf.\ in particular \eqn{zyFactLP-NLP}. Even though we used a different definition for the rapidity compared to the one used in Section \ref{sectionDY}, the conclusion is unchanged: the diphoton cross section again takes the form of \eqn{zyFactLP-NLP}. We will show in Section \ref{beyondNLP} that this factorised structure no longer holds beyond NLP LL, as indeed it did not for Drell-Yan.

We will now attempt to reproduce the 
result in \eqn{diphotonfinitecrosssection} by means of the 
method of shifted kinematics. We are interested in a cross 
section that is differential in both the invariant mass and 
the rapidity of one of the final state photons. The NLO real emission term up to 
NLP in the soft limit is expressed in terms of the LO squared 
matrix element, with initial state momenta shifted 
according to\footnote{Note that  $p_a \to p_1$ and $p_b \to p_2$, 
see figures \ref{fig:DYLO} 
and \ref{fig:diph-tree}.} \eqn{shifted_momenta}, integrated
over the \emph{full} 3-particle phase space $\int dR_3$, as 
defined in \eqn{R3Def}:
\begin{align}\label{eq:NLOShift}
\frac{d\hat{\sigma}_{q\bar{q}}^{(1)}(Q^2,z,\eta,\epsilon,\bar\mu^2)}{dzd\eta}\bigg|_{\text{real}}^{\text{NLP}} &= \frac{1}{2\hat s}
\int dR_3 \, \overline{\left|\M\right|^2_{\mathrm{NLO, NLP}}}\,,
\end{align}
where we keep only terms up to NLP on the right hand side, and where
\begin{equation}\label{M2shifted}
\overline{\left|\M\right|^2_{\mathrm{NLO, NLP}}} = g_s^2C_F\frac{2p_1\cdot p_2}{(p_1\cdot k)(p_2\cdot k)}
\overline{\left|\M(p_1+\delta p_1,p_2+\delta p_2)\right|_\text{LO}^2}.
\end{equation}
The shift of momenta in \eqn{M2shifted} implies 
the partonic centre of mass energy shifts as $\hat{s}\to z\hat{s}$. It also induces a shift 
in the rapidity: using momentum conservation within 
the Born approximation, the shift gives
\begin{align}\label{eq:etashift}
\eta = \frac{1}{2}\log\left(\frac{p_3^+}{p_3^-}\right)
&=\frac12\log\left(\frac{(p_1+p_2)^+-p_4^+}{(p_1+p_2)^--p_4^-}\right) \nonumber \\
&\to\frac12\log\left(\frac{(p_1+\delta p_1+p_2+\delta p_2)^+-p_4^+}{(p_1+\delta p_1+p_2+\delta p_2)^--p_4^-}\right) \nonumber \\ 
&=\frac12\log\left(\frac{(p_1+p_2)^+-p_4^+-k^+}{(p_1+p_2)^--p_4^--k^-}\right).
\end{align}
On the other hand, the delta function in $\int dR_3$
in the phase space \eqn{R3Def} fixes the rapidity to 
$\eta = 1/2\log\left(p_3^+/p_3^-\right)$. However,
within $\int dR_3$ we need to use momentum 
conservation for the $2\to 3$ process 
$q(p_1) + \bar{q}(p_2) \rightarrow \gamma(p_3) + \gamma(p_4) + g(k)$, which implies 
\begin{align}\label{eq:etaNLO}
\eta &= \frac{1}{2}\log\left(\frac{p_3^+}{p_3^-}\right)
= \frac12\log\left(\frac{(p_1+p_2)^+-p_4^+-k^+}{(p_1+p_2)^--p_4^--k^-}\right).
\end{align}
The two expressions do indeed coincide! Furthermore, 
upon closer inspection, we see that the rapidity can 
be expanded for soft $k$, with the leading power term
being equal to its Born value:
\begin{align}\label{eq:etaexpansion}
\eta &= \frac{1}{2}\log\left(\frac{(p_1+p_2-p_4)^+}{(p_1+p_2-p_4)^-}\right)+\frac12\left(\frac{k^-}{p_3^-}-\frac{k^+}{p_3^+}\right)+\mathcal{O}(k^2)\nonumber \\
&\equiv\bar{\eta}+\delta\eta+\mathcal{O}\left((1-z)^2\right),
\end{align}
where $p_1$, $p_2$ and $p_4$ are now the Born momenta, 
and where we defined $\bar{\eta}$ to be the rapidity 
at lowest order, and $\delta\eta \sim \mathcal{O}(1-z)$ 
its NLP correction. Both the fact that the shifted rapidity 
coincides with its NLO exact value, and that upon expansion 
in powers of $1-z$ the first term of the NLO rapidity is given by its Born value are relevant for our
discussion. Indeed, we can use \eqn{eq:etaexpansion} 
and expand the shifted squared matrix element around $\bar\eta$:
\begin{align}
\left|\M(p_1+\delta p_1,p_2+\delta p_2,\bar{\eta}+\delta\eta)\right|_\text{LO}^2 &=\left|\M(z\s,\bar{\eta})\right|_\text{LO}^2+\frac{\partial\left|\M(p_1,p_2,\bar{\eta})\right|_\text{LO}^2}{\partial\bar{\eta}}\delta\eta+\mathcal{O}(\delta\eta^2)\label{Mderivativeeta}.
\end{align}
Furthermore, following ref.~\cite{Laenen:1992ey}, we 
can expand the Dirac delta function defining the 
rapidity in the phase space $\int dR_3$ around its Born 
value, according to \eqn{eq:etaexpansion}:
\be \label{Rderivativeeta}
\delta\left(\eta-\frac12\log\left(\frac{p_3^+}{p_3^-}\right)\right) = \delta(\eta- \bar{\eta})+\delta\eta
\frac{\partial}{\partial \eta}\delta(\eta-\bar{\eta}) +\mathcal{O}(\delta\eta^2).
\ee
Next, we insert both \eqns{Mderivativeeta}{Rderivativeeta}
into \eqn{eq:NLOShift}. Focusing on the integration over $\eta$, we integrate by parts the term involving the 
derivative of the Dirac delta in \eqn{Rderivativeeta}, 
and arrive at 
\begin{align} \nn
\int & d\eta  \left[|\mathcal{M}(z\hat{s},\bar{\eta})|_\text{LO}^2+\frac{\partial |\mathcal{M}(p_1,p_2,\bar{\eta})|^2_\text{LO}}{\partial \bar{\eta}}\delta\eta+\mathcal{O}(\delta\eta)^2\right]
\left[\delta(\eta- \bar{\eta})+\delta\eta
\frac{\partial}{\partial \eta}\delta(\eta-\bar{\eta}) +\mathcal{O}(\delta\eta^2)\right] \\[0.2cm]
&=\int d\eta\, \bigg[|\mathcal{M}(z\hat{s},\bar \eta)|_\text{LO}^2+\delta\eta\left(\frac{\partial |\M(\hat{s},\eta)|_\text{LO}^2}{\partial \eta}-\frac{\partial |\M(z\hat{s},\eta)|_\text{LO}^2}{\partial \eta}\right)+\mathcal{O}(\delta\eta^2)\bigg] \delta(\eta-\bar{\eta}). \label{shiftderivative0}
\end{align}
The difference of the derivatives in the second term of \eqn{shiftderivative0} is at least $\mathcal{O}(1-z)$. Since it is multiplied by $\delta\eta$, this whole term is beyond NLP accuracy and can therefore be neglected. We can now 
proceed with the complete phase space integration from 
\eqn{eq:NLOShift}. Using the results in 
\eqn{shiftderivative0}, it reads
\begin{align}\label{eq:NLOShiftB} \nn
\frac{d\hat{\sigma}_{q\bar{q}}^{(1)}(Q^2,z,\eta,\epsilon,\bar\mu^2)}{dzd\eta}\bigg|_{\text{real}}^{\text{NLP}} &= 
\frac{1}{2\hat s}
\int dQ^2 \, d\Phi_3(p_1+p_2; p_3,p_4,k)
\, \delta\bigg(z - \frac{Q^2}{\hat s} \bigg) \\
&\hspace{-1.0cm}\times\,
\delta\left[Q^2-(p_3+p_4)^2\right] \, 
\delta(\eta-\bar{\eta})\, g_s^2C_F\frac{2p_1\cdot p_2}{(p_1\cdot k)(p_2\cdot k)}\, \overline{|\mathcal{M}(z\hat{s},\bar \eta)|_\text{LO}^2}.
\end{align}
The phase space integration involves only the 
eikonal factor $2p_1\cdot p_2/(p_1\cdot k \, p_2\cdot k)$,
which, using the result for the phase space integration of \eqn{eq:phaseintnlo}, gives
\begin{align} \nn
\int dR_3 \frac{\hat{s}}{(p_1\cdot k)(p_2\cdot k)} 
&= \int dR_3 \frac{2}{k^+k^-} 
=\int dz\, d\eta\,\frac{1}{32\pi^3}
\left(\frac{\bar\mu^2}{\s}\right)^{2\epsilon}
\frac{1}{\cosh^2\eta}(4\cosh^2\eta)^\epsilon  \\
&\hspace{2.0cm}\times \frac{e^{2\eps\gamma_E}\Gamma(-\epsilon)}{\Gamma(1-2\epsilon)\Gamma(1-\epsilon)}(1-z)^{-1-2\epsilon} (1+\epsilon(1-z)+...),
\end{align}
such that \eqn{eq:NLOShiftB} finally gives
\begin{align} \nn
\frac{d\hat{\sigma}_{q\bar{q}}^{(1)}(Q^2,z,\eta,\epsilon,\bar\mu^2)}{dzd\eta}\bigg|_{\text{real}}^{\text{NLP}} & = 
g_s^2\,C_F\,\frac{\alpha^2}{2\pi N_c}\left(\frac{\bar\mu^2}{\s}\right)^{\epsilon}\frac{e^{2\eps\gamma_E}\Gamma(-\epsilon)}{\Gamma(1-2\epsilon)\Gamma(1-\epsilon)} \frac{1}{z\s}\left(\frac{\bar\mu^2}{z\s}\right)^{\epsilon} \\[0.2cm]
&\hspace{-2.0cm}\times\left[(1+\tanh^2\eta)(1-\epsilon)^2-\frac{\epsilon(1-\epsilon)}{\cosh^2\eta}\right](4\cosh^2\eta)^\epsilon z(1-z)^{-1-2\epsilon}+....
\end{align}
The NLO cross section up to NLP LL accuracy is then given by 
\begin{align}\label{universality-result} \nn
\frac{d\hat{\sigma}_{q\bar{q}}^{(1)}(Q^2,z,\eta,\epsilon,\bar\mu^2)}{dzd\eta}\bigg|_{\text{real}}^{\text{NLP}} &= \frac{\alpha_s}{4\pi}C_F\frac{ 4\pi\alpha^2}{ N_cz\hat{s}}\left(\frac{\bar\mu^2}{\s}\right)^{\epsilon}\left(\frac{\bar\mu^2}{z\s}\right)^{\epsilon}\frac{2e^{2\eps\gamma_E}\Gamma(-\epsilon)}{\Gamma(1-2\epsilon)\Gamma(1-\epsilon)} 
\\[0.1cm] \nn
&\hspace{1.0cm}\times\, (1+\tanh^2\eta)
z(1-z)^{-1-2\epsilon} \\[0.2cm]
&= C_F K_\text{NLP}(z,\epsilon)
\frac{d\bar{\sigma}_{q\bar{q}}^{(0)}}{dzd\eta}(z\s,\eta),
\end{align}
which is the same as \eqn{eq:NLObare}. We introduced here again a $K$-factor, as in \eqn{KNLP}. We note that this cross section is precisely as was advocated in ref.~\cite{DelDuca:2017twk}, even though the cross section is now double differential, whereas that paper only considered cross sections that are differential in the threshold variable $z$. The divergences in the form of poles in $\epsilon$ can now be removed by adding the virtual contribution and by means of mass factorisation. Starting from \eqn{eq:massfactorisation}, we perform the same steps to arrive at the desired finite cross section, namely \eqn{diphotonfinitecrosssection}. 

Summarising, from the first calculation, i.e.\ \eqn{diphotonfinitecrosssection}, we have already concluded that the differential distribution is of the form of \eqn{zyFactLP-NLP}, i.e.\ that the dependence on the rapidity variable for the NLP LL contribution is the same as in the LO result. We then noted that there exists another prescription to obtain the NLP NLO contribution, namely the shifted kinematics method of \cite{DelDuca:2017twk}, which, before phase space integration, is given by \eqn{M2shifted}. This presents a problem however, since the shift in kinematics also affects the rapidity dependence, while we know that this should stay the same as the LO contribution. We could show explicitly that the effect of the shift on the rapidity variable is actually of NNLP accuracy, and therefore does not affect the general NLO-NLP formula eq.\ (4.17) of \cite{DelDuca:2017twk}. We can hence safely apply this formula and obtain \eqn{universality-result}, which is now also differential in the rapidity. The advantage of rephrasing the derivation in terms of this general NLO-NLP formula is that we can profit from the result of \cite{Bahjat-Abbas:2019fqa} and upgrade the NLO result to a resummed result quite straightforwardly. This will be discussed in detail in Section \ref{resum}.

\subsubsection{NLO result beyond NLP}
\label{beyondNLP}
Beyond NLP, the contribution is suppressed by factors of $(1-z)$ in the limit $z \to 1$. A logarithmic term $\log(1-z)$ multiplied by $(1-z)$ appears at N$^2$LP, while the cross section is free of this logarithm starting from N$^3$LP, which can be seen from the splitting kernel function in \eqn{eq:splitting}. While we have investigated the universal structure up to NLP LL at NLO in Section~\ref{sec:nlophoton}, it is still of interest to examine contributions beyond NLP at NLO. 
Due to the simple structure of the NLO squared amplitude in \eqn{eq:ampnlogeneral} and our generalised phase space integration formula \eqn{eq:phaseintnlo} in the soft limit, we can straightforwardly calculate the NLO cross section up to arbitrary powers of $(1-z)$. In this section we present the full analytic results valid up to N$^3$LP at NLO. Recalling the mass factorisation formula \eqn{eq:massfactorisation}, the finite NLO differential cross section receives contributions from two parts: the NLO correction and the subtraction term which is a convolution of the splitting kernel function and the LO cross section. Up to N$^3$LP, the LO cross section entering the subtraction term is
\begin{align}\label{eq:LOnnnlp}
   \frac{d\hat{\sigma}_{q\bar{q}}^{(0)}(\xi\hat{s},\xi,z,\eta,\epsilon,\bar\mu^2)}{dzd\eta}\bigg|^{\text{N}^3\text{LP}}
   &=2\frac{d\bar{\sigma}^{(0)}_{q\bar{q}}}{dzd\eta}(Q^2,\eta,\epsilon,\bar\mu^2)\delta(\xi-z) \nonumber \\ &\hspace{-1.5cm}+\frac{\pi\alpha^2e^{\epsilon\gamma_E}}{N_c Q^2\Gamma(1-\epsilon)}\left(\frac{\bar\mu^2}{Q^2}\right)^{\epsilon}(4\cosh^2\eta)^{\epsilon} \frac{(1-\epsilon)^2}{8\cosh^4\eta}(1-z)^2\left[1+(1-z)\right] \nonumber \\
    &\hspace{-1.5cm}\times\left[4(2-\cosh2\eta)+8\epsilon\cosh2\eta+2\epsilon^2(\cosh4\eta-1)\right]\delta(\xi-z)\,.
\end{align}
We note that the N$^2$LP and N$^3$LP contributions have the same structure. Combining \eqn{eq:LOnnnlp} and \eqn{eq:splitting}, the subtraction term is given by
\begin{align}\label{eq:NLOsubtractionnnnlp}
\int_0^1d\xi\,\Gamma_{qq}^{(1)}(\xi)\frac{d\hat{\sigma}_{q\bar{q}}^{(0)}(\xi\hat{s},\xi,z,\eta,\bar\mu^2)}{dzd\eta} &= -\frac{\alpha_sC_F}{\pi}\left(\frac{\bar\mu^2}{Q^2}\right)^{\epsilon}\nonumber \\
    &\hspace{-2.5cm}\times\frac{1}{\epsilon}\left[2\frac{1}{1-z}\bigg|_+-2+(1-z)+\frac{3}{2}\delta(1-z)\right] \bigg\{\frac{d\bar{\sigma}^{(0)}_{q\bar{q}}}{dzd\eta}(Q^2,\eta,\epsilon,\bar\mu^2)  \nonumber \\ 
    &\hspace{-2.5cm}+ \frac{\pi\alpha^2}{ N_c Q^2\Gamma(1-\epsilon)}\left(\frac{\bar\mu^2}{Q^2}\right)^{\epsilon}(4\cosh^2\eta)^{\epsilon}\frac{(1-\epsilon)^2e^{\eps\gamma_E}}{16\cosh^4\eta}(1-z)^2\left[1+(1-z)\right]\nonumber \\
    &\hspace{-2.5cm}\times\left[4(2-\cosh2\eta)+8\epsilon\cosh2\eta+2\epsilon^2(\cosh4\eta-1)\right]\bigg\}\,.
\end{align}
Note that the contributions of N$^3$LP and beyond in $\Gamma_{qq}^{(1)}(\xi)$ are zero. Up to N$^3$LP, the NLO real correction is given by
\begin{align}\label{eq:NLObarennnlp}
    \frac{d\hat{\sigma}_{q\bar{q}}^{(1)}(Q^2,z,\eta,\epsilon,\bar\mu^2)}{dzd\eta}\bigg|_{\text{real}}^{\text{N}^3\text{LP}}&= \frac{\alpha_sC_F}{\pi}\left(\frac{\bar\mu^2}{Q^2}\right)^{\epsilon}\frac{e^{\epsilon\gamma_E}\left(1+z^2\right)z^{\epsilon}(1-z)^{-1-2\epsilon}\Gamma(-\epsilon)}{\Gamma(1-2\epsilon)} \nonumber \\
    &\hspace{-2.5cm}\times\bigg\{\frac{d\bar{\sigma}^{(0)}_{q\bar{q}}}{dzd\eta}(Q^2,\eta,\epsilon,\bar\mu^2)  \nonumber \\ 
    &\hspace{-2.5cm}+ \frac{\pi\alpha^2e^{\epsilon\gamma_E}}{N_c Q^2\Gamma(1-\epsilon)}\left(\frac{\bar\mu^2}{Q^2}\right)^{\epsilon}(4\cosh^2\eta)^{\epsilon}(1-z)^2\frac{1+(1-z)}{16\cosh^4\eta}\nonumber \\
    &\hspace{-2.5cm}\times\left[4\left(2-\cosh2\eta\right)-\epsilon\left(4+11\cosh2\eta+4\cosh4\eta+\cosh6\eta\right)+\mathcal{O}(\epsilon^2)\right]\bigg\}\,,
\end{align}
Up to the chosen prefactor, the N$^2$LP and N$^3$LP contributions have the same structure. The prefactor is chosen to give the same  $1/\epsilon$ divergent terms as \eqn{eq:splitting}, namely
\begin{align}
    \frac{e^{\epsilon\gamma_E}\left(1+z^2\right)z^{\epsilon}(1-z)^{-1-2\epsilon}\Gamma(-\epsilon)}{\Gamma(1-2\epsilon)} &= \frac{1}{\epsilon^2}\delta(1-z)-\frac{1}{\epsilon}\left[2\frac{1}{1-z}\bigg|_+-2+(1-z)\right] \nonumber \\ 
    &\hspace{-3cm}+4\frac{\log(1-z)}{1-z}\bigg|_+-4\log(1-z)+2-\frac{\pi^2}{4}\delta(1-z) \nonumber \\ 
    &\hspace{-3cm}+(1-z)\left[2\log(1-z)-1\right]+\frac{2}{3}(1-z)^2+\mathcal{O}(\epsilon,(1-z)^3)\,,
\end{align}
Combining \eqns{eq:NLOsubtractionnnnlp}{eq:NLObarennnlp}, and ignoring for now the Dirac delta function $\delta(1-z)$ terms which will be given together with the NLO virtual corrections in the following, we find that the $\mathcal{O}(1/\epsilon)$ pole has been cancelled. The finite part up to N$^3$LP is then given by
\begin{align}\label{eq:finteNNNLP}
    \frac{d\hat{\sigma}_{q\bar{q}}^{(1)}(Q^2,z,\eta)}{dzd\eta}\bigg|_{\text{ren}}^{\text{N}^3\text{LP}} &= \frac{\alpha_sC_F}{\pi}\frac{\pi\alpha^2}{N_c Q^2} \bigg\{4\left(1+\tanh^2\eta\right)\frac{\log(1-z)}{1-z}\bigg|_+ \nonumber \\ 
    &\hspace{-1cm}+2\left(1+\tanh^2\eta\right)\left[1-2 \log(1-z)\right] \nonumber \\ 
    &\hspace{-1cm}+\frac{1-z}{8\cosh^4\eta}\left[4\left(5+\cosh4\eta\right)\log(1-z)-14+23\cosh2\eta+2\cosh4\eta+\cosh6\eta\right] \nonumber \\ 
    &\hspace{-1cm}+\frac{\left(1-z\right)^2}{6\cosh^4\eta}\left[7-\cosh2\eta + \cosh4\eta\right] + \mathcal{O}\left[(1-z)^3\right]\bigg\}\,.
\end{align}
As mentioned before, there is a logarithmic term $\log(1-z)$ at N$^2$LP, while there are no such logarithms at N$^3$LP and beyond. It is clear that the $\eta$-structure of the coefficients of $\log(1-z)$ at N$^2$LP and beyond in this double differential cross section is different from those at LP and NLP, such that the LL resummation of threshold logarithms is non-trivial.
In order to get the full analytic results valid up to N$^3$LP at NLO, we need to keep track of the terms proportional to $\delta(1-z)$ when combining \eqns{eq:NLOsubtractionnnnlp}{eq:NLObarennnlp}, and add the contribution from the NLO virtual correction. We combine all these contributions proportional to $\delta(1-z)$ in the form
\begin{align}\label{eq:fintedelta}
    \frac{d\hat{\sigma}_{q\bar{q}}^{(1)}(Q^2,z,\eta)}{dzd\eta}\bigg|_{\text{ren},\delta(1-z)} &= \frac{\alpha_sC_F}{\pi}\frac{\pi\alpha^2}{N_c Q^2} \delta(1-z)\bigg[\frac{1}{6}\left(2\pi^2-21\right)\left(1+\tanh^2\eta\right) \nonumber \\ 
    &\quad+\frac{1}{4}\left(1+\tanh\eta\right)\left(5+\tanh\eta\right)\log\left(\frac{1-\tanh\eta}{2}\right)\nonumber \\ 
    &\quad+\frac{1}{4}\left(1-\tanh\eta\right)\left(5-\tanh\eta\right)\log\left(\frac{1+\tanh\eta}{2}\right) \nonumber \\ 
    &\quad+\frac{1}{4}\left(6-2\tanh\eta-\frac{1}{\cosh^2\eta}\right)\log^2\left(\frac{1-\tanh\eta}{2}\right)\nonumber \\ 
    &\quad+\frac{1}{4}\left(6+2\tanh\eta-\frac{1}{\cosh^2\eta}\right)\log^2\left(\frac{1+\tanh\eta}{2}\right)\bigg]\,.
\end{align}
The combination of  \eqns{eq:finteNNNLP}{eq:fintedelta} is the analytic expression for diphoton production, up to N$^3$LP, for the finite NLO cross section that is double differential in the diphoton invariant mass and single photon rapidity.

\section{Threshold resummation for rapidity distributions}
\label{resum}

In Section \ref{sectionDY} and \ref{diphoton} we discussed fixed order computations of the Drell-Yan process and QCD-induced diphoton production. In this section we will resum the leading threshold logarithms at both leading and next-to-leading power. Our method was previously only used to obtain resummed cross sections that are differential in the threshold variable $z$. In this section we will investigate whether we can also resum cross sections that are additionally differential in the rapidity. We start by discussing the case of diphoton production, where we consider the rapidity of one of the photons, using results of Section \ref{diphoton}. Subsequently we look at the Drell-Yan process. As a direct application of the diphoton case, we first develop resummation at NLP LL accuracy for the distribution differential in the rapidity of one of the final state leptons. Finally, using results of Section \ref{sectionDY}, we 
consider the distribution differential in the total rapidity of the final state (of the off-shell photon), and develop the resummation of large logarithms of $1-z$ at NLP LL accuracy for this case, too.

\subsection{NLP resummation of diphoton production}
\label{resumdiphoton}

Let us recall that in terms of the soft power expansion the partonic cross section can be written as
\begin{equation}\label{powerexpansionMPS}
\hat{\sigma} \sim \frac{1}{2\hat{s}}\left[\int d\Phi_\text{LP}\left|\M\right|_\text{LP}^2+\int d\Phi_\text{LP}\left|\M\right|_\text{NLP}^2+\int d\Phi_\text{NLP}\left|\M\right|_\text{LP}^2+\mathcal{O}(\text{NNLP})\right],
\end{equation}
i.e. NLP large logarithms arise both from the squared matrix element expanded to NLP and from the expansion of the phase space. The results of Section \ref{diphoton} allow us to deal with the LLs in the second term of \eqn{powerexpansionMPS}. Therefore, as discussed in \cite{Bahjat-Abbas:2019fqa} for the case of invariant mass distribution, we need first to assess the influence of the third terms of \eqn{powerexpansionMPS}, namely, we need to determine whether LLs can arise from the NLP contribution of the phase space. 

To this end, we need the squared matrix element involving $n$ gluons in the final state at LP and the corresponding 
($n+2$)-particle phase space measure. For LL accuracy, the former has the 
simple eikonal form  
\begin{equation}\label{LPsqmatrixn}
|\M|^2_{\text{LP},n} = f(\alpha_s,\epsilon,\bar\mu^2,\eta)\prod_{i=1}^n\frac{p_1\cdot p_2}{p_1\cdot k_i p_2\cdot k_i},
\end{equation}
where $f(\alpha_s,\epsilon,\bar\mu^2,\eta)$ is a general 
function that collects all factors not involved in the 
phase space integration. The ($n+2$)-particle phase 
space measure is given by
\begin{align}
\int d\Phi_{n+2}&
=\int dQ^2 \int d\eta \int\frac{d^dp_3}{(2\pi)^{d-1}}\delta_+(p_3^2)\int\frac{d^dp_4}{(2\pi)^{d-1}}\delta_+(p_4^2)\left[\prod_{i=1}^n\int\frac{d^dk_i}{(2\pi)^{d-1}}\delta_+(k_i^2)\right]\nonumber \\
&\hspace{-1.0cm}\times\, 
\delta(Q^2-(p_3+p_4)^2)
\,\delta\left(\eta-\frac12\log\left(\frac{p_3^+}{p_3^-}\right)\right)(2\pi)^d
\delta^{(d)}\left(p_1+p_2-p_3-p_4-\sum_{i=1}^n k_i\right),
\end{align}
where we put any $\bar\mu$-dependence 
in the general function $f$. We can integrate 
out $p_4$ using the momentum-conserving delta 
function. We then have
\begin{align}
    &\delta(p_4^2) = \frac{1}{\s}\delta\left(1-\frac{2p_3\cdot(p_1+p_2)}{\s}-\frac{2\sum_i k_i\cdot(p_1+p_2)}{\s}+\frac{2p_3\cdot\sum_i k_i}{\s}+\frac{2\sum_{i<j}k_i\cdot k_j}{\s}\right),\nn\\
    &\delta(Q^2-(p_3+p_4)^2) = \frac{1}{\s}\delta\left(1-z-\frac{2\sum_i k_i\cdot(p_1+p_2)}{\s}+\frac{2\sum_{i<j}k_i\cdot k_j}{\s}\right).\label{deltacorrelation}
\end{align}
In order to perform all these integrals we use a representation of the delta function as a Laplace transform, given by 
\begin{equation}\label{deltatransform}
    \delta(x) = \int_{-i\infty}^{i\infty}\frac{dT}{2\pi i}e^{Tx}.
\end{equation}
Turning to light-cone coordinates, the integrated squared matrix element reads
\begin{align}\label{nplus2PSLP}
\int d\Phi_{n+2}|\M|^2_{\text{LP},n} &
= \frac{2\pi}{\s^2}\frac{\Omega_{d-2}^{n+1}}{2^{n+1}}\int\frac{dp_3^+dp_3^-}{(2\pi)^{d-1}}
(2p_3^+p_3^-)^\frac{d-4}{2}\left[\prod_{i=1}^n\int\frac{dk_i^+dk_i^-}{(2\pi)^{d-1}}
(2k_i^+k_i^-)^{\frac{d-4}{2}}
\frac{1}{k_i^+k_i^-}\right]\nonumber \\[0.2cm]
&\hspace{0.5cm}\times\int dQ^2 d\eta\frac{dT}{2\pi i}
\frac{d\tau}{2\pi i} 2p_3^+\delta(p_3^+-e^{2\eta}p_3^-)
f(\alpha_s,\epsilon,\mu^2,\eta)e^{T(1-z)+\tau} \nonumber \\[0.3cm]
&\hspace{0.5cm}\times 
\prod_{i=1}^ne^{-\sqrt{\frac{2}{\s}}(T+\tau)(k_i^++k_i^-)} e^{-\sqrt{\frac{2}{\s}}\tau(p_3^++p_3^-)}\nonumber \\
&\hspace{0.5cm}\times\left(1+\frac{2(T+\tau)}{\s}
\sum_{i<j}k_i\cdot k_j+\sum_{i=1}^n\frac{2\tau}{\s}(p_3^+k_i^-+p_3^-k_i^+)+\dots\right),
\end{align}
where $\Omega_{2d}=2\pi^{d}/\Gamma(d)$. The last line originates from expanding the exponential in the Laplace transform of the delta functions in \eqn{deltacorrelation}: this is possible because the terms proportional to $k_i\cdot k_j$ 
and $(p_3^+k_i^-+p_3^-k_i^+)$ are subleading in the small $k_i$-expansion with respect to the terms in the third line of \eqn{nplus2PSLP}. Terms that involve the perpendicular components, which would contribute at NLP, are odd and vanish upon integration. Subleading terms represented by the ellipsis in the last line of \eqn{nplus2PSLP} will be beyond NLP, as they involve higher powers of the soft momentum $k_i$. 
We can then integrate out $p_3^+$ using the delta function, and integrate over $p_3^-$ and all the $k_i^\pm$. In order to do the $T$- and $\tau$-integrals, we use that these integrals are of the form of an inverse Laplace transform, viz.\
\begin{equation}\label{inverselaplace}
    \int_{-i\infty}^{i\infty}\frac{dT}{2\pi i}e^{T(1-x)}\left(\frac{1}{T}\right)^{\alpha} = \frac{(1-x)^{\alpha-1}}{\Gamma(\alpha)}. 
\end{equation}
Collecting all the terms, we find that the phase space integral is given by
\begin{align}
    \int d\Phi_{n+2}|\M|^2_{\text{LP}, n} &= \frac{2\pi}{\s^2}\frac{\Omega_{d-2}^{n+1}}{2^{n+1}}\frac{2^{n(\frac{d-4}{2})}}{(2\pi)^{(n+1)(d-1)}}2^{\frac{d-2}{2}}\left(\frac{\s}{2}\right)^{1-(n+1)\epsilon}\frac{\Gamma\left(\frac{d-4}{2}\right)^{2n}}{\Gamma(n(d-4))}\int dQ^2\int d\eta \nonumber \\
    &\quad\times \frac{1}{4\cosh^2\eta}(4\cosh^2\eta)^\epsilon f(\alpha_s,\epsilon,\mu^2,\eta)(1-z)^{n(d-4)-1}z^{d-3}\nonumber \\
    &\quad\times \left(1+\frac{(n-1)(d-4)}{4}(1-z)+\frac{d-2}{2}(1-z)+\mathcal{O}\left((1-z)^2\right)\right)\nonumber \\
    &=\int dzd\eta\,\tilde{f}(\alpha_s,\epsilon,\mu^2,\eta,\s,n)\frac{\Gamma(\frac{d-4}{2})^{2n}}{\Gamma(n(d-4))}(1-z)^{n(d-4)-1}\nonumber \\
    &\quad\times \left(1-\epsilon\frac{n-3}{2}(1-z)+\mathcal{O}\left((1-z)^2\right)\right).
\end{align}
We immediately see that the NLP term does not contribute at LL, since it is multiplied by a factor of $\epsilon$.\footnote{We absorbed some irrelevant factors into a new function $\tilde{f}$.} We conclude that there is no NLP effect due to the phase space measure at leading logarithmic accuracy. 

We can now focus on the second term of \eqn{powerexpansionMPS}. 
It was already shown in ref.~\cite{Bahjat-Abbas:2019fqa} how to perform threshold resummation at NLP, which essentially relies on generalising the $K$-factor at next-to-leading order to a general leading power soft function. At NLO we found
\begin{align}\label{eq:diffnloknlp}
    \frac{d\hat{\sigma}^{(1)}_{q\bar{q}}}{dzd\eta}(Q^2,\eta,\epsilon) &= C_FK_\text{NLP}(z,\epsilon)\frac{d\bar{\sigma}^{(0)}_{q\bar{q}}}{dzd\eta}(Q^2,\eta,\epsilon),
\end{align}
up to NLP, where we used the fact that $z\s = Q^2$. We generalise the $K$-factor straightforwardly to a leading power soft function, and we find
\begin{align}
    \frac{d\hat{\sigma}^{(1)}_{q\bar{q}}}{dzd\eta}(Q^2,\eta,\epsilon) &= z\mathcal{S}_\text{LP}(z,\epsilon)\frac{d\bar{\sigma}^{(0)}_{q\bar{q}}}{dzd\eta}(Q^2,\eta,\epsilon).
\end{align}
This can be transformed into Mellin space 
\begin{align}
    \int dz\, z^{N-1}\frac{d\hat{\sigma}^{(1)}_{q\bar{q}}}{dzd\eta}(Q^2,\eta,\epsilon) &= \mathcal{S}_\text{LP}(N+1)\frac{d\bar{\sigma}^{(0)}_{q\bar{q}}}{dzd\eta}(Q^2,\eta,\epsilon).
\end{align}
The LP soft function at order $\mathcal{O}(\alpha_s)$, expanded consistently in $N$-space up to next-to-leading power terms in $1/N$, is given by \cite{Bahjat-Abbas:2019fqa}
\begin{equation}
\mathcal{S}_\text{LP}(N) = \left(\frac{\bar{\mu}^2}{Q^2}\right)^\epsilon\frac{2\alpha_s C_F}{\pi}\left[\frac{1}{\epsilon}\left(\log N-\frac{1}{2N}\right)+\log^2N-\frac{\log N}{N}\right].
\end{equation}
The LP soft function is known to exponentiate, 
which can be proven for instance by 
the replica trick \cite{Gardi:2010rn}. 
Setting $\bar{\mu}^2=Q^2$ we find
\begin{align}
\int dz\, z^{N-1}\frac{d\hat{\sigma}_{q\bar{q}}}{dzd\eta}(Q^2,\eta,\epsilon) &= \frac{\pi\alpha^2C_F}{N_c Q^2}(1+\tanh^2\eta)\exp\left[\frac{2\alpha_sC_F}{\pi}\frac{1}{\epsilon}\log N\right]\nonumber \\
&\quad\times \exp\left[\frac{2\alpha_s C_F}{\pi}\log^2N\right]\left(1+\frac{2\alpha_sC_F}{\pi}\frac{1}{\epsilon}\frac{1}{2N}+\frac{2\alpha_sC_F}{\pi}\frac{\log N}{N}\right).
\end{align}
The poles in $\epsilon$ are now subtracted by the parton distribution functions by defining
\begin{equation}
    q_\text{LL,NLP}(N,Q^2) = q(N,Q^2)\exp\left[\frac{\alpha_sC_F}{\pi}\frac{\log N}{\epsilon}\right]\left(1+\frac{\alpha_sC_F}{\pi}\frac{1}{\epsilon}\frac{1}{2N}\right).
\end{equation}
and similarly for the antiquark $\bar{q}$. We then find 
\begin{align}\label{diphotonresummed}
    \int_0^1dz \, z^{N-1}\frac{d\hat{\sigma}_{q\bar{q}}}{dz d\eta} &= \frac{\pi\alpha^2}{N_c Q^2}(1+\tanh^2\eta)\exp\left[\frac{2\alpha_s C_F}{\pi}\log^2N\right]\left(1+\frac{2\alpha_sC_F}{\pi}\frac{\log N}{N}\right),
\end{align}
which is the resummed cross section for the leading logarithms at both LP and NLP.

\subsection{NLP resummation of the Drell-Yan process}\label{sectionDYresum}

As a last application we consider again the Drell-Yan process. 
In Section \ref{sectionDY} we have discussed the fixed-order calculation of the distribution differential in the total 
rapidity of the final state, i.e., the rapidity of the 
produced off-shell photon. However, in the previous section we have derived the resummed distribution for the diphoton cross  section, differential in the rapidity of one of the final state photons. It is then straightforward to apply the same result to Drell-Yan, and derive the resummed distribution differential in the rapidity of one of the two leptons. We will then conclude by considering again the Drell-Yan rapidity distribution differential in the total rapidity of the final state, obtaining the corresponding resummed result at NLP, with LL accuracy.

\subsubsection{Drell-Yan process for a final state lepton-antilepton pair}

Using the definition of the rapidity from \eqn{rapidity}, one readily finds at leading order in $d=4$  dimensions that the cross section of the Drell-Yan process, producing an lepton-antilepton pair, is given by 
\begin{align}
    \frac{d\hat{\sigma}^{(0)}_{q\bar{q}}}{dzd\eta}(\s,\eta) = \frac{\hat\sigma_0(\s)}{2\cosh^2\eta}\delta(1-z)\,,
\end{align}
where $\hat\sigma_0(\s)$ is given by 
\begin{equation}
    \hat\sigma_0(\s) = \frac{4\pi\alpha^2e_q^2}{3N_c\s}\,.
\end{equation}
The squared matrix element is independent of the rapidity $\eta$, such that the shift in kinematics only induces a shift in the centre of mass energy $\s$, namely $\s\to z\s$. Using the method of shifted kinematics, one finds a NLO cross section up to next to leading power that reads
\begin{align}
    \frac{d\hat{\sigma}^{(1)}_{q\bar{q},\text{ NLP}}}{dzd\eta}(z,Q^2,\eta,\epsilon) &= \frac{\alpha_s}{4\pi}C_F\frac{\hat\sigma_0(z\s)}{2\cosh^2\eta}\Bigg[\frac{1}{\epsilon}\left(8-8\left.\frac{1}{1-z}\right\vert_+\right)\nn\\
    &\quad+16\left.\frac{\log(1-z)}{1-z}\right\vert_+-16\log(1-z)+8\Bigg].
\end{align}
Resummation is obtained as usual now, as discussed in the case of diphoton production. After removing the collinear divergences through mass factorisation, we immediately obtain
\begin{align}
    \int_0^1dz\, z^{N-1}\frac{d\hat\sigma_{q\bar{q}}}{dzd\eta}(Q^2,\eta)=\frac{\hat\sigma_0(Q^2)}{2\cosh^2\eta}\exp\left[\frac{2\alpha_sC_F}{\pi}\log^2N\right]\left(1+\frac{2\alpha_sC_F}{\pi}\frac{\log N}{N}\right).
\end{align}
We were hence again able to obtain a resummed cross section at leading logarithmic accuracy at LP and NLP.

\subsubsection{Drell-Yan process for a final state off-shell photon}\label{subsec:DYgamma}

The derivation of the resummed distribution differential in the total rapidity of the final state easily follows from the results of Section \ref{sectionDY}. First of all, as shown in Appendix \ref{app: DYPS}, integration of the LP squared matrix element against the NLP phase space does not generate LLs at NLP. Therefore, also in this case we can neglect the third term in \eqn{powerexpansionMPS} and focus on the second term. In Section \ref{sec:DYshifted} we found
\begin{align}
\frac{d\hat{\sigma}^{(1)}_{q\bar q,\,\text{NLP}}}{dzdy}(z,y,\eps,\bar\mu^2) =C_F\hat\sigma_0 \left(\frac{\delta(y)+\delta(1-y)}{2}+\mathcal{O}(\eps)\right)K_{\rm{NLP}}(z,\eps),
\end{align}
This result is already in the right form. We can proceed as before and generalise the $K$-factor to a leading power soft function, and then use the fact that this soft function can be exponentiated. The resummed result then yields, after removing the collinear divergences through mass factorisation,
\begin{equation}\label{resumRapDY}
     \int dz\,z^{N-1} \frac{d\hat\sigma_{q\bar{q}}}{dzdy}(z,y)=\hat\sigma_0\left(\frac{\delta(y)+\delta(1-y)}{2}\right)\exp\left[\frac{2\alpha_sC_F}{\pi}\log^2N\right]\left(1+\frac{2\alpha_sC_F}{\pi}\frac{\log N}{N}\right).
\end{equation}
Note that this resummed cross section has the same structure as for the other processes, even though we are considering a different rapidity variable in this case. Upon converting \eqn{resumRapDY} back to $z$-space, and expanding the result in powers of $\as$, it is possible to check that the leading logarithmic contribution up to NLP does agree with the LLs appearing in \eqns{PartonicRapidityNNLO-LP-Expanded}{PartonicRapidityNNLO-NLP-Expanded}, after PDF renormalisation has been taken into account.

\section{Conclusions}\label{conclusions}
In this paper we have extended the treatment of next-to-leading power corrections from single differential cross sections to double differential cross sections,
in particular including rapidity dependence for a number of observables. We examined the NLP structure for fixed order results and derived resummed cross sections, 
to leading logarithmic accuracy at NLP.

In Section~\ref{sectionDY} we gave explicit NNLO partonic cross section expressions for the Drell-Yan process, with dependence on both the threshold and rapidity variable up to NLP accuracy. We showed that at LL accuracy at NLP the rapidity dependence indeed factorises from the dependence on the threshold variable, as was already known for LP contributions.  Beyond LL accuracy the $z$- and $y$-dependence is more entangled. At NLO, we achieved the same result by using the method of shifted kinematics \cite{DelDuca:2017twk}. 
We then examined the case of diphoton production differential in the diphoton invariant mass and single-photon rapidity in Section \ref{diphoton}. We constructed the NLO cross section up to NLP terms by generalising a method involving momentum shifts for single-differential cross sections. 
We also presented analytical results of the NLO cross section up to N$^3$LP.
Generalising the analysis in \refr{Bahjat-Abbas:2019fqa}, we then derived in Section \ref{resum} a result for the resummed cross section for diphoton production for the leading logarithms at NLP, differential in both the threshold variable and the rapidity. Using the same methods, we also resummed the NLP leading logarithms for the Drell-Yan process, both for the lepton pair and single lepton inclusive case. 

Extension beyond leading logarithms would clearly be interesting, though our results show that this is very challenging due to the non-factorising structure of rapidity and threshold logarithms for that case. Extension to different differential observables would likewise be desirable, as would the inclusion of
off-diagonal channels \cite{Beneke:2020ibj,vanBeekveld:2021mxn}. Our present results are, we believe, useful to further the insight into and use of next-to-leading power corrections for phenomenological studies.

\acknowledgments

L.V. is supported by Fellini - Fellowship for Innovation 
at INFN, funded by the European Union's Horizon 2020 
research programme under the Marie Sk\l{}odowska-Curie 
Cofund Action, grant agreement no. 754496 and by 
Compagnia di San Paolo through grant 
TORP\_S1921\_EX-POST\_21\_01. The research of G.W. was supported in part by the International Postdoctoral Exchange Fellowship Program from China Postdoctoral Council under Grant No. PC2021066.

\appendix

\section{Phase space integrals for fixed-order Drell-Yan}
\label{PhaseSpace-DY}
In this appendix we compute the relevant phase space integrals for doubly differential distributions in both the invariant mass and the rapidity that we use for the Drell-Yan process. 
For the three-body phase space integrals at NNLO, it is hard to obtain complete results at NLP with this approach. The result in this appendix for the three-body phase space integral can only be used to calculate the contributions proportional to $C_F^2$ up to LL at NLP. However, this approach can be used in other processes, e.g.\ pure quantum electrodynamics corrections. Note that we have used a more powerful method with a different parametrisation of external momenta to calculate the three-body phase space integrals in Section~\ref{sec:NNLO1r1v}. 
 
Before we give the results, let us first introduce 
a pair of light-like vectors
\begin{align} 
n^{\mu} = \frac{1}{\sqrt{2}}(1,0,0,1)\,, 
\quad \bar{n}^{\mu} = \frac{1}{\sqrt{2}}(1,0,0,-1) \,.
\end{align}
For any momenta $p$ and $q$, we have 
\begin{align}
\begin{split}\label{pppmdef}
p^+ = \bar{n}\cdot p=\frac{1}{\sqrt{2}}&(p^0+p^3)\,, 
\quad p^- = n\cdot p=\frac{1}{\sqrt{2}}(p^0-p^3) \,, \\
p\cdot q &= p^+q^-+p^-q^++\bm{p}_{\perp}\cdot \bm{q}_{\perp} \,.
\end{split}
\end{align}
In the centre of mass frame of 
initial states, we then have 
\begin{align}\label{p1p2cm}
p_a^{\mu} = \sqrt{\frac{\hat{s}}{2}}n^{\mu}\,, 
\quad p_b^{\mu} =  \sqrt{\frac{\hat{s}}{2}}\bar{n}^{\mu} \,.
\end{align}
In this coordinate system the integration 
measure reads  
\begin{equation}
  d^dp = dp^+dp^-d^{d-2}\bm{p}_\perp.
\end{equation} 
\subsection{One-body phase space integral}
\label{1PhaseSpace-DY}
The one-body phase space integral is trivial, we have
\begin{align}\label{eq:phaseDY1invmy}
    \int d \Phi_{\gamma^*} &= \int \frac{d^dq}{(2\pi)^{d-1}} \, (2\pi)^d\delta^{(d)}(p_a + p_b - q) \, \delta(q^2 - Q^2)\, \delta\bigg[y - \frac{p_a \cdot q - z \, p_b \cdot q}{(1-z)
(p_a \cdot q + p_b \cdot q)} \bigg] \nn \\
          &= \frac{2\pi}{Q^2} \, \delta\left(1-z\right) \delta\left(y-\frac{1}{2}\right)\,.
\end{align}

\subsection{Two-body phase space integral}
\label{2PhaseSpace-DY}

We define the two-body phase space integral in both invariant
mass and rapidity $y$ as
\begin{align}\label{eq:phaseDY2invmy}
    \int d \Phi_{\gamma^*g} &= \bigg(\frac{\bar \mu^2 e^{\gamma_E}}{4\pi}\bigg)^\eps\int \frac{d^dq}{(2\pi)^{d-1}} \frac{d^d k}{(2\pi)^{d-1}}
\,(2\pi)^d\, \delta^{(d)}(p_a + p_b - q - k) \, \delta_+(k^2) \, \delta(q^2 - Q^2) \nonumber \\
&\hspace{4.0cm}\times \, \delta\bigg[y - \frac{p_a \cdot q - z \, p_b \cdot q}{(1-z)
(p_a \cdot q + p_b \cdot q)} \bigg]\,,
\end{align}
and $\delta_+\left(k^2\right) = 
\delta\left(k^2\right) \theta\left(k^{(0)}\right)$. We first use the momentum-conserving delta function to integrate out $q$. We obtain
\begin{align}
    \delta(q^2 - Q^2) 
    &= \frac{1}{\hat{s}}\delta\left[(1-z)-\frac{2}{\hat{s}}(p_a+p_b)\cdot k \right]\,, \label{eq:deltafunc21} \\
    \delta\bigg[y - \frac{p_a \cdot q - z \, p_b \cdot q}{(1-z) (p_a \cdot q + p_b \cdot q)} \bigg] 
    &= (1-z) \, \delta\left[(1-z)(1-y)-\frac{2}{\hat{s}}p_a\cdot k\right]\,,\label{eq:deltafunc22}
\end{align}
where we have used \eqn{eq:deltafunc21} to remove $p_b\cdot k$ in \eqn{eq:deltafunc22}.
When we consider NLO real corrections and NNLO real-virtual corrections up to NLP, the squared amplitudes depends on $k$ in the following general form 
\begin{align}\label{eq:squaredamp1}
   \frac{\hat{s}^{\alpha+\beta}}{(2p_a\cdot k)^{\alpha}(2p_b\cdot k)^{\beta}} = \left(\sqrt{\frac{\hat{s}}{2}}\right)^{\alpha+\beta}\frac{1}{(k^-)^{\alpha}(k^+)^{\beta}} \,.
\end{align}
After inserting \eqn{eq:deltafunc21} and \eqn{eq:deltafunc22} into \eqn{eq:phaseDY2invmy} and combining it with \eqn{eq:squaredamp1}, we can perform the integration straightforwardly, such that we have 
\begin{align}\label{eq:phaseDY2invmyr}
    \int d \Phi_{\gamma^*g}\frac{\hat{s}^{\alpha+\beta}}{(2p_a\cdot k)^{\alpha}(2p_b\cdot k)^{\beta}} &=\frac{1}{8\pi}\left(\frac{\bar\mu^2}{\s}\right)^\epsilon\frac{e^{\eps\gamma_E}(1-y)^{-\epsilon-\alpha}y^{-\epsilon-\beta}}{\Gamma(1-\epsilon)}\left(1-z\right)^{1-2\epsilon-\alpha-\beta} \,.
\end{align}

\subsection{Three-body phase space integral}
\label{3PhaseSpace-DY}
We give an alternative approach to calculate the three-body phase space integral, instead of the novel method of parametrisation of external momenta used in Section~\ref{sec:NNLO1r1v}. The three-body phase space integral for distributions differential in both the invariant mass and rapidity $y$ is defined as
\begin{align}\label{eq:phaseDY3invmy}
    \int d \Phi_{\gamma^*(gg+q\bar q)} &= \bigg(\frac{\bar \mu^2 e^{\gamma_E}}{4\pi}\bigg)^{2\eps}\int \frac{d^dq}{(2\pi)^{d-1}} 
\frac{d^d k_1}{(2\pi)^{d-1}} \frac{d^d k_2}{(2\pi)^{d-1}}
\,(2\pi)^d\, \delta^{(d)}(p_a + p_b - q - k_1 - k_2) \nonumber \\
&\hspace{2.0cm}\times  \,\delta_+(k_1^2)\, \delta_+(k_2^2)  \, \delta(q^2 - Q^2)\, \delta\bigg[y - \frac{p_a \cdot q - z \, p_b \cdot q}{(1-z)
(p_a \cdot q + p_b \cdot q)} \bigg]\,.
\end{align}
We integrate out $q$ by using the momentum-conserving delta function, and obtain
\begin{align}
\delta(q^2 - Q^2) &= \frac{1}{\hat{s}}\delta\left[(1-z)-\frac{2}{\hat{s}}(p_a+p_b)\cdot (k_1+k_2)+\frac{2}{\hat{s}}k_1\cdot k_2 \right]\,, \label{eq:deltafunc31} \\
    \delta\bigg[y - \frac{p_a \cdot q - z \, p_b \cdot q}{(1-z) (p_a \cdot q + p_b \cdot q)} \bigg] 
    &= (1-z)\left[(1+z)-\frac{2}{\hat{s}}k_1\cdot k_2\right] \nonumber \\
    &\hspace{-4.0cm}\times\delta\left[(1+z)(1-z)(1-y)-\frac{2}{\hat{s}}(1+z)p_a\cdot(k_1+k_2)+\frac{2}{\hat{s}}(y+z-yz)k_1\cdot k_2\right]\,,\label{eq:deltafunc32}
\end{align}
where we have used \eqn{eq:deltafunc31} to remove $p_b\cdot k_1$ in \eqn{eq:deltafunc32}. Note that $k_1\cdot k_2$ is a power suppressed term in \eqn{eq:deltafunc31}. We apply \eqn{deltatransform} to \eqn{eq:deltafunc31} and expand in powers of $T$, which yields 
\begin{align}\label{eq:deltafuncexp1}
    \delta(q^2 - Q^2) &= \frac{1}{\hat{s}}\int \frac{dT}{2\pi i}\exp\left\{T\left[(1-z)-\frac{2}{\hat{s}}(p_a+p_b)\cdot (k_1+k_2)\right] \right\}\nonumber \\
    &\times\left[1+T\frac{2}{\hat{s}}k_1\cdot k_2+\mathcal{O}\left(T^2\right)\right].
\end{align}
The term $k_1\cdot k_2$ is also power suppressed in \eqn{eq:deltafunc32}. We apply \eqn{deltatransform} to \eqn{eq:deltafunc32} and expand in powers of $\tau$, which yields 
\begin{align}\label{eq:deltafuncexp2}
    \delta\bigg[y - & \frac{p_a \cdot q - z \, p_b \cdot q}{(1-z) (p_a \cdot q + p_b \cdot q)} \bigg] = (1-z)\left[(1+z)-\frac{2}{\hat{s}}k_1\cdot k_2\right] \nonumber \\
    &\times\int\frac{d\tau}{2\pi i}\exp\left\{\tau\left[(1+z)(1-z)(1-y)-\frac{2}{\hat{s}}(1+z)p_a\cdot(k_1+k_2)\right] \right\}\nonumber \\
    &\times\left[1+\tau\frac{2}{\hat{s}}(y+z-yz)k_1\cdot k_2+\mathcal{O}\left(\tau^2\right)\right]\,,
\end{align}
We now substitute \eqn{eq:deltafuncexp1} and \eqn{eq:deltafuncexp2} into \eqn{eq:phaseDY3invmy} and find that the phase space measure depends on $k_1\cdot k_2=k_1^-k_2^+ + k_1^+k_2^-+k_{1\perp}\cdot k_{2\perp}$ linearly at NLP. 

If we only consider the contributions proportional to $C_F^2$ up to NLP LL in \eqn{PartonicRapidityDefB-2realB}, the squared amplitudes do not involve the term $k_1\cdot k_2$. As a result, the $k_{1\perp}\cdot k_{2\perp}$ term does not contribute by symmetry. The remaining integrations over $k_1^+,k_1^-,k_2^+$ and $k_2^-$, as well as the inverse Laplace transformations over $T$ and $\tau$, are straightforward. Up to LL, at LP, we have
\begin{align}
    \int d \Phi_{\gamma^*(gg+q\bar q)} \, \frac{\hat{s}^2}{(2p_a\cdot k_1)(2p_a\cdot k_2)(2p_b\cdot k_1)(2p_b\cdot k_2)}&= \frac{1}{32\pi^3 \s}\left(\frac{\bar\mu^2}{\s}\right)^{2\eps}\frac{e^{2\eps\gamma_E}\Gamma^2(-\epsilon)}{\Gamma^2(1-2\epsilon)} \nonumber \\
    &\times y^{-1-2\epsilon}(1-y)^{-1-2\epsilon}(1-z)^{-1-4\epsilon}\,,
\end{align}
and at NLP we have
\begin{align}
    \int d \Phi_{\gamma^*(gg+q\bar q)} \, \frac{\hat{s}(2p_a\cdot k_1+2p_a\cdot k_2+2p_b\cdot k_1+2p_b\cdot k_2)}{(2p_a\cdot k_1)(2p_a\cdot k_2)(2p_b\cdot k_1)(2p_b\cdot k_2)}&= \frac{1}{32\pi^3 \s}\left(\frac{\bar\mu^2}{\s}\right)^{2\eps}\frac{e^{2\eps\gamma_E}\Gamma^2(-\epsilon)}{\Gamma^2(1-2\epsilon)} \nonumber \\
    &\times y^{-1-2\epsilon}(1-y)^{-1-2\epsilon}(1-z)^{-4\epsilon}\,.
\end{align}
We have checked that the above results are consistent with the corresponding ones given by using the method in Section~\ref{sec:NNLO1r1v}.

\section{Phase space integral for fixed-order diphoton production}
\label{PSintegral}

In Section \ref{diphoton} we calculate the
diphoton rapidity distribution at LO and NLO,
for which we need respectively the two- and 
three-particle phase space. 
\subsection{Two-body phase space integral}\label{PSintegral2}
The two-body 
phase space is defined as follows:
\begin{align}\label{R2Def}
\int dR_2 &= \int dQ^2 \, d\eta \, d\Phi_2(p_1+p_2;p_3,p_4) 
\, \delta\left[Q^2-(p_3+p_4)^2\right] \, \delta\left[\eta
-\frac12\log\left(\frac{p_3^+}{p_3^-}\right)\right],
\end{align}
where 
\begin{align} \nn
\int d\Phi_2(p_1+p_2;p_3,p_4) &= 
\bigg(\frac{\bar \mu^2 e^{\gamma_E}}{4\pi}\bigg)^\eps 
\int \frac{d^dp_3}{(2\pi)^{d-1}}\frac{d^dp_4}{(2\pi)^{d-1}} \\ 
&\hspace{2.0cm} \times \, \delta_+\left(p_3^2\right) 
\, \delta_+\left(p_4^2\right) 
\, (2\pi)^d\delta^{(d)}(p_1+p_2-p_3-p_4)\,,
\end{align}
and the decomposition into light-cone momentum 
components $p_{i\pm}$ follows the definition in 
\eqn{pppmdef}. The two-body phase space is easy 
to evaluate, and one gets 
\begin{align}\label{R2Res}
\int dR_2 &= \frac{e^{\eps \gamma_E}}{4\pi\Gamma(1-\epsilon)}
\left(\frac{\bar \mu^2}{\s}\right)^\epsilon 
\int dz \, d\eta \, \left(2\cosh\eta\right)^{-2+2\epsilon}
\delta(1-z).
\end{align}
\subsection{Three-body phase space integral}\label{PSintegral3}
The three-body phase space reads 
\begin{align}\label{R3Def} 
\int dR_3 &= \int dQ^2\, d\eta \, d\Phi_3(p_1+p_2; p_3,p_4,k) 
\, \delta\left[Q^2-(p_3+p_4)^2\right] \, \delta\left[\eta
-\frac12\log\left(\frac{p_3^+}{p_3^-}\right)\right],
\end{align}
where 
\begin{align} \nn
\int d\Phi_3(p_1+p_2; p_3,p_4,k) &= 
\bigg(\frac{\bar \mu^2 e^{\gamma_E}}{4\pi}\bigg)^{2\eps} 
\int \frac{d^dp_3}{(2\pi)^{d-1}} \frac{d^dp_4}{(2\pi)^{d-1}}
\frac{d^dk}{(2\pi)^{d-1}}\,\delta_+(p_3^2) \, \delta_+(p_4^2) 
\, \delta_+(k^2) \\ 
&\hspace{4.0cm}\times \, 
(2\pi)^d\delta^{(d)}(p_1+p_2-p_3-p_4-k)\,.
\end{align}
We start by performing the integration over 
$p_4$ with the momentum-conserving delta function,
obtaining
\begin{align}
\int dR_3 &= (2\pi) \bigg(\frac{\bar \mu^2 e^{\gamma_E}}{4\pi}\bigg)^{2\eps} 
\int dQ^2 \,d\eta\, \frac{dp_3^+dp_3^-d^{d-2}\bm{p}_{3,\perp}}{(2\pi)^{d-1}} 
\frac{dk^+dk^-d^{d-2}\bm{k}_\perp}{(2\pi)^{d-1}} \nonumber \\
&\hspace{0.0cm} \times 
\,\delta_+\left(2p_3^+p_3^--\bm{p}_{3,\perp}^2\right) 
\, \delta_+\left(2k^+k^--\bm{k}_\perp^2\right) 
\, \delta\left[\hat{s}-2p_3\cdot(p_1+p_2)-
2k\cdot(p_1+p_2)+2p_3\cdot k\right] \nonumber \\
&\hspace{1.0cm} \times 
\, \delta\left[Q^2-\hat{s}+2k\cdot(p_1+p_2)\right] \,
\delta\left[\eta-\frac12\log\left(\frac{p_3^+}{p_3^-}\right)\right]
\nonumber \\
&= \bigg(\frac{\bar \mu^2 e^{\gamma_E}}{4\pi}\bigg)^{2\eps}
\frac{\Omega_{d-2}\Omega_{d-3}}{2(2\pi)^{2d-3}\hat{s}}
\int dz d\eta \, dp_3^+dp_3^-dp^2_{3,T}\,
dk^+dk^-dk^2_T\, d\cos\alpha\, p_3^+ 
\left(p^2_{3,T}\right)^{\frac{d-4}{2}}
\left(k^2_T\right)^{\frac{d-4}{2}} \nonumber \\
&\quad\times 
\left(1-\cos^2\alpha\right)^{\frac{d-5}{2}} \,
\delta_+\left(2p_3^+p_3^--p_{3,T}^2\right) \, 
\delta_+\left(2k^+k^--k_T^2\right) \, 
\delta\left[(1-z)-\sqrt{\frac{2}{\hat{s}}}(k^++k^-)\right] \nonumber \\
&\quad\times 
\delta\left(p_3^+-p_3^-e^{2\eta}\right) \, 
\delta\left[z-\sqrt{\frac{2}{\hat{s}}}(p_3^++p_3^-)
+\frac{2}{\hat{s}}(p_3^+k^-+p_3^-k^
+-p_{3,T}k_T\cos\alpha)\right] \,,
\end{align}
where $\Omega_{2d}=2\pi^{d}/\Gamma(d)$. Applying \eqn{deltatransform} to the last delta function, we can integrate out the angle $\alpha$ by using
\begin{align}\label{eq:integrateangle}
&\int_{-1}^1d\cos\alpha \left(1-\cos^2\alpha\right)^{\frac{-1-2\epsilon}{2}} \exp\left(-\frac{2}{\hat{s}}Tp_{3,T}k_T\cos\alpha\right) \nonumber \\
&=\frac{4^\epsilon\pi\Gamma(1-2\epsilon)}{\Gamma^2(1-\epsilon)}{}_0F_1\left(1-\epsilon,\frac{T^2p^2_{3,T}k^2_T}{\hat{s}^2}\right) \nonumber \\
&=\frac{4^\epsilon\pi\Gamma(1-2\epsilon)}{\Gamma(1-\epsilon)}\sum_{j=0}^{\infty}\frac{1}{\Gamma(1-\epsilon+j)}\left(\frac{T^2p^2_{3,T}k^2_T}{\hat{s}^2}\right)^j\frac{1}{j!}\,,
\end{align}
where we have used the series representation of the hypergeometric function ${}_0F_1\left(a,z\right)$ in the last line. Now, we have
\begin{align}
    \int dR_3 &= \frac{\bar\mu^{4\epsilon}e^{2\eps \gamma_E}}{(4\pi)^2\pi\hat{s}\Gamma(1-\epsilon)}\int dz d\eta\sum_{j=0}^{\infty}\int\frac{dT}{2\pi i}\frac{T^{2j}}{\Gamma(1-\epsilon+j)\hat{s}^{2j}j!}\int dp_3^+dp_3^-dp^2_{3,T}dk^+dk^-dk^2_T\, p_3^+\nonumber \\
    &\quad\times \left(p^2_{3,T} \, k^2_T\right)^{j-\epsilon}\delta_+\left(2p_3^+p_3^--p_{3,T}^2\right) \, \delta_+\left(2k^+k^--k_T^2\right) \, \delta\left[(1-z)-\sqrt{\frac{2}{\hat{s}}}(k^++k^-)\right] \nonumber \\
    &\quad\times \delta\left(p_3^+-p_3^-e^{2\eta}\right) \, \exp\left\{T\left[z-\sqrt{\frac{2}{\hat{s}}}(p_3^++p_3^-)+\frac{2}{\hat{s}}(p_3^+k^-+p_3^-k^+)\right]\right\} \,.
\end{align}
Combining the above $\int dR_3$ with the squared amplitudes in \eqn{eq:ampnlogeneral}, we use the four delta functions to integrate over $p_{3}^+,\,p_{3,T},\,k^+$ and $k_T$. The integration over $p_{3}^-$ and $k^-$ and the inverse Laplace transformation would be straightforward. Finally, we have
\begin{align}\label{eq:phaseintnlo}
    \int dR_3(&k^+)^{\alpha}(k^-)^{\beta}(p_3^+)^{\gamma}(p_3^-)^{\kappa} = \bar\mu^{4\epsilon}\int dz d\eta\frac{2^{-6-\frac{1}{2}(\alpha+\beta+\gamma+\kappa)}z^{1+\gamma+\kappa-2\epsilon}}{\pi^{3}\hat{s}^{-1+2\epsilon-\frac{1}{2}(\alpha+\beta+\gamma+\kappa)}} \nonumber \\
    &\times\sum_{j=0}^{\infty}\sum_{n=0}^{\infty}(1-z)^{1+\alpha+\beta-2\epsilon+2j+n}(e^{2\eta})^{1+\gamma-\epsilon+j}(e^{2\eta}-1)^{n}(e^{2\eta}+z)^{-2-\gamma-\kappa+2\epsilon-2j-n}\nonumber \\
    &\times \frac{e^{2\eps\gamma_E}\Gamma(1+\alpha-\epsilon+j)\Gamma(1+\beta-\epsilon+j+n)\Gamma(2+\gamma+\kappa-2\epsilon+2j+n)}{j!n!\Gamma(1-\epsilon)\Gamma(2+\gamma+\kappa-2\epsilon)\Gamma(1-\epsilon+j)\Gamma(2+\alpha+\beta-2\epsilon+2j+n)} \,.
\end{align}
Note that this result is valid up to arbitrary power of $1-z$. Only the $j=0, n=0$ and $j=0,n=1$ terms are necessary at NLP because $\alpha+\beta\geq-2$ in the NLO squared amplitudes of \eqn{eq:ampnlogeneral}.

\section{NLP phase space contribution for Drell-Yan}\label{app: DYPS}
In this appendix we discuss the NLP contribution of the phase space for the Drell-Yan production of a virtual photon, i.e.\ $q(p_a)\bar q(p_b)\to \gamma^*(q)$. This result is needed in Section \ref{subsec:DYgamma}. The $(n+1)$-particle phase space integral is given by 
\begin{align}
    \int d\Phi_{n+1} &= \int \frac{d^dq}{(2\pi)^{d-1}}\delta_+(q^2-Q^2)\int \prod_{i=1}^n\left[\frac{d^dk_i}{(2\pi)^{d-1}}\delta_+(k_i^2)\right] \nonumber \\
    &\quad\times \int d\eta\, \delta\left(\eta-\frac12\log\left(\frac{q^+}{q^-}\right)\right)(2\pi)^d\delta^{(d)}\left(p_a+p_b-q-\sum_{i=1}^nk_i\right) \nonumber \\
    &= \frac{2\pi}{\s}\int \prod_{i=1}^n\left[\frac{d^dk_i}{(2\pi)^{d-1}}\delta_+(k_i^2)\right]\delta\left(1-z-\sqrt{\frac{2}{\s}}\sum_i(k_i^++k_i^-)+\frac{2}{\s}\sum_{i<j}k_i\cdot k_j\right)\nonumber \\
    &\quad\times \int d\eta \, \delta\left(\eta-\frac12\log\left(\frac{\sqrt{\s/2}-\sum_i k_i^+}{\sqrt{\s/2}-\sum_i k_i^-}\right)\right),\label{DYNLPPSint}
\end{align}
where we used the rapidity variable $\eta$ for the moment, and that $(p_a+p_b)^+=(p_a+p_b)^- = \sqrt{\s/2}$. We can manipulate the delta function of the rapidity $\eta$ as follows:
\begin{align}
    \delta\left(\eta-\frac12\log\left(\frac{\sqrt{\s/2}-\sum_i k_i^+}{\sqrt{\s/2}-\sum_i k_i^-}\right)\right) &= 2\left(\sqrt{\frac{\s}{2}}-\sum_i k_i^+\right)\nonumber\\
    &\quad\times\delta\left(\sum_i k_i^+-\sqrt{\frac{\s}{2}}-e^{2\eta}\left(\sum_i k_i^--\sqrt{\frac{\s}{2}}\right)\right),
\end{align}
likewise as was done in the analogous derivation for diphoton production, where we had $p_3^+$ and $p_3^-$ as inputs of the logarithm. 
As was done in Section \ref{subsec:DYgamma}, we introduce the rapidity variable $y$. It is related to $\eta$ via
\begin{equation}\label{ydefeta}
    y = \frac{1-e^{2\eta}z}{(e^{2\eta}+1)(1-z)}.
\end{equation}
Using the expression for the squared matrix element at leading power from \eqn{LPsqmatrixn}, we write
\begin{align}
    \int d\Phi_{n+1}\left|\M\right|_{\text{LP},n}^2&=\frac{4\pi}{\s}\frac{\Omega_{d-2}^n}{2^n}\int\left[\prod_{i=1}^n\frac{dk_i^+dk_i^-}{(2\pi)^{d-1}}(2k_i^+k_i^-)^{-\epsilon} \frac{1}{k_i^+k_i^-}\right]\nonumber \\
    &\hspace{-2cm}\times \int dy \left(-\frac{2(1-z)}{1+z}\right)\frac{dT}{2\pi i} \frac{d\tau}{2\pi i}\, \left(\sqrt{\frac{\s}{2}}-\sum_{i=1}^n k_i^+\right)e^{T(1-z)}e^{\tau\sqrt{\s/2}(e^{2\eta(y)}-1)}f(\alpha_s,\epsilon,y,\bar\mu^2)\nonumber \\
    &\hspace{-2cm}\times \prod_{i=1}^n e^{k_i^+(-\sqrt{2/\s}T+\tau)}e^{k_i^-(-\sqrt{2/\s}T-e^{2\eta(y)}\tau)}\left(1+\frac{2T}{\s}\sum_{i<j}(k_i^+k_j^-+k_i^-k_j^+)+\mathcal{O}(T^2)\right).
\end{align}
For the moment, we will compactly write $\eta(y)$ instead of the expression in \eqn{ydefeta}. Using the expression for the squared matrix element at leading power from \eqn{LPsqmatrixn}, we write
\begin{align}
    \int d\Phi_{n+1}\left|\M\right|_{\text{LP},n}^2&=\frac{4\pi}{\s}\frac{\Omega_{d-2}^n}{2^n}\int\left[\prod_{i=1}^n\frac{dk_i^+dk_i^-}{(2\pi)^{d-1}}(2k_i^+k_i^-)^{-\epsilon} \frac{1}{k_i^+k_i^-}\right]\nonumber \\
    &\hspace{-1cm}\times \int dy \left(-\frac{2(1-z)}{1+z}\right)\frac{dT}{2\pi i} \frac{d\tau}{2\pi i}\, \left(\sqrt{\frac{\s}{2}}-\sum_{i=1}^n k_i^+\right)e^{T(1-z)}e^{\tau\sqrt{\s/2}(e^{2\eta(y)}-1)}f(\alpha_s,\epsilon,v,\mu^2)\nonumber \\
    &\hspace{-1cm}\times \prod_{i=1}^n e^{k_i^+(-\sqrt{2/\s}T+\tau)}e^{k_i^-(-\sqrt{2/\s}T-e^{2\eta}\tau)}\left(1+\frac{2T}{\s}\sum_{i<j}(k_i^+k_j^-+k_i^-k_j^+)+\mathcal{O}(T^2)\right).
\end{align}
The first contribution we will evaluate is given by
\begin{align}
  I_1&\equiv\int\left[\prod_{i=1}^n dk_i^+dk_i^-(k_i^+k_i^-)^{\frac{d-6}{2}}\right] \int dy \left(-\frac{2(1-z)}{1+z}\right)\frac{dT}{2\pi i} \frac{d\tau}{2\pi i}\, \sqrt{\frac{\s}{2}}e^{T(1-z)}e^{\tau\sqrt{\s/2}(e^{2\eta(y)}-1)}\nonumber \\
    &\quad\times \prod_{i=1}^n e^{k_i^+(-\sqrt{2/\s}T+\tau)}e^{k_i^-(-\sqrt{2/\s}T-e^{2\eta(y)}\tau)}.
\end{align}
We then integrate over all the momenta $\{k_i\}_{i=1}^n$, which all become Gamma functions, i.e.\ 
\begin{align}
    I_1 &= \left(-\frac{2(1-z)}{1+z}\right)\sqrt{\frac{\s}{2}}\int dy \frac{dT}{2\pi i} \frac{d\tau}{2\pi i}\, e^{T(1-z)}e^{\tau\sqrt{\s/2}(e^{2\eta(y)}-1)} \nn\\
    &\quad\times \left(\frac{1}{\sqrt{\frac{2}{\s
    }}T-\tau}\right)^{n\left(\frac{d-4}{2}\right)}\left(\frac{1}{\sqrt{\frac{2}{\s
    }}T+e^{2\eta(y)}\tau}\right)^{n\left(\frac{d-4}{2}\right)}\Gamma\left(\frac{d-4}{2}\right)^{2n}.
\end{align}
Upon subsequent coordinate transformations $$\tilde{T}=\sqrt{\frac{2}{\s}}T-\tau,\qquad \tilde{\tau} = \tilde{T}+\tau(1+e^{2\eta(y)}),$$ one finds the following expression for $I_1$, namely
\begin{equation}
    I_1 = -2^{-1+n\eps}\s^{-n\eps}\frac{\Gamma^{2n}(-\eps)}{\Gamma^2(-n\eps)}\int dy\, y^{-1-n\eps}(1-y)^{-1-n\eps}\bigg(1+y(1-z)+\mathcal{O}((1-z)^2)\bigg),
\end{equation}
where we now explicitly used \eqn{ydefeta} for the rapidity $y$.
Using similar methods, the second contribution that we calculate reads 
\begin{align}
    I_2 &\equiv  \int\left[\prod_{i=1}^n dk_i^+dk_i^-(k_i^+k_i^-)^{\frac{d-6}{2}}\right] \int dy \left(-\frac{2(1-z)}{1+z}\right)\frac{dT}{2\pi i} \frac{d\tau}{2\pi i}\, \left(-\sum_{i=1}^nk_i^+\right)\nonumber \\
    &\quad\times e^{T(1-z)}e^{\tau\sqrt{\s/2}(e^{2\eta(y)}-1)} \prod_{i=1}^n e^{k_i^+(-\sqrt{2/\s}T+\tau)}e^{k_i^-(-\sqrt{2/\s}T-e^{2\eta(y)}\tau)} \nonumber \\
    &= \left(\frac{2(1-z)}{1+z}\right)\int dy\frac{dT}{2\pi i}\frac{d\tau}{2\pi i}e^{T(1-z)}e^{\tau\sqrt{\s/2}(e^{2\eta(y)}-1)}\nonumber \\
    &\quad\times n\left[\int dk^+dk^-(k^+)^{\frac{d-4}{2}}(k^-)^{\frac{d-6}{2}}e^{k^+(-\sqrt{2/\s}T+\tau)}e^{k^-(-\sqrt{2/\s}T-e^{2\eta(y)}\tau)}\right]\nonumber \\
    &\quad\times \left[\int dk^+dk^-(k^+k^-)^{\frac{d-6}{2}}e^{k^+(-\sqrt{2/\s}T+\tau)}e^{k^-(-\sqrt{2/\s}T-e^{2\eta(y)}\tau)}\right]^{n-1}\nonumber \\
    &=2^{-1+n\eps}\s^{-n\eps}\frac{\Gamma^{2n}\left(-\epsilon\right)}{\Gamma^2\left(-n\epsilon\right)}\int dy\,y^{-1-n\eps}(1-y)^{-1-n\eps}\left(y(1-z)^{-2n\epsilon}+\mathcal{O}((1-z))\right),
\end{align}
where we rewrote the expression in such a way that it has a recognisable common prefactor with $I_1$. We now turn our attention to the third integral, which reads
\begin{align}
    I_3&\equiv\int\left[\prod_{i=1}^n dk_i^+dk_i^-(k_i^+k_i^-)^{\frac{d-6}{2}}\right] \int dy\left(-\frac{2(1-z)}{1+z}\right)\frac{dT}{2\pi i} \frac{d\tau}{2\pi i}\, \sqrt{\frac{\s}{2}}e^{T(1-z)}e^{\tau\sqrt{\s/2}(e^{2\eta(y)}-1)}\nonumber \\
    &\quad\times \prod_{i=1}^n e^{k_i^+(-\sqrt{2/\s}T+\tau)}e^{k_i^-(-\sqrt{2/\s}T-e^{2\eta(y)}\tau)}\left(\frac{2T}{\s}\sum_{i<j}(k_i^+k_j^-+k_i^-k_j^+)\right)\nonumber \\
    &= \left(-\frac{2(1-z)}{1+z}\right)\sqrt{\frac{2}{\s}}\int dy\frac{dT}{2\pi i}\frac{d\tau}{2\pi i}e^{T(1-z)}e^{\tau\sqrt{\s/2}(e^{2\eta(y)}-1)}T\nonumber \\
    &\quad\times n(n-1)\left[\int dk^+dk^-(k^+)^{\frac{d-4}{2}}(k^-)^{\frac{d-6}{2}}e^{k^+(-\sqrt{2/\s}T+\tau)}e^{k^-(-\sqrt{2/\s}T-e^{2\eta(y)}\tau)}\right]\nonumber \\
    &\quad\times \left[\int dk^+dk^-(k^+)^{\frac{d-6}{2}}(k^-)^{\frac{d-4}{2}}e^{k^+(-\sqrt{2/\s}T+\tau)}e^{k^-(-\sqrt{2/\s}T-e^{2\eta(y)}\tau)}\right]\nonumber \\
    &\quad\times \left[\int dk^+dk^-(k^+k^-)^{\frac{d-6}{2}}e^{k^+(-\sqrt{2/\s}T+\tau)}e^{k^-(-\sqrt{2/\s}T-e^{2\eta(y)}\tau)}\right]^{n-2}\nonumber \\
    &= 2^{-1+n\eps}\s^{-n\eps}\frac{\Gamma^{2n}\left(-\epsilon\right)}{\Gamma^2\left(-n\epsilon\right)}\frac{(n-1)}{2}\epsilon\int dy\, y^{-1-n\eps}(1-y)^{-1-n\eps}\left((1-z)^{-2n\eps}+\mathcal{O}((1-z))\right).
\end{align}
We again extracted a common prefactor with $I_1$ and $I_2$ and we observe that the expression is proportional to $\epsilon$, which means that this expression is subleading in the logarithmic expansion. 

We potentially have a fourth integral that contributes and it is given by
\begin{align}
    I_4 &\equiv \int\left[\prod_{i=1}^n dk_i^+dk_i^-(k_i^+k_i^-)^{\frac{d-6}{2}}\right] \int dy \left(-\frac{2(1-z)}{1+z}\right)\frac{dT}{2\pi i} \frac{d\tau}{2\pi i}\, \left(-\sum_{i=1}^nk_i^+\right)\nonumber \\
    &\quad\times e^{T(1-z)}e^{\tau\sqrt{\s/2}(e^{2\eta(y)}-1)}\prod_{i=1}^n e^{k_i^+(-\sqrt{2/\s}T+\tau)}e^{k_i^-(-\sqrt{2/\s}T-e^{2\eta(y)}\tau)}\left(\frac{2T}{\s}\sum_{i<j}(k_i^+k_j^-+k_i^-k_j^+)\right).
\end{align}
Looking at the delta function in \eqn{DYNLPPSint}, which reads
\begin{equation}
    \delta\left(1-z-\sqrt{\frac{2}{\s}}\sum_i(k_i^++k_i^-)+\frac{2}{\s}\sum_{i<j}k_i\cdot k_j\right),
\end{equation}
we argue that both $\sum_i k_i^+$ and $\sum_{i<j}k_i\cdot k_j$ are at least of order $\mathcal{O}(1-z)$, which means that the result from integral $I_4$ is of order $\mathcal{O}\left((1-z)^2\right)$, and hence that it does not contribute at the required accuracy.\footnote{Upon carrying out the calculation, we indeed found a contribution starting from NNLP.} Adding the three NLP contributions, one obtains
\begin{align}
    I_1+I_2+I_3 &= -2^{-1+n\epsilon}\s^{-n\epsilon}\frac{\Gamma^{2n}\left(-\epsilon\right)}{\Gamma^2\left(-n\epsilon\right)}\int dy\,y^{-1-n\eps}(1-y)^{-1-n\eps}(1-z)^{-1-2n\epsilon}\nonumber\\
    &\quad\times\bigg(1+y(1-z)-y(1-z)-\epsilon\frac{(n-1)}{2}(1-z)+\mathcal{O}\left((1-z)^2\right)\bigg).
\end{align}
We immediately see that the NLP contribution of the phase space integral vanishes for the leading logarithms, which is the reason that we can resum the cross section in Section \ref{subsec:DYgamma} using the steps performed there.

\bibliography{spire.bib}
\bibliographystyle{JHEP.bst}
\end{document}